\documentclass[11pt]{book}
\usepackage{graphicx,amsmath,amssymb,amsfonts,fullpage,subfigure,color}
\usepackage{multirow,bigdelim}
\usepackage[font=small,labelfont=bf]{caption}

\providecommand{\HHH}{\mathbf{H}}

\providecommand{\GGG}{\mathbf{G}}
\providecommand{\III}{\mathbf{I}}

\providecommand{\MMM}{\mathbf{M}}

\providecommand{\QQQ}{\mathbf{Q}}
\providecommand{\RRR}{\mathbf{R}}

\providecommand{\hhh}{\mathbf{h}}

\providecommand{\kkk}{\mathbf{k}}

\providecommand{\qqq}{\mathbf{q}}
\providecommand{\rrr}{\mathbf{r}}

\providecommand{\vvv}{\mathbf{v}}

\providecommand{\0}{\mathbf{0}}
\providecommand{\vare}{\varepsilon}

\providecommand{\dd}{\partial}
\providecommand{\percent}{\%{}}

\providecommand{\up}{\uparrow}
\providecommand{\dn}{\downarrow}

\providecommand{\cccc}{\hat{c}^{\phantom\dag}}
\providecommand{\cdag}{\hat{c}^\dag}
\renewcommand{\Re}{\mathop{\mathrm{Re}}}
\renewcommand{\Im}{\mathop{\mathrm{Im}}}

\DeclareMathOperator*{\Tr}{Tr}

\DeclareMathOperator*{\sgn}{sgn}

\providecommand{\half}{\tfrac{1}{2}}
\providecommand{\abs}[1]{\left\lvert{#1}\right\rvert}
\providecommand{\mean}[1]{\left\langle #1 \right\rangle}

\providecommand{\ket}[1]{\left|#1\right>}
\providecommand{\bra}[1]{\left<#1\right|}

\providecommand{\pmat}[1]{\begin{pmatrix} #1 \end{pmatrix}}
\providecommand{\psmat}[1]{\left(\begin{smallmatrix} #1 \end{smallmatrix}\right)}

\providecommand{\red}[1]{{\color{red}{#1}\color{black}}}
\providecommand{\blue}[1]{{\color{blue}{#1}\color{black}}}

\providecommand{\todo}[1]{}             
\providecommand{\optional}[1]{}
\providecommand{\editorcheck}[1]{}

\providecommand{\todo}[1]{\red{#1}}     
\providecommand{\optional}[1]{\blue{#1}}
\providecommand{\editorcheck}[1]{\blue{#1 [Editor: please verify] }}

\providecommand{\onlinecite}[1]{\cite{#1}}

\newcommand{\eone}[1]{\left\langle #1 \right\rangle}
\newcommand{\expect}[3]{\left\langle #1 \left| #2 \right| #3 \right\rangle}

\newcommand{\lr}[1]{\left\langle#1\right\rangle}

\newcommand{\be}{\begin{equation}}
\newcommand{\ee}{\end{equation}}

\newcommand{\vc}{\mathbf}

\newcommand{\hatad}{\hat{a}^{\dag}}
\newcommand{\hataa}{\hat{a}^{\phantom\dag}}
\newcommand{\hatn}{{\hat{n}}}
\newcommand{\hatx}{{\hat{x}}}

\newcommand{\hatH}{{\hat{H}}}
\newcommand{\bea}{\begin{eqnarray}}
\newcommand{\eea}{\end{eqnarray}}

\newcommand{\calh}{{h}}
\newcommand{\calhhat}{{\hat{h}}}

\begin{document}

%

\begin{center}
{
\LARGE\textbf{
Optical Lattice Emulators: \\
Bose and Fermi Hubbard Models
}
}\\

\bigskip
{
\Large
Eric Duchon$^1$, Yen Lee Loh$^2$, and Nandini Trivedi$^1$, 
}\\

\medskip

Chapter in the book ``Novel Superfluids'', Part B, ed. Ketterson and Bennemann, OUP.
\\
\medskip

$^1$Department of Physics, The Ohio State University, 191 W
Woodruff Avenue, Columbus, OH  43210
\\$^2$Department of Physics and Astrophysics, University of North Dakota,
101 Cornell St Stop 7129, Grand Forks, ND  58202
\\
\end{center}

\begin{center}
\textbf{Abstract}
\end{center}

The grand challenge of condensed matter physics is to understand the emergence of novel phases arising from the organization of many degrees of freedom, especially in regimes where particle interactions dominate over the kinetic energy. Brought to the forefront by the discovery of high-temperature superconductivity in complex copper-based oxides, it is absolutely astounding that a simple model, like the Hubbard model, is able to capture so many essential aspects of the physics of these materials. However, whether the repulsive fermion Hubbard model really contains a d-wave superconducting ground state is still an open question after decades of study. 

\begin{figure*}[htb]
  \centering
  \includegraphics[width=0.7\textwidth]{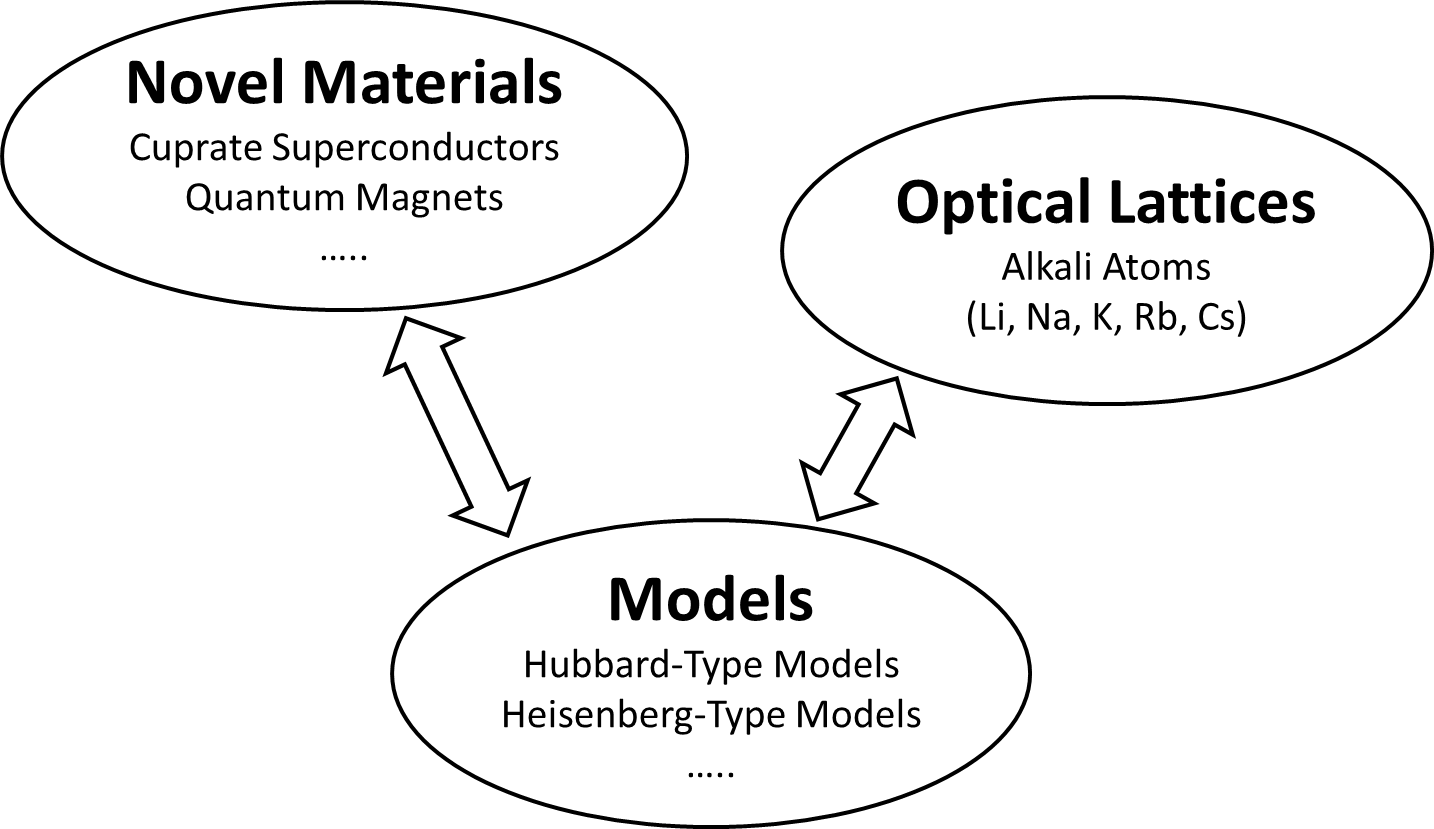}
  \caption{Quantum materials showing novel phenomena such as high Tc superconductivity or spin liquids can often be described by simple looking but hard to solve models with charge, spin and orbital degrees of freedom.
  Such models can be emulated by neutral atoms in optical lattices
  thereby opening up an unexpected connection between condensed matter physics and cold atoms.}
  \label{fig:materialsModelsAtoms}
\end{figure*}

Optical lattices, with a unique ability to tune interactions and density, have emerged as unusual laboratories for making realizations of such Bose and Fermi Hubbard and 
Heisenberg-type models\cite{bloch2005,blochReview2008,xu2006,greiner2002,chin2006,gunter2006,ospelkaus2006,schneider2008,jordens2008} and observing phase transitions without the uncertainty posed by complex materials. Against this backdrop, quantum Monte Carlo simulations in strongly interacting regimes are emerging as an important bridge between materials-based condensed matter physics and cold atoms, highlighted in Figure \ref{fig:materialsModelsAtoms}.

With up to $10^6$ bosons, the scale of modern numerical simulations is able to match the scale of experimental cold atom systems thereby allowing a direct comparison.
In this chapter we report on the progress made in mapping out the finite temperature phase diagram for strongly correlated Bose and Fermi Hubbard models, both textbook examples of interacting Hamiltonians. These models 
show quantum phase transitions as a function of tuning the interaction and density. The primary challenges are:
 (i) clear diagnostics for phase identification; (ii) extracting information from coexisting phases in the same confining potential; and (iii) thermometry and equilibration in the optical lattice. 

The Bose Hubbard model is the quintessential model to describe the depletion of a Bose-Einstein condensate with increasing repulsion, which ultimately causes the condensate to vanish at a critical point and the Mott insulator to emerge at stronger repulsion. The Fermi Hubbard model with increasing attractive interactions is the prototypical model to capture the BCS-BEC crossover from a regime of large overlapping Cooper pairs to tightly bound bosonic pairs, and to describe the formation of a pseudogap at finite temperature that signifies pairing correlations without long range phase coherence. With population imbalance, the same model can capture the physics of modulated superfluids -- the long-sought-after Fulde-Ferrell-Larkin-Ovchinnikov phase. Repulsive interactions 
bring in the richness of antiferromagnetism, the evolution from spin-density wave insulator at half-filling due to band structure nesting to Heisenberg interactions between local moments,
the possibility of unusual d-wave superfluids upon doping, and connections with high-$T_c$ superconductivity. There is a lot of life in Hubbard models!

This chapter is a pedagogical review of the Hubbard model for bosons with repulsion and for fermions with attraction and repulsion primarily using
two methods, one chosen for its simplicity and insights (mean field theory) and the other chosen for its accuracy and reliability (quantum Monte Carlo methods). From a comparison of the two 
methods we glean valuable information into the effects of fluctuations that dominate quantum phase transitions. The chapter includes an in-depth comparison with experiments.
We conclude with a discussion of future developments where the technical methods expounded on here, mean field theory and quantum Monte Carlo, could be useful.

\tableofcontents

\setcounter{chapter}{0}
\chapter{Optical Lattices}

\section{Introduction}

Important experimental developments in atomic and molecular physics, such as laser cooling using magneto-optical traps \cite{phillips1997}, laser based precision spectroscopy such as optical frequency combs \cite{glauber2005}, 
and the ability to manipulate individual entangled atoms and photons \cite{haroche2012}, have opened up many new connections with condensed matter physics and with the emerging field of quantum information. 

\begin{figure*}[b]
  \centering
  \includegraphics[width=0.7\textwidth]{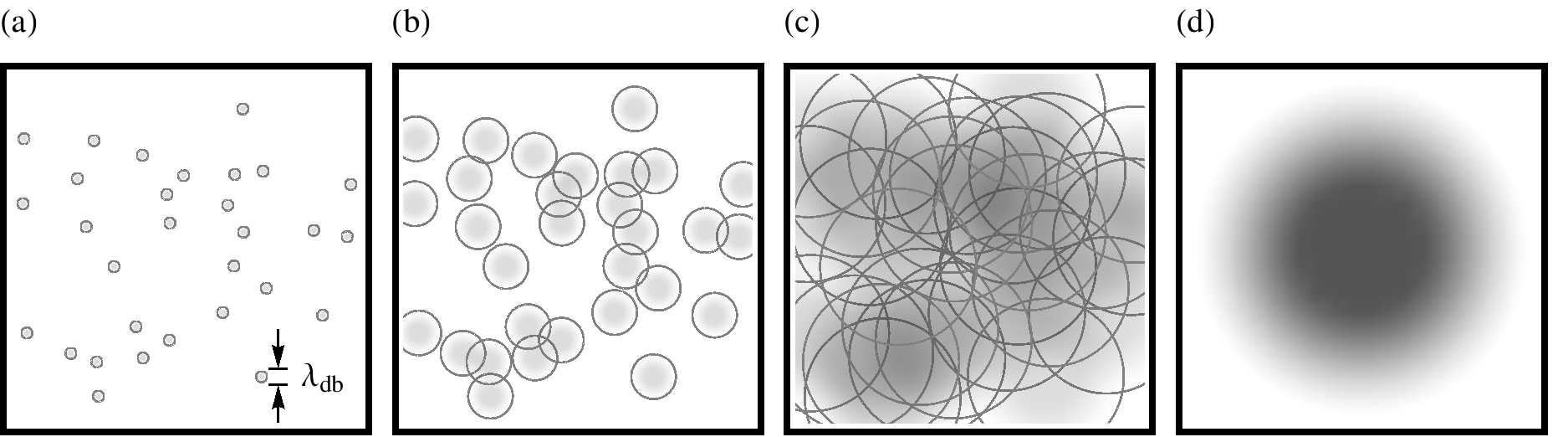}
  \caption{The importance of particle statistics and quantum interactions increase as the temperature is decreased, shown schematically in (a-d). At low temperature, the de Broglie wavelength $\lambda_{\rm db}$ becomes comparable to the interparticle spacing and eventually (d) the bosons condense into a single state.}
  \label{fig:BECschematic}
\end{figure*}

A major breakthrough was the observation of Bose-Einstein condensation (BEC) in 1995 by Cornell, Ketterle, and Wieman \cite{cornell2001}. As shown schematically in Figure \ref{fig:BECschematic}, for a fixed density $n$ of atoms in a trap with an average inter-particle spacing 
$r_o\sim n^{-1/3}$ , the de Broglie wavelength 
$\lambda_{\rm db} =h/\sqrt{2\pi mk_B T}$ increases as the temperature is lowered. The BEC temperature $T_c$ is defined by the condition $\lambda_{\rm db} \approx r_o$ and signals the temperature at which wave functions of different particles just start to overlap and 
statistics becomes important (see Figure \ref{fig:BECschematic}(b)). For Rb$^{87}$, the very first system to show a BEC, $T_c\approx 1\,\mu$K for about $10^5$ atoms in a magnetic trap with frequency $\omega=2\pi \times 24\,$Hz.
The stringent requirement of sub-microkelvin temperatures gives an idea of why it took so long to produce an experimental realization of a BEC. It necessitated a paradigm shift 
from $^3$He-based dilution refrigerators and adiabatic demagnetization cooling methods to laser cooling techniques.
Upon lowering the temperature further, more and more atoms condense into the same state (see Figure \ref{fig:BECschematic}(c)) and ultimately form a giant ``atom'' described by a single wave function, a coherent state of a million atoms (see Figure \ref{fig:BECschematic}(d)). One could think of 
this giant atom as a coherent laser of matter waves.
While the phenomena of BEC had been predicted by Bose and Einstein in the 1920s,
the experimental observation almost 70 years later opened up new avenues for research into the effect of interactions in depleting the condensate and even more interestingly into the possibility of driving an interaction-tuned quantum phase transition from a superfluid to a Mott insulator. 

\begin{figure*}[t]
  \centering
  \includegraphics[width=0.7\textwidth]{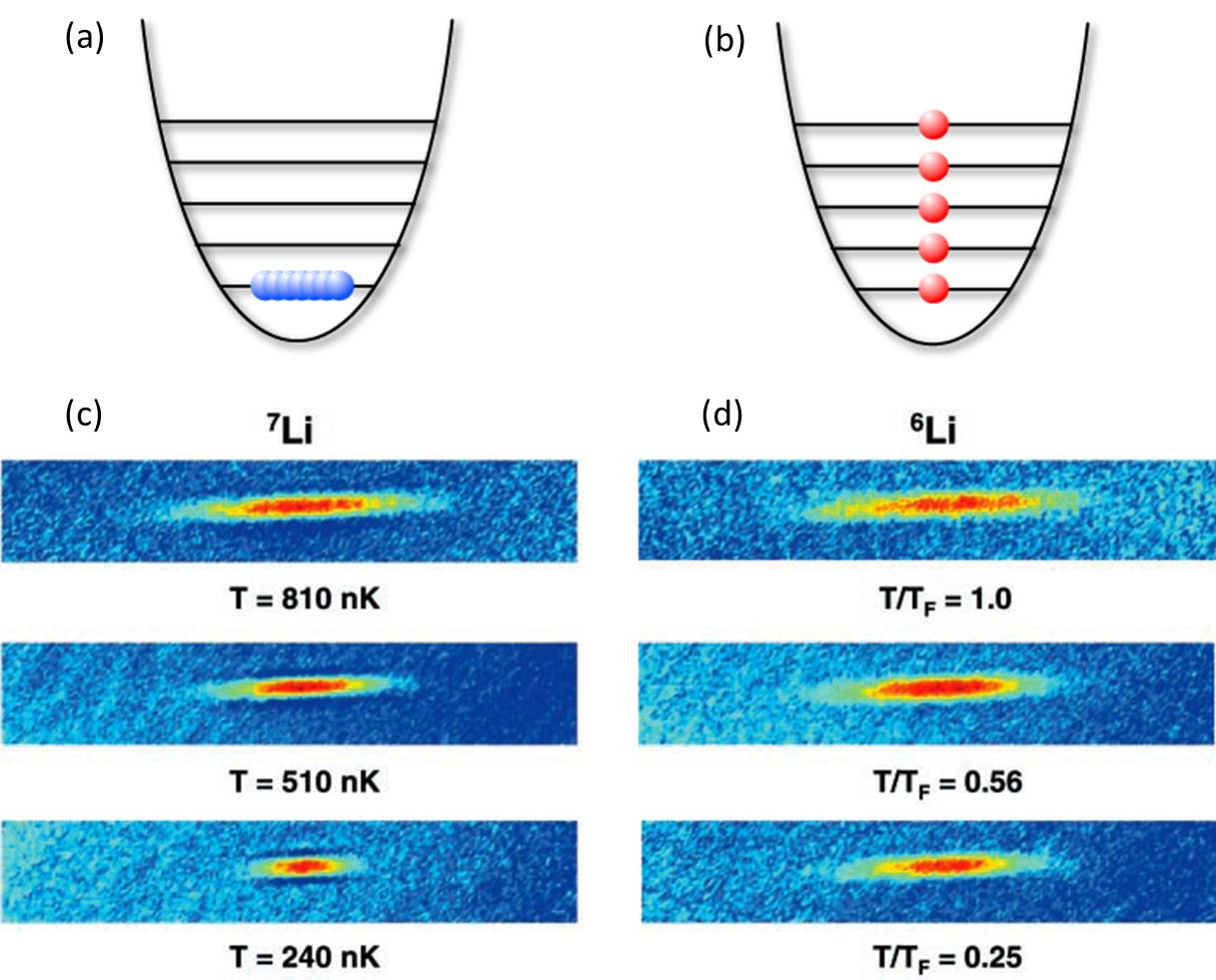}
  \caption{Comparison of bosons ($^7$Li) and fermions ($^6$Li) near the condensation and Fermi temperature, respectively. The atom density is imaged after a fixed expansion time and the quantum statistics of each species revealed.}
  \label{fig:HuletExptPic}
\end{figure*}

The ultimate fate of neutral atoms is decided by whether the number of neutrons in the nucleus is even or odd. For example, $^{87}$Rb has 37 electrons, 37 protons,
and 50 neutrons, so that the composite atom is a boson. Similarly $^7$Li (3 electrons, 3 protons, 4 neutrons) is a boson whereas the isotope $^6$Li, with one less neutron, is a fermion.
The difference shows up beautifully as two atomic clouds of $^7$Li and $^6$Li are cooled below their degeneracy temperatures. 
The velocity distributions of the bosonic and fermionic atomic clouds are captured within time-of-flight images in
Figure~\ref{fig:HuletExptPic} and clearly show that whereas the bosonic cloud size keeps shrinking because of a large pile-up of atoms forming a condensate at zero velocity in the center of the cloud,
the fermionic cloud does not compress below a certain size due to the Fermi pressure, whose magnitude is determined by the number of atoms. This is the canonical behavior of non-interacting bosonic and fermionic systems.

The next important question is: what is the role of interactions?  This question has been addressed for continuum gases with interactions tuned by magnetic field across a Feshbach resonance. 
In this chapter we will address it in the context of an optical lattice, represented by a single-band Hubbard model for bosons or fermions, where it is possible to drive quantum phase transitions by tuning the intensity of the optical lattice.

In the following sections we describe how the parameters of the Hubbard model, the tunneling and interaction parameters, can be obtained quantitatively in terms of the strength 
and periodicity of the optical lattice potential, tuned by the laser intensity and wavelength. Armed with this direct link between the parameters of a theoretical model and the 
actual experimental optical lattice, we then turn to the phase diagram of the Bose Hubbard model. We discuss the 
definition of the order parameter as an expectation value of the
annihilation operator in a coherent state 
and contrast this phase-coherent superfluid state with the phase-incoherent Mott state
naturally defined in terms of number states. Therein lies the tussle between coherent states and number states that ultimately drives a quantum phase transition as interactions are increased.
Following the repulsive Bose Hubbard model, we then discuss the phase diagram of the Fermi Hubbard model with both attractive and repulsive interactions. 


\subsection{Linking Experiment to Theory: Calculating the Effective Hubbard parameters}\label{sec:parameters}

\begin{figure*}[h]
  \centering
  \includegraphics[width=0.5\textwidth]{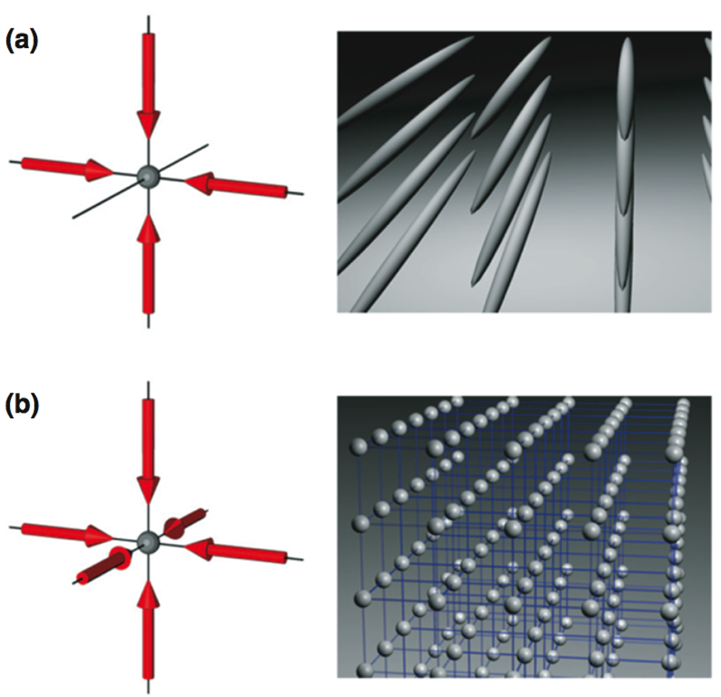}
  \caption{Counterpropagating laser beams impose periodic potentials on cold atom clouds. The atoms can be arranged into two-dimensional pancakes, (a) one-dimensional tubes, or (b) lattice structures.}
  \label{fig:latticeSchematic}
\end{figure*}
Cold atom clouds can be diced into systems of reduced dimension by laser standing waves.
For example, two counter-propagating laser beams create a stack of 2D potentials, 4 beams create an array of 1D tubes, and 6 beams create a 3D optical lattice of point-like potentials 
(see Figure \ref{fig:latticeSchematic}).
Here, consider a 3D optical lattice formed by counterpropagating linearly polarized laser beams of wavelength $\lambda=2\pi/k$, which form a standing wave with nodes separated by a distance $a_{\text{latt}}=\lambda/{2}$.  The laser's electric field acts on atoms to produce a quasi-static potential according to the ac Stark effect.  If the laser frequency is slightly lower than the transition frequency between two atomic levels (red-detuned), the atoms are attracted towards high intensities, i.e., the potential has minima at the intensity maxima; conversely, for a blue-detuned laser, the potential has minima at intensity minima.  In either case one has a sinusoidal potential of spatial period 
$a_{\text{latt}}=\lambda/{2}$.  
It is conventional to work in units of the recoil energy 
$E_R = \frac{\hbar^2k^2}{2m} = \frac{2\pi^2 \hbar^2}{m\lambda^2}
= \frac{\pi^2 \hbar^2}{2m a_{\text{latt}}^2}
$, which is the change in kinetic energy of the atom associated with the emission or absorption of a photon with momentum $k$.
\cite{blochReview2008}

\begin{figure}[!htbp]
	\subfigure[
		Dispersion relation
		]{
		\includegraphics[width=0.48\textwidth]{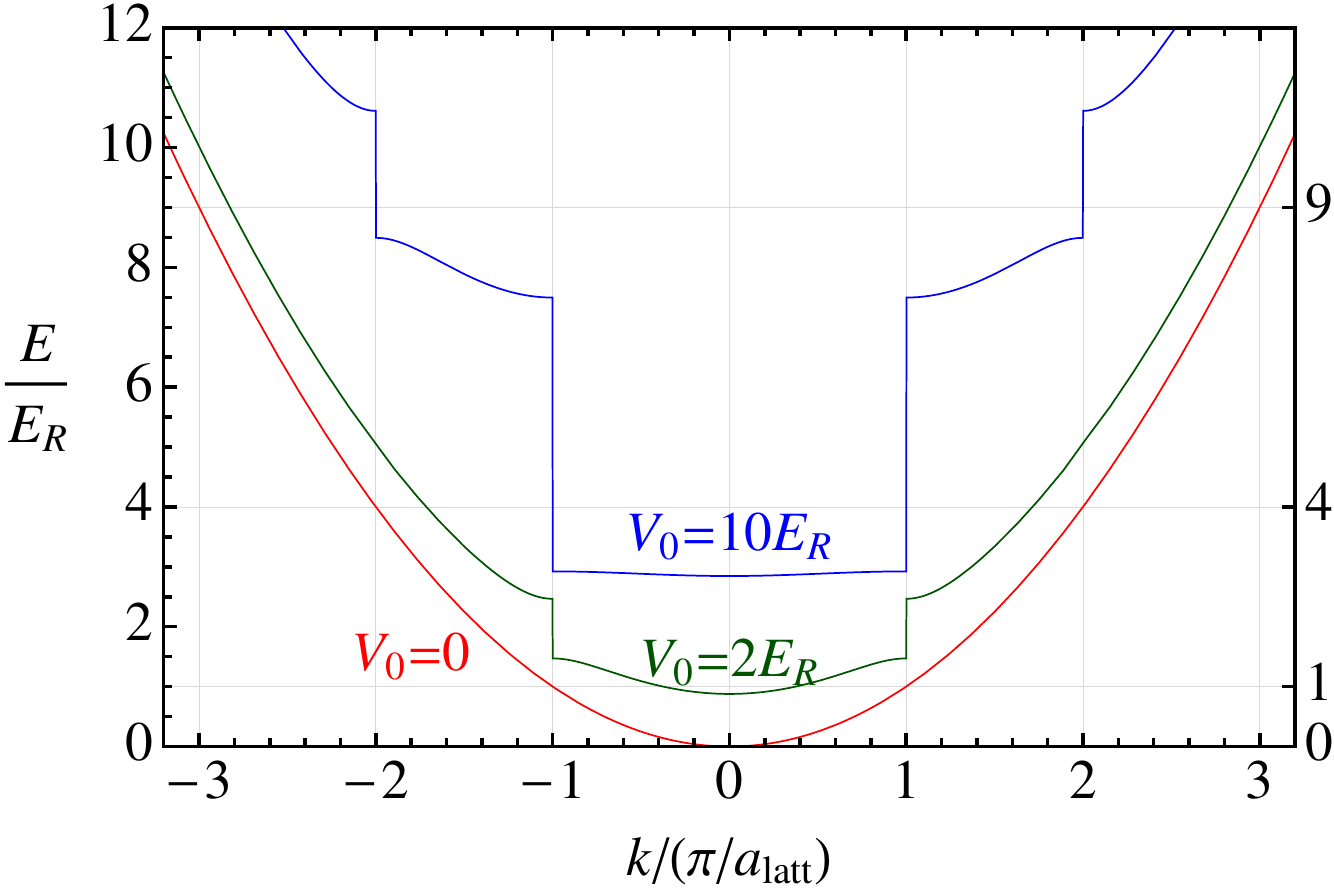} \label{MathieuDispersion}
	}
	\subfigure[
		Bandstructure as a function of lattice depth $V_0$
		]{
		\includegraphics[width=0.48\textwidth]{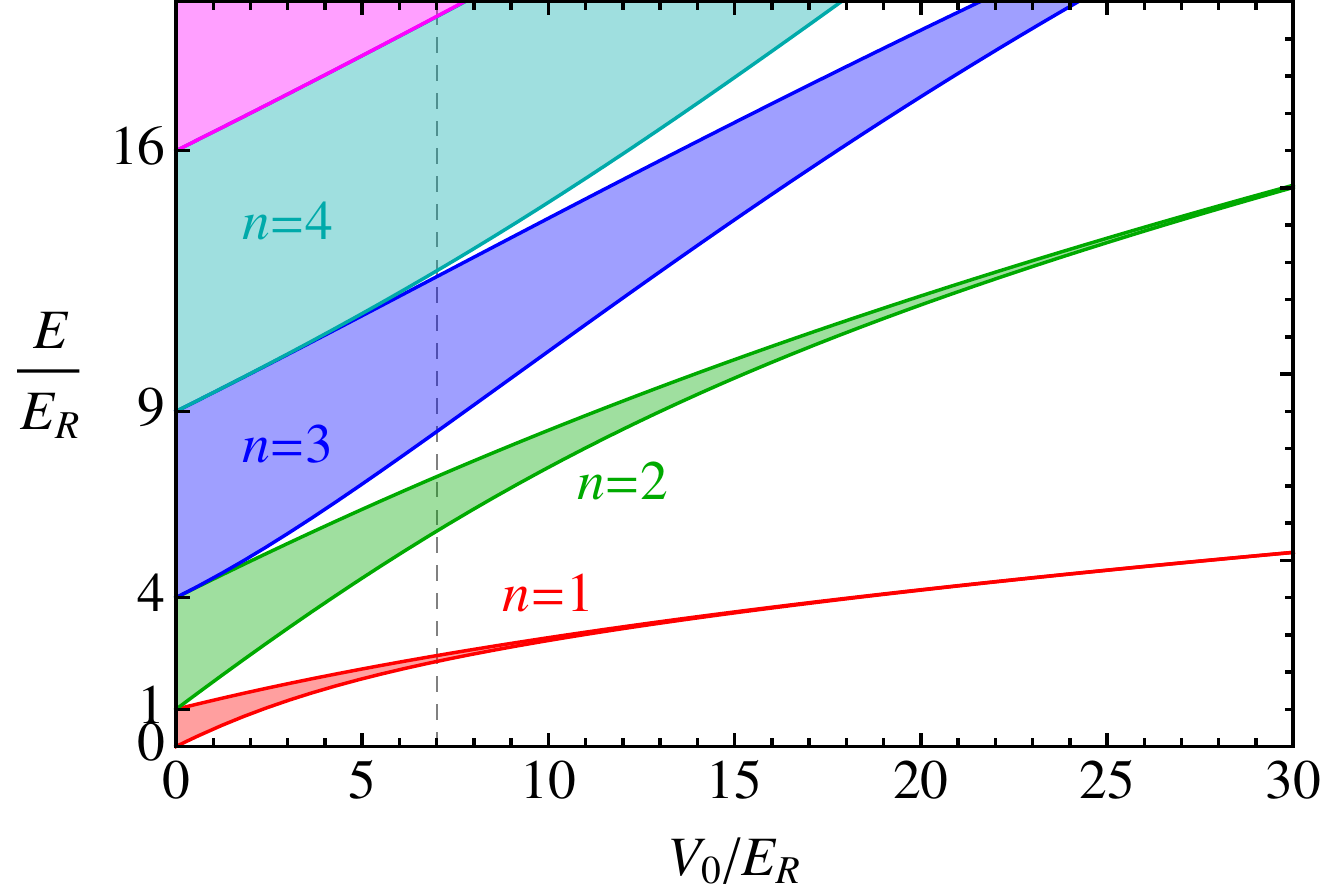} \label{MathieuBandgaps}
	}
	\subfigure[
		Bandstructure relative to potential
		]{
		\includegraphics[width=0.48\textwidth]{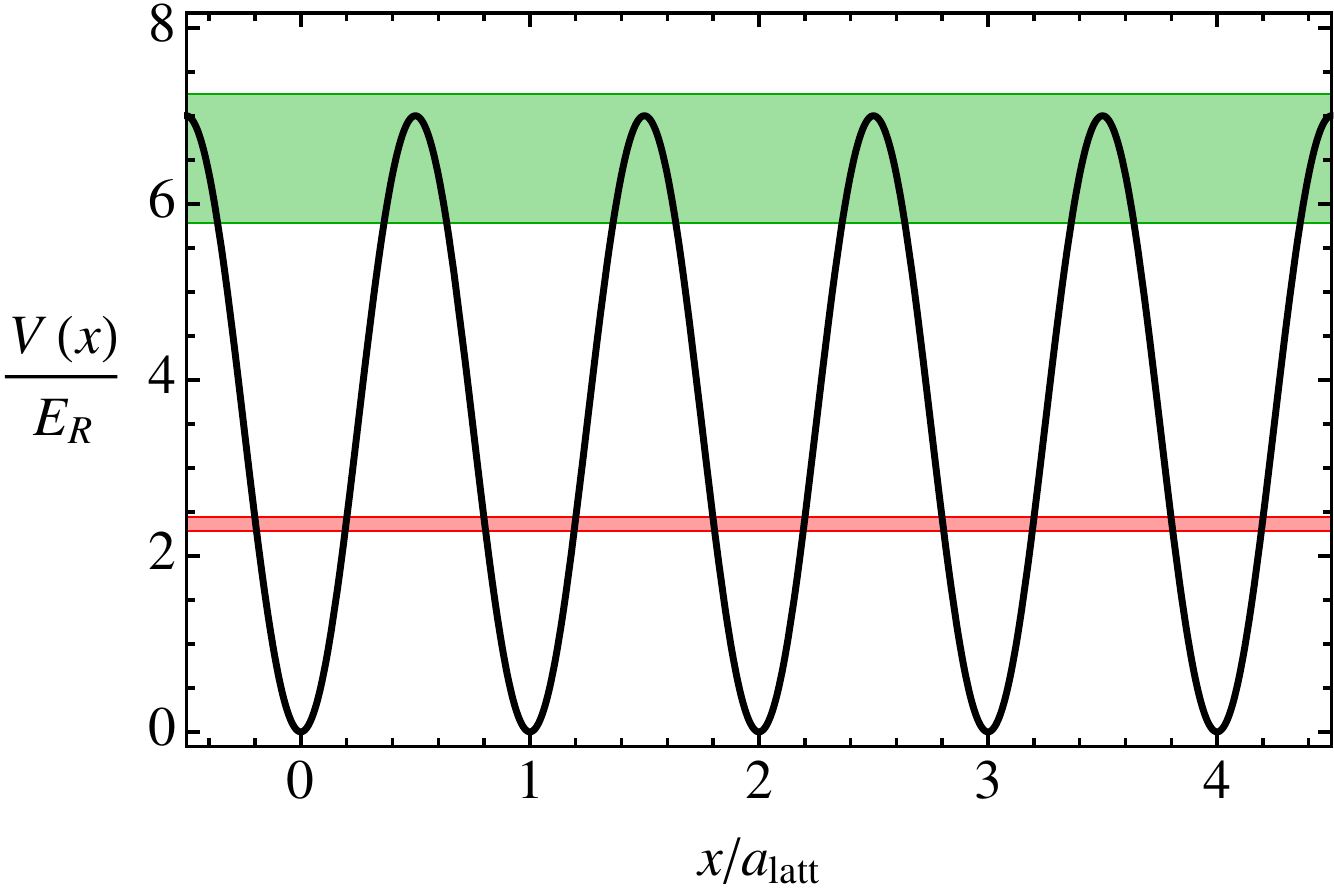} \label{MathieuBandsInPotential}
	}
	\subfigure[
		Bloch and Wannier functions
		]{
		\includegraphics[width=0.48\textwidth]{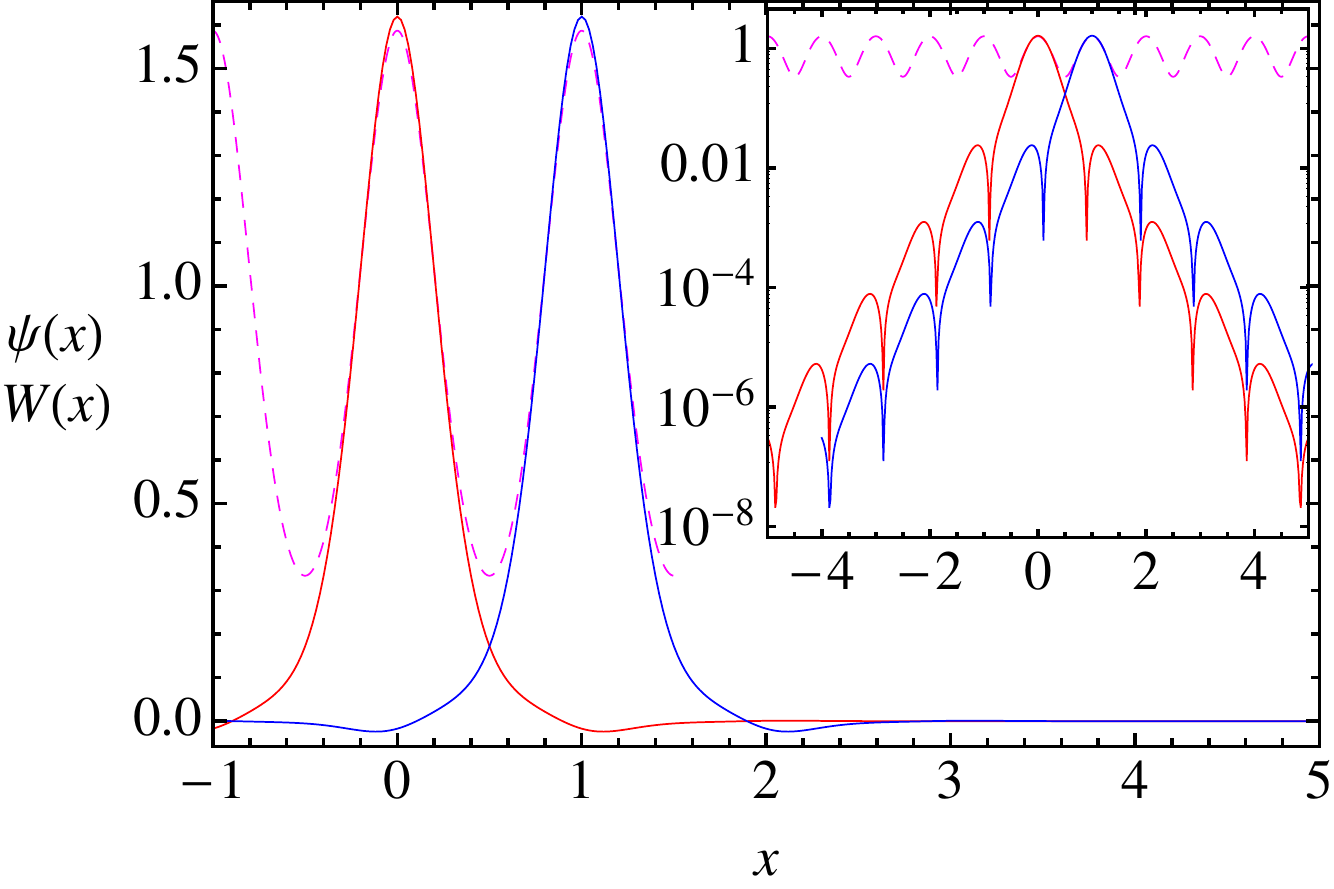} \label{WannierFunctions}
	}
	\subfigure[
		Bloch functions for various $k$
		]{
		\includegraphics[width=0.31\textwidth]{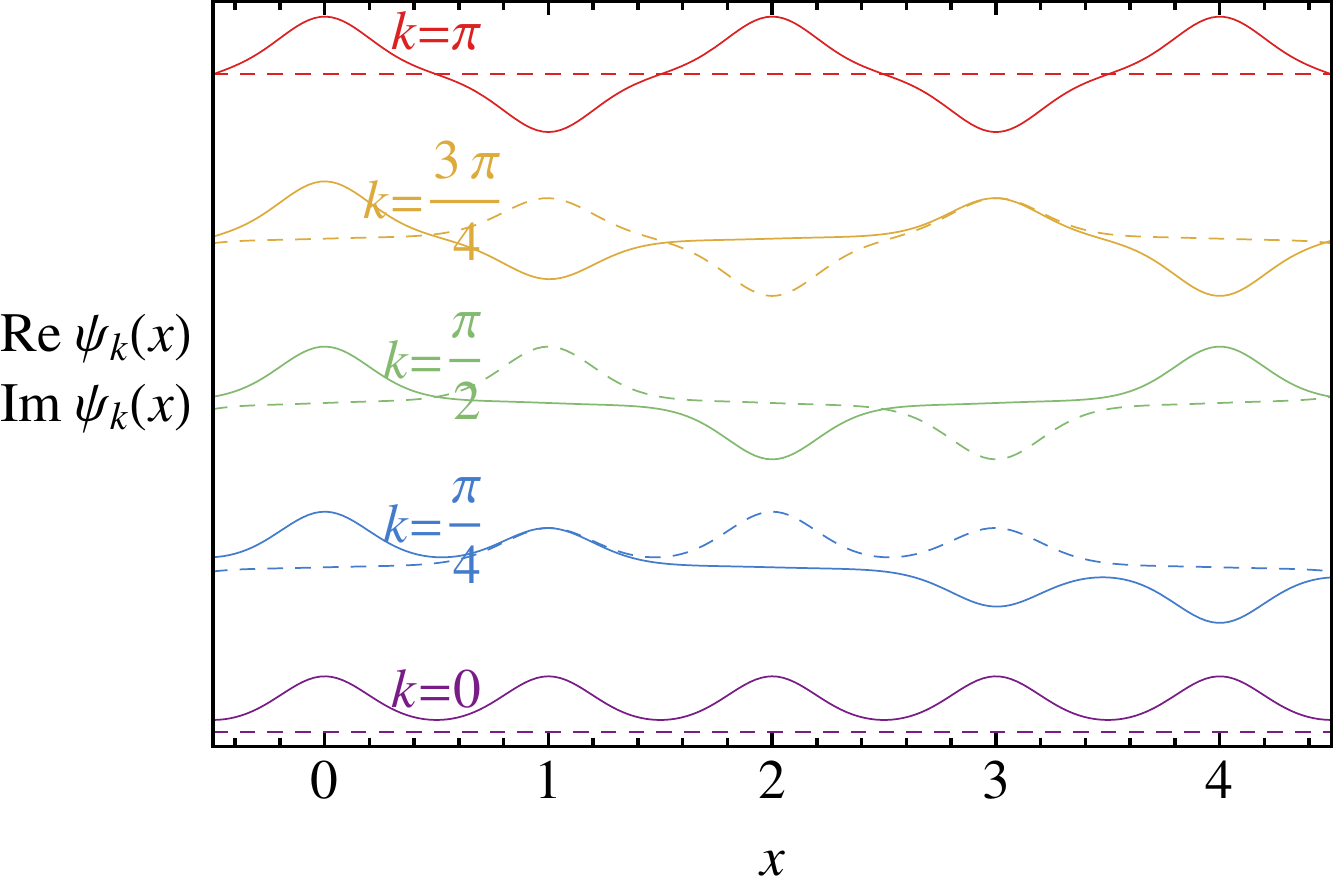} \label{BlochFunctions}
	}
	\subfigure[
		Amplitude
		]{
		\includegraphics[width=0.31\textwidth]{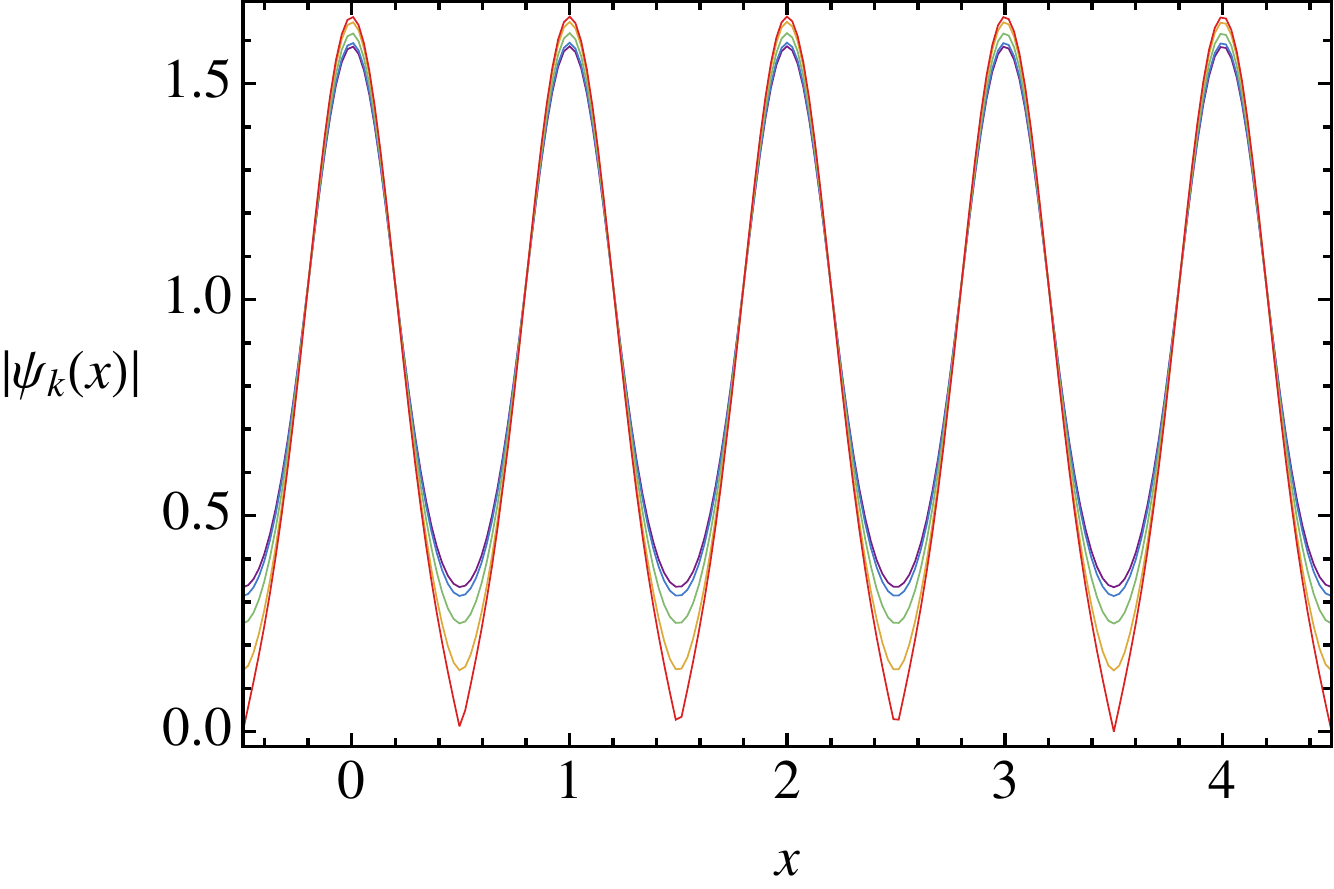} \label{BlochFunctionsAmplitude}
	}
	\subfigure[
		Phase
		]{
		\includegraphics[width=0.31\textwidth]{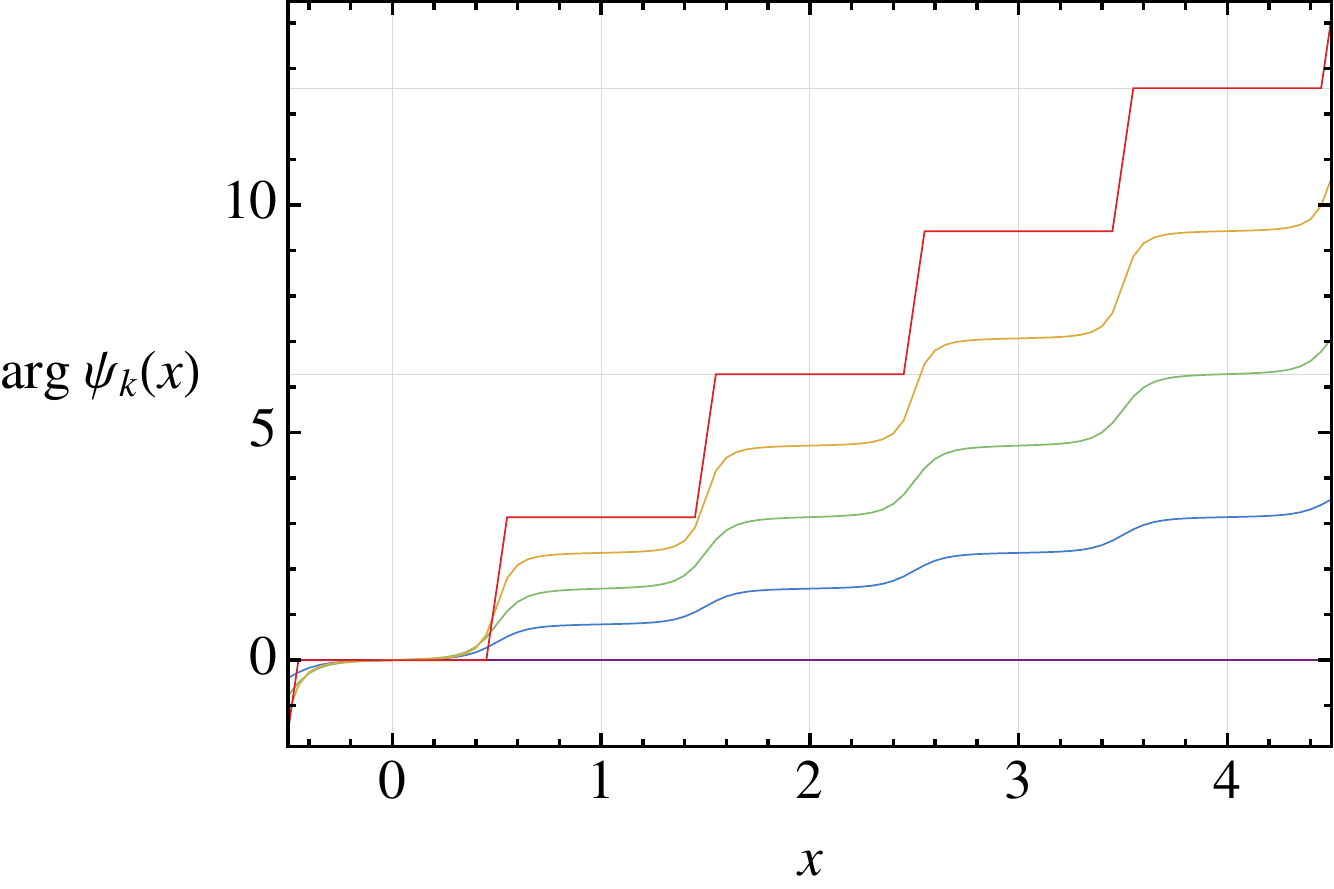} \label{BlochFunctionsPhase}
	}
	\caption{
		\label{Mathieu}
		Quantum mechanical description of a particle moving in a 
			1D sinusoidal optical lattice potential.
		(a) Dispersion relation $E(k)$ in the extended zone scheme.
		(b) Development of energy bands and bandgaps with increasing lattice depth $V_0$.
		(c) Energies of the two lowest bands for $V_0=7E_R$.
		(d) Lowest Bloch function $\psi_0(x)$ (dashed) 
				and two adjacent Wannier functions $W(x)$ and $W(x-a_{\text{latt}})$
			  for $V_0=7E_R$.  Inset shows logarithmic plot.
		(e,f,g) Bloch functions $\psi_k(x)$ 
			at crystal momenta $k a_{\text{latt}}=0, \frac{\pi}{4}, \frac{\pi}{2}, \frac{3\pi}{4}, \pi$
			 for $V_0=7E_R$.
	}
	\end{figure}

In the analysis below, for simplicity we consider a particle of mass $m$ in a one-dimensional sinusoidal potential of depth $V_0$ and spacing $a_{\text{latt}}$.  The wavefunction satisfies the Schr{\"o}dinger equation
	\begin{align}
	-\frac{\hbar^2}{2m} \frac{\dd^2 \psi}{\dd x^2}	
	+ \left( V_0 \sin^2 \frac{\pi x}{a_{\text{latt}}}   \right) \psi(x)
	&= E \psi(x).
	\end{align}	
This is a form of Mathieu's differential equation.  The eigenenergies can be written in terms of Mathieu characteristic value functions as
	\begin{align}
	\tfrac{E_k}{E_R}
	&=
		\alpha
		+ \tfrac{V_0}{2E_R}
	\end{align}
where $\alpha = \mathtt{MathieuCharacteristicA}[\tfrac{k a_{\text{latt}}}{\pi}, -\tfrac{V_0}{4E_R}]  $
(for concreteness we refer to the Mathieu functions as defined in \textit{Mathematica}).
This bandstructure is illustrated in Figure~\ref{MathieuDispersion} and Figure~\ref{MathieuBandsInPotential} as a function of crystal momentum $k$ (in the extended zone scheme) and lattice depth $V_0$.  Eigenfunctions of crystal momentum (Bloch functions) can be written as linear combinations of even and odd Mathieu functions:
	\begin{align}
	\psi_{k} (x)
	&=
		\mathtt{MathieuC}[\alpha, -\tfrac{V_0}{4E_R}, \tfrac{\pi x}{a_{\text{latt}}}]  
	+ (i \sgn k)
		\mathtt{MathieuS}[\alpha, -\tfrac{V_0}{4E_R}, \tfrac{\pi x}{a_{\text{latt}}}] 
		. 
	\end{align}
These Bloch functions are illustrated in Figure~\ref{BlochFunctions}.  It can be verified that they satisfy $\psi_{k} (x+a_{\text{latt}}) = e^{i k a_{\text{latt}}} \psi_{k} (x)$, and that they are normalized such that
	\begin{align}
	\int_0^1 dx~  \abs{ \psi_{k} (x)  }^2 &= 1 .
	\end{align}

The Schr{\"o}dinger equation can be approximated by a tight-binding model by representing the Hilbert space using a basis of localized Wannier functions.  The lowest-band Wannier function $W(x)$ can be constructed by taking a linear combination of Bloch functions, with phase factors chosen to give constructive interference near $x=0$ and destructive interference elsewhere.  As proved in Ref.~\onlinecite{kohn1959}, there is a unique form for the ``correct'' Wannier function; in this case it is
	\begin{align}
	W(x) 
	&=\int_{-\pi}^\pi \frac{dk}{2\pi}~  	\psi_{k} (x).
	\end{align}
The Wannier functions decay exponentially, and they are orthonormal such that
	\begin{align}
	\int_{-\infty}^\infty dx ~ W(x - x_1)~ W(x - x_2)
	&=\delta_{x_1 x_2}
	\end{align}
where $x_1,x_2=0,\pm a_{\text{latt}},\pm 2a_{\text{latt}},\dotsc$.  See Figure~\ref{WannierFunctions}.  There is nevertheless a finite tunneling matrix element between adjacent sites,
	\begin{align}
	t &= \bra{W_{x_1}} \hat{H} \ket{W_{x_2}} .
	\end{align}
The value of $t$ can be computed	 by evaluating the matrix element as integrals over Wannier functions with the real-space Hamiltonian.  However, it is easier to obtain $t$ by equating the bandwidth of the one-dimensional tight-binding model to the bandwidth in terms of Mathieu functions (see Figure~\ref{MathieuBandgaps}):
	\begin{align}
	4t &= E_{n,\pi} - E_{n,0}
	\nonumber\\
	\therefore\quad
	\frac{t}{E_R} &= \tfrac{1}{4}
		\left(
			\mathtt{MathieuCharacteristicB}[1, \tfrac{V_0}{4E_R}]  
		-	\mathtt{MathieuCharacteristicA}[0, \tfrac{V_0}{4E_R}]  
		\right)
		.
	\label{tMathieu}
	\end{align}

For a 3D cubic lattice formed by the sum of three sinusoidal potentials, the wavefunction is separable, and the tunneling amplitude between adjacent sites can be found from the depth of the sinusoidal potential in the relevant direction.  The concepts can obviously be generalized to any lattice, although the eigenfunctions may have to be computed numerically for a general potential.

Now consider pairwise interactions.
We will focus on cold atom systems dominated by $s$-wave scattering (between bosons or between different fermion species), so that we can ignore detailed features of the interatomic potential and work solely with the $s$-wave scattering length $a_s$.  The interatomic potential can then be replaced by a pseudopotential 
$U(\rrr)=g \delta(\rrr)$
where $g = \frac{4\pi\hbar^2 a_s}{2M_r}$ is a point interaction strength
and $M_r$ is the reduced mass.
In the Wannier function basis one thus has an effective Hubbard interaction 
	\begin{align}
	U &=	g \int d^3r~  \abs{  W(\rrr)  }^4
		\\\therefore\quad
	\frac{U}{E_R} 
	&=	 \frac{8 a_{\text{latt}}{}^2 a_s}{\pi} \int d^3r~  \abs{  W(\rrr)  }^4
	=	 \frac{8 a_{\text{latt}}{}^2 a_s}{\pi} \left[  \int dx~  \abs{  W(x)  }^4  \right]^3
	.
	\label{UWannier}
	\end{align}

	\begin{figure}[htb]
	\subfigure[
		]{
		\includegraphics[width=0.48\textwidth]{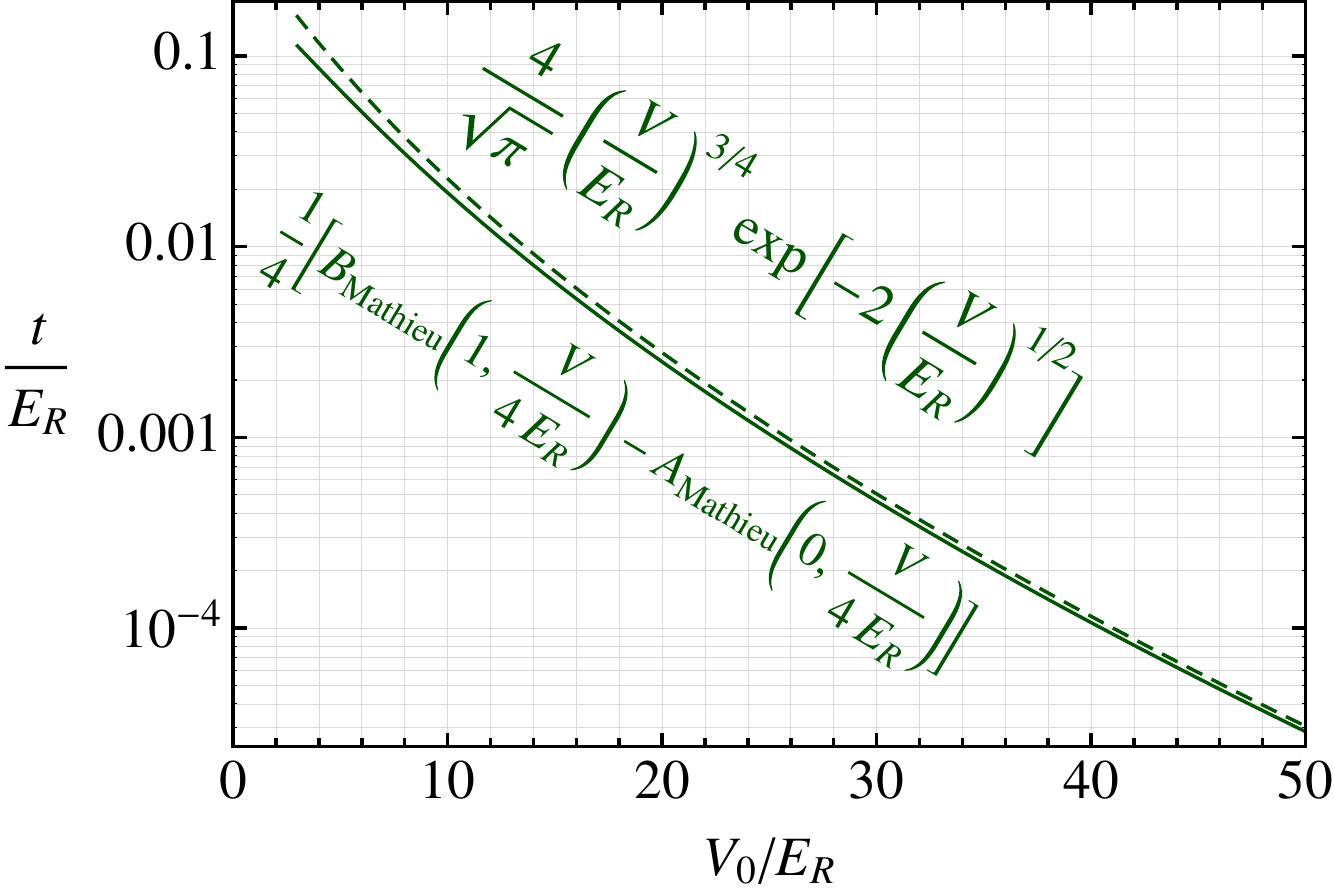} \label{Effectivet}
	}
	\subfigure[
		]{
		\includegraphics[width=0.48\textwidth]{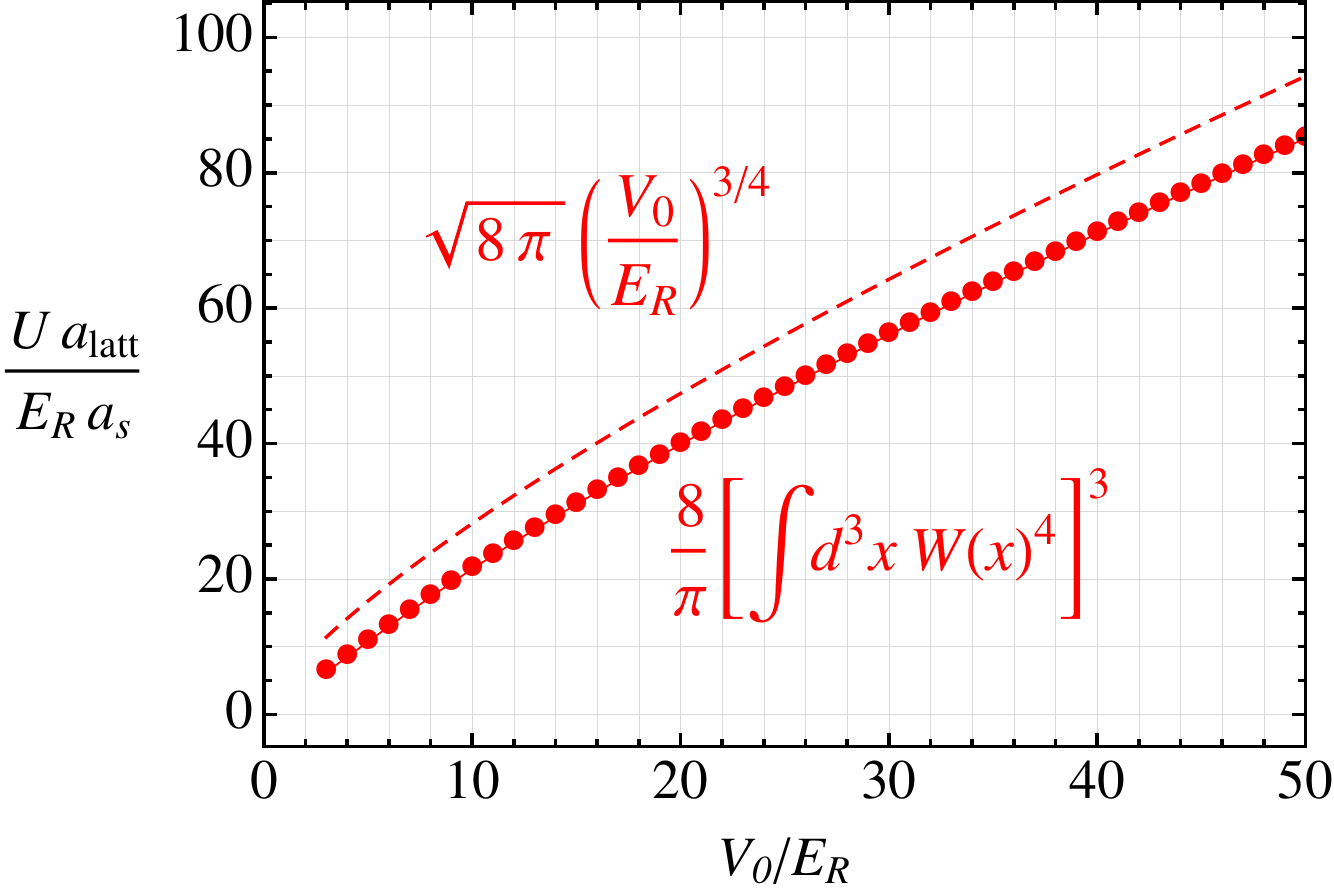} \label{EffectiveU}
	}
	\caption{
		\label{Jaksch}
		Effective hopping amplitude $t$ and Hubbard interaction $U$ for 
		particles in a 3D optical lattice with a sinusoidal potential of depth $V_0/E_R$.
		Upper curves show asymptotic forms.  
		Lower curves are calculated from Mathieu functions and Wannier functions.
	}
	\end{figure}

\paragraph{Asymptotic forms:}
We will focus on optical lattices that are sufficiently deep ($V_0 \gtrsim 6E_R$) so that the system can be treated within a single-band Hubbard description.  It is instructive to study the limit $V_0/E_R \gg 1$.
To find the asymptotic form of $U$, it is sufficient to approximate the sinusoidal potential by a harmonic potential near $x=0$, so that the Bloch functions and Wannier functions resemble a Gaussian near $x=0$:
	\begin{align}
	V(x) &= V_0 \sin^2 \tfrac{\pi x}{a_{\text{latt}}} 
	\approx \tfrac{V_0 \pi^2}{a_{\text{latt}}^2} x^2
	\approx \half m\Omega^2 x^2
		,\qquad
	\Omega^2 = \tfrac{2\pi^2 V_0}{m a_{\text{latt}}^2},
	\nonumber\\
	W(x) &= \pi^{-1/4} {x_0}^{-1/2} \exp \left(  -\tfrac{x^2}{2x_0^2}  \right)
		,\qquad
	x_0 = \left(  \frac{\hbar^2 a_{\text{latt}}^2}{2\pi^2 mV}  \right)^{1/4}
		,\qquad
	\nonumber\\
	\frac{U}{E_R} &= \sqrt{8\pi} \ \frac{a_s}{a_{\text{latt}}} 
			\left(  		  \frac{V_0}{E_R}  \right)^{3/4}   
			.
	\label{UGaussian}
	\end{align}
The tunneling amplitude, however, is strongly influenced by the shape of the tails.  Since the sinusoidal potential is lower than the approximating harmonic potential, the actual wavefunction has more significant tails than a Gaussian (this can be seen from the JWKB approximation).  The correct asymptotic form of $t$ can be obtained from the asymptotic behavior of Eq.~\eqref{tMathieu}:
	\begin{align}
	\frac{t}{E_R}   &\approx \frac{4}{\sqrt{\pi}} 
		\left(  
		  \frac{V_0}{E_R}  \right)^{3/4}   
			\exp  \left( -2\sqrt{\frac{V_0}{E_R}}
	 \right).
 \label{tAsymptotic}
	\end{align}
Equations \eqref{UWannier}, \eqref{UGaussian}, \eqref{tMathieu}, and \eqref{tAsymptotic} are plotted in Figure~\ref{Jaksch}.

\paragraph{Example:}
Consider fermionic ${}^{40}$K atoms ($m=40m_p$), which have D1 and D2 transitions
(${}^1 \mathrm{S} _{1/2}   \rightarrow {}^1 \mathrm{P} _{1/2}$ and
${}^1 \mathrm{S} _{1/2}   \rightarrow {}^1 \mathrm{P} _{3/2}$) 
at 769 nm (390 THz).
Consider a 3D optical lattice formed by six lasers at wavelength $\lambda=830 \mathrm{nm}$,
which is red-detuned from the D transitions.
The lattice spacing is $a_{\text{latt}} = \lambda/2 = 415 \mathrm{nm}$, and  
the recoil energy is $E_R \approx 7240~h\; \mathrm{Hz}$, where $h$ is Planck's constant.

For a laser intensity coresponding to a lattice depth $V_0 = 7 E_R$, 
Eq.~\eqref{tMathieu} gives 
$t = 0.039 E_R \approx 286~h \mathrm{Hz} \approx 14~k_B~\mathrm{nK}$.
This corresponds to hopping on the timescale of milliseconds.

In a magnetic field the two lowest hyperfine states are 
$\ket{F=\tfrac{9}{2}, m_F=-\tfrac{9}{2}}$ and 
$\ket{F=\tfrac{9}{2}, m_F=-\tfrac{7}{2}}$.
For $s$-wave 
scattering 
between two ${}^{40}$K atoms in these two states,
there is a Feshbach resonance at 202.1 G (where $1~\mathrm{G} = 10^{-4}~\mathrm{T}$).
At a field 220 G, the s-wave scattering length is $a_s \approx +110 a_0$ where $a_0$ is the Bohr radius.  
Thus, if the two atoms are in the same well of the optical lattice described above, they experience a Hubbard repulsion given by $\frac{U a_{\text{latt}}}{E_R a_s} \approx 15$, so 
$U \approx 0.21E_R ~h \mathrm{Hz} \approx 74~k_B~\mathrm{nK}$, corresponding to $U/t \approx 5.4$.

There are far more details and subtleties than we can address here; the reader is advised to consult Refs.~\onlinecite{blochReview2008,ulrichSchneiderThesis} for more information.

\section{Bosons in Optical Lattices: Bose Hubbard Model}

The protons, neutrons, and electrons that make up the majority of our everyday surroundings are fermions.  However, they can form composite bosonic objects, such as neutral atoms with even numbers of neutrons or Cooper pairs, which can then Bose condense at low temperatures.  Remarkably, many of these condensed states can be described by the relatively simple Bose Hubbard lattice model (BHM). Some of the initial impetus for its study in the 1980s was provided by helium adsorbed into a Vycor lattice, and by Cooper pairs tunneling between superconducting islands in Josephson arrays \cite{reppy1984,finotello1988,takahide2000}. The achievement of Bose-Einstein condensation in dilute, ultracold atomic gases using magneto-optical techniques and the development of optical lattices significantly broadened the number of physical systems described by the BHM and enabled tunable experimental studies in one, two, and three dimensions \cite{bloch2008}. 
The BHM may even be a suitable low-energy effective theory \cite{fisher1990}
for certain superconductor-insulator transitions where disorder destroys the phase coherence between Cooper pairs without breaking apart individual pairs, leaving a ``bosonic insulator'' \cite{stewart2007,hollen2011,sacepe2011,bouadim2011}.

The development of several statistically-exact numerical techniques for bosonic systems gives physicists an unprecedented opportunity to fully explore a quantum many-body system. Meanwhile, experiments using optical lattices have precise control over all parameters in the system, even disorder, although the tradeoff is the need for a confining potential.
This convergence has enabled direct comparisons between experiment and theory, which are especially important for determining universal exponents and charting the system's behavior in the quantum critical regime.

Detailed phase diagrams and other investigations of the Bose Hubbard model have been driven by its realization in experiments of ultracold atoms trapped on optical lattices, and this overview will be guided by this fact. 
The model and its phases are described in some detail, followed by a tutorial on a basic, single-site mean field theory that qualitatively characterizes the phases in the BHM at both zero and finite temperatures. A brief description of the statistically exact, efficient and large-scale worldline quantum Monte Carlo precedes a detailed discussion of the interplay between experiments and theory in the development of knowledge about the BHM.
The main body of the text is confined to systems in two and three dimensions, and to the single-species, single-band BHM in a square or cubic lattice with either a uniform or radially symmetric potential.

The many-body physics of bosonic atoms on a lattice with $z$ nearest neighbors can be captured by the relatively simple Bose Hubbard Hamiltonian,
\begin{equation}
\hatH=-\frac{t}{z}\sum_{\lr{ij}}\left(\hatad_i\hataa_j+\hataa_i \hatad_j\right)+\frac{U}{2}\sum_i \hatn_i(\hatn_i-1)-\mu\sum_i\hatn_i
,
\label{eq:BHM}
\end{equation}
where $t$ is the boson hopping amplitude between neighboring sites $\left\langle ij\right\rangle$, $U>0$ is the on-site repulsive interaction, and $\mu$ is the chemical potential that controls the boson density. The number operator for particles on site $i$ is $\hat{n}_i=\hat{a}_i^{\dagger}\hat{a}_i$, where $\hat{a}\;(\hat{a}^{\dagger})$ is the bosonic annihilation (creation) operator.  As discussed in Sec.~\ref{sec:parameters}, the energies $t$, $U$, and $\mu$ are set by the optical lattice depth $V_0$  and the lattice spacing $a_{\text{latt}}$, such that $t/U$ decreases as $V$ or $a_{\text{latt}}$ increases. Note that this Hamiltonian has $U(1)$ symmetry since it is invariant under $\hataa\rightarrow\hataa e^{i\theta}$ and $\hatad\rightarrow\hatad e^{-i\theta}$.

\subsection{Ground States}
\subsubsection{Mott Insulator}
Consider the limiting case of the BHM with average boson density $n=N_b/L^d$ ($N_b$ bosons in a $d$-dimensional system with $L$ sites along each side). In the limit $t\rightarrow0$ the Hamiltonian $\hat{H}_U=\frac{U}{2}\sum_i\hatn_i(\hatn_i-1)-\mu\sum_i\hatn_i$ is site-decoupled. Since $\hat{H}_U$ is diagonal in the Fock (boson number) basis, the ground state has $n$ bosons on every site where

\be
\begin{array} { c c }
\mu/U<0  & n=0 \\
0<\mu/U<1 & n=1 \\
1<\mu/U<2 & n=2 \\
\vdots & \vdots \\
\end{array}
\label{eq:zerot}
\ee
This state requires the number of particles to be commensurate with the number of lattice sites so the density is quantized. The ground state of this ideal Mott insulator with $n$ bosons per site is the product state
\be
\ket{\Phi^{MI}}=\prod_i^{L^d}\left(\hatad_i\right)^n\ket{0}.
\label{eq:MIproductstate}
\ee
In some sense, this state can be thought of as
\be
\ket{\Phi^{MI}}=\prod_j^{n L^d} \hatad_{j} e^{-i\theta_{j}} \ket{0}
\ee
where a phase $\theta_{j}$ is explicitly identified with each boson. Since the various $\theta_j$ are completely uncorrelated, the overall gauge symmetry is maintained in this state.

 Excitations in a Mott insulator with density $n$ are gapped by the finite energy cost to add a particle $(Un-\mu)$ or to remove a particle $(U(n-1)-\mu)$, so the Mott insulator is incompressible. 
 The energy to add a particle in a Mott state with $n$ particles per site is degenerate with the energy to remove a particle from a Mott state with $(n+1)$ particles per state.
 Thus at small tunneling $t$, superfluid order emerges precisely near integer values of $\mu/U$, where the gap to add or remove particles is on the order of $t$. Any such transition point between the Mott insulator and the superfluid at zero temperature is a quantum critical point (QCP).

\begin{figure*}[t]
  \centering
  \includegraphics[width=0.7\textwidth]{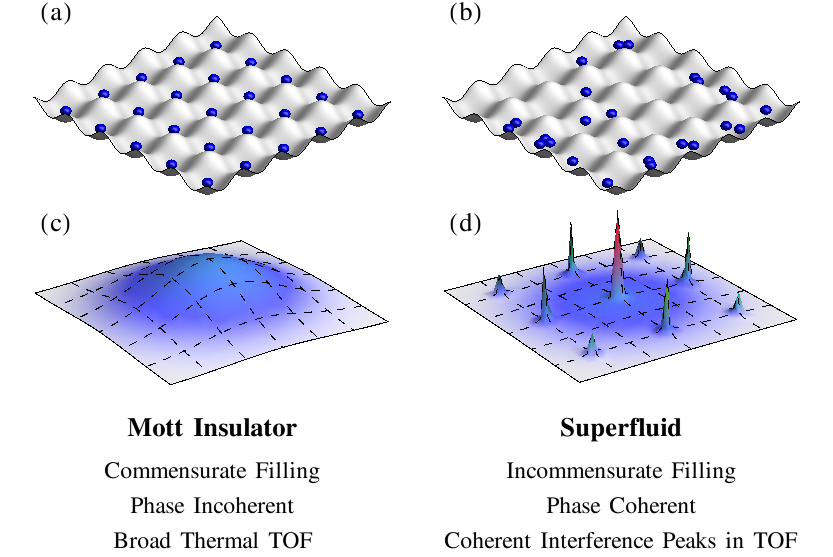}
  \caption{Schematic of a Mott insulator state and a superfluid state in a lattice (a,b) and after time of flight (TOF) expansion (c,d). During TOF, the lattice potential is switched off and the atoms' crystal momenta are mapped onto the continuum momenta that govern the atom cloud expansion. (c) The atoms in the Mott insulator have random relative phases so they destructively interfere during expansion and result in a broad distribution. (d) Each atom in the condensate has the same phase, so sharp peaks emerge at momenta that are multiples of the inverse lattice spacing. Uncondensed atoms contribute to a broad background. Adapted from \cite{bloch2005}.}
  \label{fig:BHMphaseSchematic}
\end{figure*}

\subsubsection{Superfluid}
The competing superfluid state is described by the order parameter 
$\lr{\hat{a}}=\sqrt{n_0}e^{i\theta}$ where $n_0$ is the condensate fraction and $\theta$ is the phase of the complex order parameter. 
In the Mott insulator each site has an arbitrary phase, but the superfluid breaks gauge symmetry and picks the {\em same} phase for all the sites, thereby generating a phase-coherent state.
In a non-interacting Bose-Einstein condensate all the $N_b$ bosons occupy the same $k=0$ state described by
\be
\ket{\Phi^{BEC}}=\left( {\hatad_{k=0}}\right ) ^{N_b}\ket{0}=\left( {\sum_i\hatad_{i}}\right )^{N_b}\ket{0}      
\ee
In real space, the BEC is described as a linear {\em coherent} superposition of configurations with fluctuating numbers of particles at a site with the {\em same} phase. 
Since the phase in a BEC is definite, by the uncertainty principle for the conjugate variables number and phase, the number of bosons at a site is completely uncertain.
As interactions are increased, one sees the depletion of the condensate fraction $n_0$ as well as increased phase fluctuations, even at $T=0$. These observations about phase coherent and incoherent states are nicely summarized in the time-of-flight type of experiment that allows the atom cloud to expand according to its intial momentum and phase distributions, shown in Figure~\ref{fig:BHMphaseSchematic}.

The kinetic part of the Hamiltonian $\hat{H}_t=-t/z\sum_{\lr{ij}}\left(\hatad_i\hataa_j+\hataa_i\hatad_j \right)-\mu\sum_i \hatn_i$ 
can be diagonalized in momentum space through the transformation $\hataa_i=(1/L^{d/2})\sum_\vc{k} \hataa_\vc{k} e^{i \vc{k}\cdot\vc{r}_i}$, which yields
\begin{flalign}
\hatH_t&=-\frac{t}{z}\frac{1}{L^d}\sum_{i,\vc{{\bf\delta}}}\sum_{\vc{k}\vc{q}} \left(\hatad_\vc{k}\hataa_\vc{q}e^{-i(\vc{k}-\vc{q})\cdot\vc{r}_i}e^{i\vc{q}\cdot\vc{{\bf \delta}}}+
\hatad_\vc{q}\hataa_\vc{k}e^{i(\vc{k}-\vc{q})\cdot\vc{r}_i}e^{-i\vc{q}\cdot\vc{{\bf \delta}}}\right) -\frac{\mu}{L^d}\sum_i\sum_\vc{kq}\hatad_\vc{k}\hataa_\vc{q}e^{-i(\vc{k}-\vc{q})\cdot\vc{r}_i} \\
&=-\frac{t}{z}\sum_\vc{k}\hatad_\vc{k}\hataa_\vc{k}\sum_\vc{{\bf \delta}}\left(e^{i\vc{k}\cdot\vc{{\bf \delta}}}+e^{-i\vc{k}\cdot\vc{{\bf \delta}}}\right)-\mu\sum_\vc{k}\hatn_\vc{k} \\
&=\sum_\vc{k}\hatn_\vc{k}\left(\epsilon(\vc{k})-\mu\right)
\label{eq:zeroU}
\end{flalign}
where $\vc{{\bf \delta}}/a_{\text{latt}}=\pm\vc{e}_1,\pm\vc{e}_2,\pm\vc{e}_3$ denotes the unit vector to one of the nearest neighbors, the dispersion is $\epsilon(\vc{k})=-(2t/z)\sum_{\alpha=1}^d\cos\left(k_\alpha a_{\text{latt}}\right)$, and $k_{\alpha}=\vc{k}\cdot\vc{e}_{\alpha}$. The expectation $\langle \hatad_\vc{k}\hataa_\vc{k}\rangle=n(\vc{k})$ is the momentum distribution function 
and will prove to be an important quantity of interest. As mentioned above, in a non-interacting BEC, $n(\vc{k})=N_b\, \delta(\vc{k})$ with all the bosons collected in the state corresponding to the lowest energy.
We will discuss the behavior of the momentum distribution as a function of both temperature and interactions.

The condensate fraction $n_0=N_0/N_b$ is defined as the fraction of the total bosons that occupy the ${\bf k}=0$ state. 
It can be obtained from the correlation function $\lim_{r\rightarrow\infty}\lr{\hatad_r \hataa_0}=n_0$ at large separations.
A non-zero value of $n_0$ implies that the amplitude to remove a particle from the $N_0$-boson condensate at site $r=0$ is phase coherent with 
the amplitude to add a particle to the condensate very far away. Such off-diagonal long range order is characteristic of a superfluid and 
in a non-interacting BEC, $n_0=1$. 
In the limit of $r\rightarrow\infty$, the contributions to $\lr{\hatad_r\hataa_0}$ come exclusively from the condensate because the weight of transitions into or out of all other states add incoherently and are suppressed to zero at long distances. Repulsive interactions between bosons and/or thermal effects can excite bosons into higher momentum states thereby depleting the condensate.
In superfluid $^4$He, only a fraction of atoms are in the condensate $(n_0\sim 0.1)$.

Another quantity of interest is the superfluid density $\rho_s$, which is distinct from the condensate fraction $n_0$, and is a measure of the rigidity of the system under a twist of the order parameter's phase.
In the ground state a superfluid is phase coherent with a fixed phase $\theta$ and does not have any net velocity. If a gradient in the phase is produced by, for example, rotating the superfluid, 
the superfluid velocity ${\bf v}_s={{\hbar}\over{m}}\nabla \theta$. The corresponding increase in free energy due to the kinetic energy of the superflow is 
\be
\Delta F={1\over 2} \int d^dx~ \rho_s v_s^2={1\over 2} \rho_s \left ({{\hbar}\over {m}} \right )^2 \int d^dx~ (\nabla \theta)^2
\ee
and provides a definition of $\rho_s$. In 2 dimensions, $\rho_s$ has units of energy.
At $T=0$ in superfluid $^4$He, although the fraction of bosons in the lowest eigenstate is only $10\%$, the superfluid density is equal to the the total boson density: $\rho_s=\rho$.
On a lattice with broken Galilean invariance it can be shown that $ \rho_s\le\rho$ and is in fact bounded by the kinetic energy. 

\subsection{Mean Field Theory\label{BoseMFT}}
\subsubsection{Formalism}
The Mott transition of the BHM is captured by the single-site mean field theory introduced by Sheshadri {\it et al} \cite{sheshadri1993}. 
Using the definition of the order parameter $\phi=\eone{\hat{a}}$, in the kinetic energy term of the Hamiltonian
the annihilation and creation operators can be trivially rewritten as $\hataa_j=\phi + (\hataa_j-\phi)$ and  $\hatad_i=\phi^\ast + (\hatad_i-\phi^\ast)$.
Thus,
\begin{equation}
\hatad_i\hataa_j=\phi^*\hataa_j+\hatad_i\phi-\left|\phi\right|^2 + [(\hatad_i-\phi)(\hataa_j-\phi)]
\end{equation}
Within this mean field approximation, it is assumed that the phase fluctuations denoted by the terms in the square bracket are small and can be neglected. 
The Hamiltonian reduces to a single-site mean field Hamiltonian as a function of the variational parameter $\phi$, given by
\begin{equation}
\hatH^{MF}_i\left[\phi\right]=\frac{U}{2}\hatn_i(\hatn_1-1)-\mu\hatn_i-t\left(\phi^*\hataa_i+\hatad_i\phi\right)+t\left|\phi\right|^2.
\label{eq:MFThamiltonian}
\end{equation}

The mean field Hamiltonian can be solved by writing Eq.~\eqref{eq:MFThamiltonian} in the Fock basis with $n=0,1,\ldots,n_{m}$ bosons. The maximum number of bosons on a site, $n_m$, should be increased until the variational energy converges; $n_m\sim 10$ is usually sufficient for $\mu/U\sim 4$. By assumption, either $\phi$ is zero (in the MI) or $\phi$ is fixed at a finite value because the $U(1)$ symmetry is broken (in the SF). In both cases, we are free to fix the phase so that $\phi$ is real.

\be
\begin{tabular}{c c c c c c c c}
 &   & $\ket{0}$ & $\ket{1}$ & $\ket{2}$ & $\cdots$ & $\ket{n_{m}}$ &  \\
$\bra{0}$ & \ldelim({7}{3mm} & $t\left|\phi\right|^2$ & $-t\phi$ & 0 & $\cdots$ & 0 & \rdelim){7}{3mm} \\
$\bra{1}$ &   & $-t\phi$ & $t\left|\phi\right|^2-\mu$ & $-\sqrt{2}t\phi$ & $\cdots$ & 0\\
$\bra{2}$ &   & 0 & $-\sqrt{2}t\phi$ & $t\left|\phi\right|^2+U-2\mu$ & $\cdots$ & 0\\
$\vdots$ &   & $\vdots$ & $\vdots$ & $\vdots$ & $\ddots$ & $\vdots$ \\
$\bra{n_{m}-1}$ &   & 0 & 0 & 0 & $\cdots$ & $-\sqrt{n_{m}}t\phi$\\
$\bra{n_{m}}$ &   & 0 & 0 & 0 & $\cdots$ & $t\left|\phi\right|^2+\left(\frac{U}{2}(n_{m}-1)-\mu \right)n_m$
\end{tabular} 
\label{eq:MatMF}
\ee

\begin{figure*}[b!]
  \centering
  \includegraphics[width=0.4\textwidth]{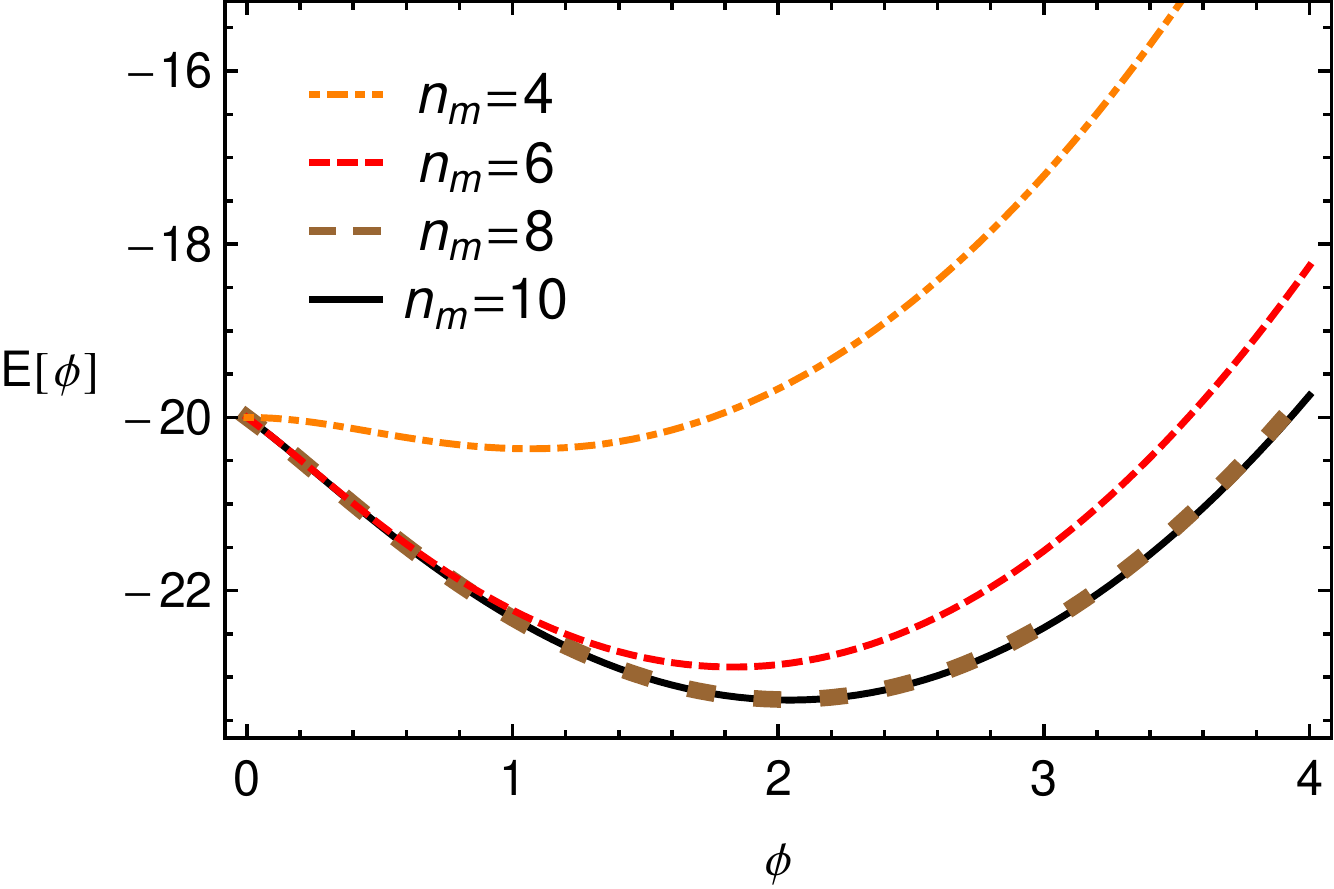}
  \caption{Variational energy $E^{MF}[\phi]$ for several values of boson number cutoff $n_m$ for $t/U=0.5$ and $\mu/U=4$. The variational energy and the optimum value of $\phi$ converges for a number cutoff near $n_m=10$. Note $\phi>0$ at the minimum of $E^{MF}[\phi]$, so this point is in the superfluid state.}
  \label{fig:MFTEpsi}
\end{figure*}

The variational energy $E^{MF}[\phi]$ [see Figure \ref{fig:MFTEpsi}] is self-consistently minimized by the following procedure: (a) construct the mean field Hamiltonian matrix in Fock space (Eq. \ref{eq:MatMF}) using a trial mean field $\phi$, (b) diagonalize the matrix and find the eigenstates $\ket{\psi_i}$ and eigenenergies $E_i$, (c) find the eigenstate $\ket{\psi_0}$ with lowest energy $E_0$, (d) update $\eone{\hat{a}}\rightarrow\phi$, and (e) repeat until the self-consistent condition $\phi=\expect{\psi_0}{\hat{a}}{\psi_0}$ is satisfied at the optimal $\tilde{\phi}$. The desired local properties are then evaluated within the lowest-energy eigenstate of $\hat{H}^{MF}_i[\tilde{\phi}]$. Local observables like the density can be estimated by $n=\expect{\psi_0}{\hatn}{\psi_0}$. By modifying the self-consistency condition to 
\be
\phi=\frac{\sum_i\expect{\psi_i}{\hataa}{\psi_i}e^{-E_i/T}}{\sum_i e^{-E_i/T}},
\ee
 properties at finite temperature $T>0$ can be accessed with the estimator 
\be
\lr{\hat{O}}=\frac{\sum_i\expect{\psi_i}{\hat{O}}{\psi_i}e^{-E_i/T}}{\sum_i e^{-E_i/T}}.
\ee

\begin{figure*}[htb!]
  \centering
  \includegraphics[width=0.7\textwidth]{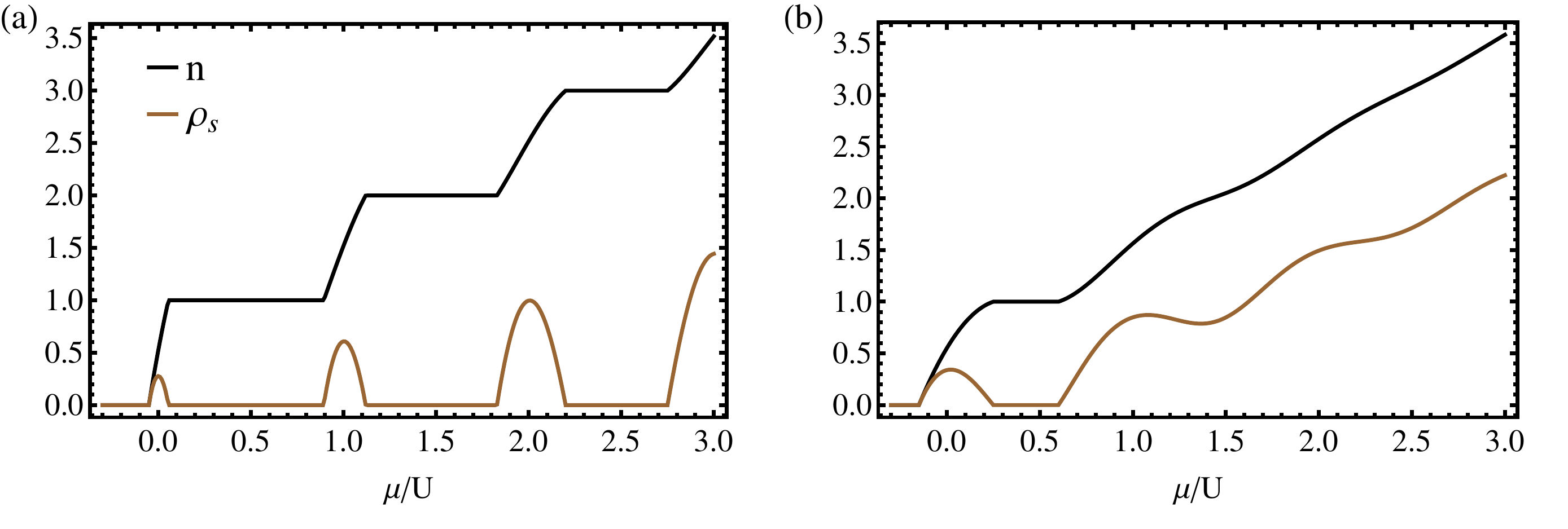}
  \caption{The superfluid density $\rho_s=\phi^2$ (within MFT) and density profiles $n$ at zero temperature and (a) $t/U=0.05$, (b) $t/U=0.15$. The stair-step increase in the density $n$ is characteristic for small values of $t/U$. At larger values of $t/U$, the superfluid density $\rho_s$ closely tracks $n$.}
  \label{fig:MFTprofile}
\end{figure*}

\begin{figure*}[htb!]
  \centering
  \includegraphics[width=0.8\textwidth]{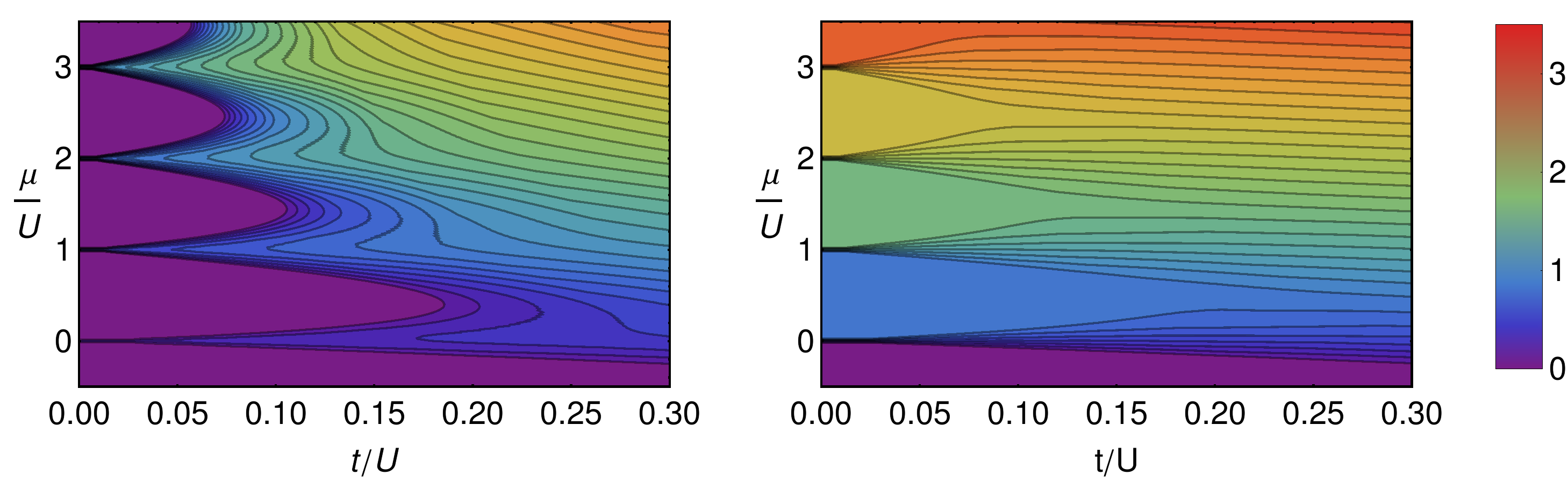}
  \caption{Superfluid density and density profiles from MFT at zero temperature. (a) The superfluid density $\rho_s$ is zero in the vacuum $\mu/U<t/U$ and in the Mott insulator. (b) The density $n$ is quantized in the Mott insulator and the particle-hole symmetric QCP occurs at tip of each Mott lobe where the transition to the superfluid takes place at constant density. The plots in Figure \ref{fig:MFTprofile}(a,b) are traces from $\mu/U=-0.3$ to $\mu/U=3$ at their respective values of $t/U$. The color scales are the same in both panels so it is clear that as $t/U$ increases, the values of $n$ and $\rho_s$ converge.}
  \label{fig:MFTrhorhos}
\end{figure*}

\begin{figure*}[htb!]
  \centering
  \includegraphics[width=0.8\textwidth]{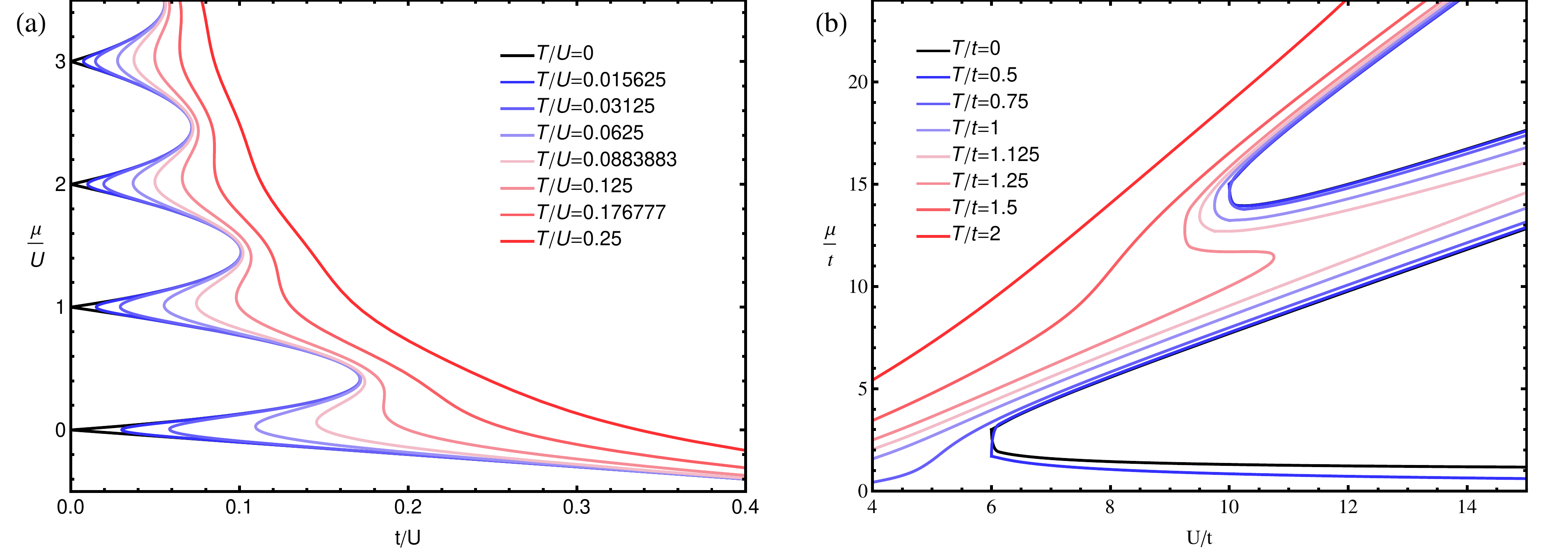}
  \caption{ These finite temperature phase diagrams calculated within MFT qualitatively show the superfluid-Mott boundary at $T=0$ and the superfluid-normal boundary at finite temperature
  determined by the vanishing of the superfluid density $\rho_s$. In both (a) the $t/U-\mu/U$ plane and (b) the $U/t-\mu/t$ plane, there is a vacuum with no particles at small $\mu$.  
 It is also evident that temperature affects the lowest density superfluid phases the most. The classic picture of the tower of Mott lobes with increasing density and decreasing area is shown in (a).  }
  \label{fig:MFTphasediagram}
\end{figure*}

\subsubsection{Mean Field Theory Results}

In the absence of inter-site tunneling the density should increase in a stairstep manner as a function of $\mu/U$ as outlined in Eq.~\eqref{eq:zerot}. Introducing a small tunneling $t/U$ shrinks each step and opens a range of $\mu/U$ over which the density deviates from integer values and the superfluid emerges in these regions [see Fig.~\ref{fig:MFTprofile}(a)]. At still higher coupling, number fluctuations cause the Mott plateaus to vanish entirely [see Figure \ref{fig:MFTprofile}(b)]. These observations are summarized in Fig.~\ref{fig:MFTrhorhos}. 
Within mean field theory, the superfluid density $\rho_s=| \phi |^2$. At $T=0$ for large $t/U$, the superfluid density $\rho_s$ approaches the density $n$, as for a Galilean invariant system.
With decreasing tunneling $t/U$, especially near a QCP, $\rho_s$ is significantly suppressed from $n$ even at zero temperature by strong correlations and within the Mott phase $\rho_s$ vanishes. 

Upon raising the temperature, the order parameter $\phi$ and consequently also $\rho_s$ is suppressed and vanishes at a critical temperature $T_c\approx t\phi$.
As shown in Figure \ref{fig:MFTphasediagram}, the superfluid order is first destroyed near the Mott insulator and at low densities.

Several extensions to this basic mean field theory have been studied. Instead of decoupling to a single site model, several sites are kept in variational cluster perturbation theory \cite{koller2006}. This retains information about the lattice geometry and dimension and is sensitive to phases with multi-site ordering, which is essential for mean field investigations of the BHM extended to include multiple species or to other lattices. 
Perturbative approaches describe the quantum phase transition from the superfluid \cite{rey2003} or from the Mott insulator \cite{freericks1994} phase.  Dynamical mean field theory models the BHM as a single impurity self-consistently coupled to a bath and works well in higher dimensions \cite{anders2011,kauch2012}.
Alternatively, the excitation spectrum can be evaluated within the random-phase approximation\cite{altman2002,menotti2008}. In some situations, explicitly time-dependent extrapolations of MFT  and time-dependent density matrix renormalization group (DMRG) methods have proven informative\cite{krutitsky2010,navez2010,natu2011,bernier2011}. These calculations hint that the gapped modes in the Mott insulator are still present at the QCP and even survive into the superfluid phase with a finite weight.

\subsection{Quantum Monte Carlo}
\subsubsection{Beyond Mean Field Theory}
Going beyond mean field theory for the BHM, several methods attempt to include
some of the quantum fluctuations discarded within MFT. In $d=1$ where fluctuation effects are most dominant, Bethe ansatz \cite{krauth1991,caux2007} and  
DMRG methods (exact up to truncation errors \cite{kuhner1998}) are options. 
Unlike in fermion models where the sign problem stymies calculations, bosonic systems can be simulated using quantum Monte Carlo (QMC) methods
in any dimension and incur only statistical errors. We discuss below the worldline QMC algorithm applied to the BHM.

\subsubsection{World-line QMC Background\label{BoseQMC}}
Over the last two decades, several path-integral-based algorithms of the ``worldline'' type have been developed \cite{ceperley1995,prokofev1998,syljuasen2002,kawashima2004}. This class of algorithms efficiently samples the partition function $Z$ of the BHM in imaginary time and has been applied to several other models. All worldline QMC simulations take place in the boson occupation number (Fock) basis in $d+1$ dimensions -- $d$ spatial dimensions and one imaginary time dimension whose extent is dictated by the inverse temperature $\beta=1/T$. The algorithms employ an auxiliary field $F$ that is updated in turn with the boson configuration $\Psi$. The auxiliary field enables the construction of global updates and enables efficient measurement of several observables. The best implementations are capable of simulating experimental-scale systems with up to $10^6$ bosons and $64^3$ lattice sites \cite{kato2009}.

A plethora of measurements including density $n$, compressibility $\kappa$, imaginary time Green function $G(r,\tau)$, momentum distribution, and superfluid density $\rho_s$ are relatively straightforward to implement. On the other hand, information about phase structures and defects (e.g., vortices) can only be obtained indirectly. Nevertheless, worldline QMC simulations have played a very important role in confirming that ultracold bosons in optical lattices emulate the BHM and have aided in the interpretation of key experiments like time-of-flight imaging. These algorithms 
have also played a role in theoretical investigations of the accuracy of the local density approximation, in developing new methods with which to identify phases in experiments and in classification of universality classes. The ensuing discussion of worldline QMC algorithms becomes somewhat technical and can be skipped if the reader is focused on the broader BHM picture.

We briefly describe motivating ideas behind worldline QMC algorithms. In this discussion, it is helpful to break the Hamiltonian into its smallest constituent pieces -- single-site terms $\calhhat_{onsite,i}$ and link terms $\calhhat_{link,ij}$ that operate on neighboring sites $i$ and $j$:
\begin{flalign}
\hatH&=\hat{H}_{onsite}+\hat{H}_{link}\\
&=\sum_i\left(\frac{U}{2} \hat{n}_i(\hat{n}_i-1)-\mu \hat{n}_i\right)+\sum_{\lr{i,j}}\left(\frac{-t}{z}\right) \left(\hatad_i \hataa_j+\hataa_i\hatad_j\right)\label{eq:MtermsinH}\\
&=\sum_i\hat{\calh}_{onsite,i}+\sum_{\lr{i,j}}\hat{h}_{link,ij}=\sum_{m=1}^M \hat{\calh}_m
\label{eq:MtermsinHb}
\end{flalign}
There are $M=(d+1)L^d$ distinct terms in Eqs. \ref{eq:MtermsinH} and \ref{eq:MtermsinHb}. These basic constituents are denoted by $\hat{h}_m$ where $m$ is shorthand for both the local Hamiltonian term type and the site(s) it operates on.

\subsubsection{Partition Function Expansion and QMC}
\begin{figure*}[h]
  \centering
  \includegraphics[width=0.9\textwidth]{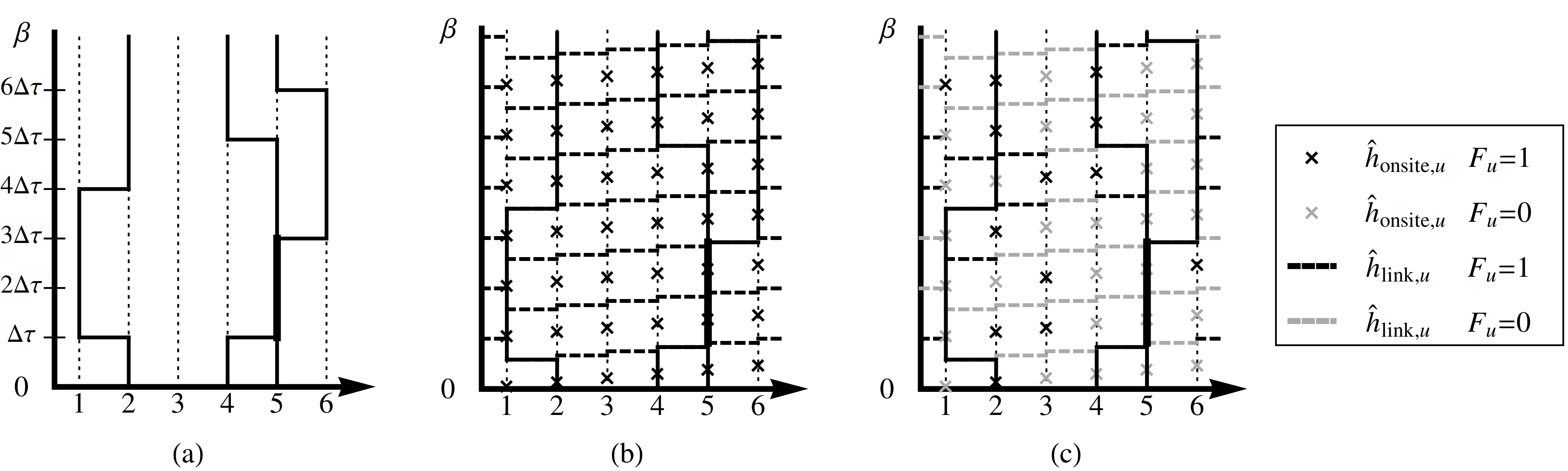}
  \caption{Example worldline configurations drawn from the partition function expansions. (a) Only the boson configuration $\Psi$ at each time $\Delta\tau$ is sampled in eq. \ref{eq:PIMCbasicZ}. (b) The boson configuration is expanded at each time $\Delta\tau$ into the local single-site and link operators $\calhhat_u$ in eq. \ref{eq:PIMClocalZ}. (c) The local terms $\calhhat_u$ are sampled using the auxiliary field $F_u$ in eq. \ref{eq:PIMCZexp}. In each of these panels, $L=6$, $N_b=3$ and $L_\tau=7$. In panel (c), the terms $\calhhat_u$ are coded by black or grey depending on the auxiliary field $F_u=0$ (identity operator) or $F_u=1$, as shown in the legend.}
  \label{fig:loopQMC}
\end{figure*}

The partition function $Z$ describes the evolution of a state $\ket{\Psi}$ over a long imaginary time $\beta$. Discretization of the imaginary time into cells of width $\Delta\tau=\beta/L_\tau$ breaks $Z$ into evolution operators over smaller imaginary time intervals. We will consider this expansion within the path integral formalism \cite{kawashima2004}, though the stochastic series expansion can be used instead \cite{syljuasen2002}. The matrix elements are evaluated in the Fock basis where each boson configuration $\Psi$ records $\psi_{\tau,i}=0,1,2,\ldots,n_m$, the number of bosons on each site $i$ and imaginary time $\tau$ with a maximum boson number $n_m$. Expanding $Z$ into the local operators in Eq. \ref{eq:MtermsinHb}, we obtain
\begin{flalign}
Z&=\sum_{\{\Psi\}}\expect{\Psi}{e^{-\beta\hatH}}{\Psi}=\sum_{\{\Psi\}}\expect{\Psi}{\lim_{L_\tau \rightarrow\infty}\prod_{\ell=1}^{L_\tau} e^{-\Delta\tau\hatH}}{\Psi}\\
&=\lim_{L_\tau\rightarrow\infty}\sum_{\{\Psi\}}\expect{\Psi}{\prod_{\ell=1}^{L_\tau} \prod_{m=1}^M e^{-\Delta\tau\calhhat_m}}{\Psi}\label{eq:PIMCbasicZ}\\
&=\lim_{L_\tau\rightarrow\infty}\sum_{\{\Psi\}}\prod_u^{L_\tau M}\expect{\psi_u'}{e^{-\Delta\tau\calhhat_u}}{\psi_u}\label{eq:PIMClocalZ}.
\end{flalign}
In the final line, the $m$ and $\ell$ indices have been rolled into the single index $u$ that specifies the type of Hamiltonian term, the site or link it acts on, and the imaginary time it acts at. The state $\psi_u\;(\psi_u')$ is the state directly before (after) the operator $\calhhat_u$. This expansion is schematically illustrated in Figure \ref{fig:loopQMC}(a). 

Next the exponential is expanded and truncated at second order in $\Delta \tau$ and an auxiliary field $F$ is introduced at each space-imaginary time coordinate $u$ with value 0 or 1
\begin{flalign}
Z&=\lim_{L_\tau\rightarrow\infty} \sum_{\{\Psi\}} \prod_u^{L_\tau M} \expect{\psi_{u}'}{ 1-\Delta\tau \calhhat_u + O((\Delta\tau E_M)^2)}{\psi_u}\label{eq:PIMCZnoverts}\\
&\approx\lim_{L_\tau\rightarrow\infty} \sum_{\{\Psi\}} \prod_u^{L_\tau M} \sum_{F_u=0,1} \expect{\psi_{u}'}{ (-\Delta\tau \calhhat_u)^{F_u}}{\psi_u}\\
&=\lim_{L_\tau\rightarrow\infty} \sum_{\{\Psi\}} \sum_{\{F\}}\prod_u^{L_\tau M}  w(\psi_u,F_u)\label{eq:PIMCZexp}
\end{flalign}
where $E_M$ is the leading order contribution to the energy and the weight $w(\psi_u,F_u)=\expect{\psi_{u}'}{ (-\Delta\tau \calhhat_u)^{F_u}}{\psi_u}$. 

Monte Carlo techniques enable efficient sampling of the large configuration space $\{\Psi\}$ and $\{F\}$ in Eq.~\eqref{eq:PIMCZexp}. A basic implementation updates a single $\psi_u$ or $F_u$ at a time by randomly proposing a new value $\tilde{\psi}_u,\tilde{F}_u$ and accepting it based on Metropolis importance sampling of the relative weight $w(\tilde{\psi}_u,\tilde{F}_u)/w(\psi_u,F_u)$. However, these algorithms slow down dramatically near critical points where the correlation length diverges because generating an independent configuration requires many more updates \cite{ceperley1995}. To overcome these difficulties, the update step is accomplished through a loop update in a fashion analogous to the Wolff or Swendsen-Wang algorithms in classical spin systems \cite{swendsen1987,wolff1989}.
The loop, worm, and directed-loop algorithms based on Eq.~\eqref{eq:PIMCZexp} probabilistically assign an auxiliary field $F_u$ for each term $\calhhat_u$, generate a new configuration $\tilde{\psi}_u$ using the $F_u$, and choose to accept $\tilde{\psi}_u$ based on the relative weights $w(\tilde{\psi}_u,F_u)/w(\psi_u,F_u)$ of the two configurations, given in Eq. \ref{eq:PIMCZexp}. The two configurations $\Psi$ and $F$ are exemplified in Figure \ref{fig:loopQMC}(a,b). 
In practice, statistically exact results are obtained by picking a large truncation $L_\tau$ or by explicitly taking the continuous time limit. 

The precise formulation of the update step defines the differences between the loop, worm, and directed-loop algorithms \cite{prokofev1998,syljuasen2002,kawashima2004}. Each type of simulation moves through configuration space in a similar way. At the start of each update step, the auxiliary field $F$ is sampled. Then two singularities, the head ($\hataa_h$ or $\hatad_h$) and tail ($\hatad_t$ or $\hataa_t$), are inserted into the worldline configuration. The head constructs a loop in the worldline configuration by probabilistically changing direction at $F_u=1$ matrix elements, closing the loop when it meets the tail. The boson configuration is updated as the head moves.

Observables like energy and density are measured after a loop closes. On the other hand, observables like the single-particle Green function 
\be
G(\vc{r},\tau)=\lr {T_\tau \hat{a}(\vc{r},\tau)\hatad (0,0)}
\label{eq:gf}
\ee
are sampled within the loop update step, where $T_\tau$ stands for time ordering.

The momentum distribution $n(\vc{k})=\lr{\hatad_\vc{k} \hataa_\vc{k}}$ is easily extracted from the equal-time Green function using
\be
n(\vc{k})=\sum_{i,j}\lr{\hatad_i \hataa_j}e^{i\vc{k}\cdot(\vc{r}_i-\vc{r}_j)}=L^{-d/2}\sum_j G(\vc{r}_j,0)e^{i\vc{k}\cdot\vc{r}_j}
\label{eq:nk}
\ee
in a uniform system. 


Two very important observables are the superfluid density $\rho_s$ and the compressibility $\kappa$. As seen from the effective action in Eq.~\eqref{eq:BHMeffectiveaction}, these quantities are the response of the free energy to a boson phase angle twist imposed by spatial and imaginary time boundary conditions, respectively. The estimator for $\rho_s$ is
\be
\rho_s=\frac{1}{L^d\beta(t/z)}\left.\frac{\partial^2 F}{\partial \theta^2}\right|_{\theta=0}=\frac{L^{2-d}\lr{W^2}}{\beta t}
\ee
Use of the average squared winding number $\lr{W^2}$ in a periodic simulation eliminates the need to explicitly impose a twist to the spatial boundary conditions (see Ref.~\cite{pollock1987} for a rigorous derivation). $\langle W^2 \rangle$ is a measure of the net boson current obtained by counting the number of bosons that have taken advantage of spatial periodic boundary conditions to wind all the way around the simulation box. For example, the 1D boson configuration in Figure \ref{fig:loopQMC}(a) has zero winding number because an equal number of bosons hop from site $i$ to a neighboring site $j$ as hop from $j$ to $i$ during the imaginary time interval from 0 to $\beta$. On the other hand, the thermodynamic compressibility is
\be
\kappa=\frac{\partial \rho}{\partial \mu}=\beta\left(\lr{\hat{N}^2}-\lr{\hat{N}}^2\right)=\frac{1}{L^{2d} \beta}\sum_{i,j}^{L^d}\int_0^\beta\int_0^\beta d\tau d\tau'\left[\lr{\hatn_i(\tau)\hatn_j(\tau')}-\lr{\hatn_i(\tau)}\lr{\hatn_j(\tau')}\right].\label{eq:localcompfirst}
\ee
where $\hat N=\sum_i \hat n_i$ is the total number operator.
Both trap curvature and disorder can be naturally included in this scheme. The trapping potential can be incorporated into this QMC scheme through a site-dependent chemical potential $\mu_i$ which modifies the local sampling weight $\hat{h}_{onsite,i}$. In the same way, onsite- and bond-disorder can be included through the $\hat{h}_{onsite,i}$ and $\hat{h}_{bond,ij}$ terms, respectively.



\begin{figure*}[t!]
  \centering
  \includegraphics[width=0.9\textwidth]{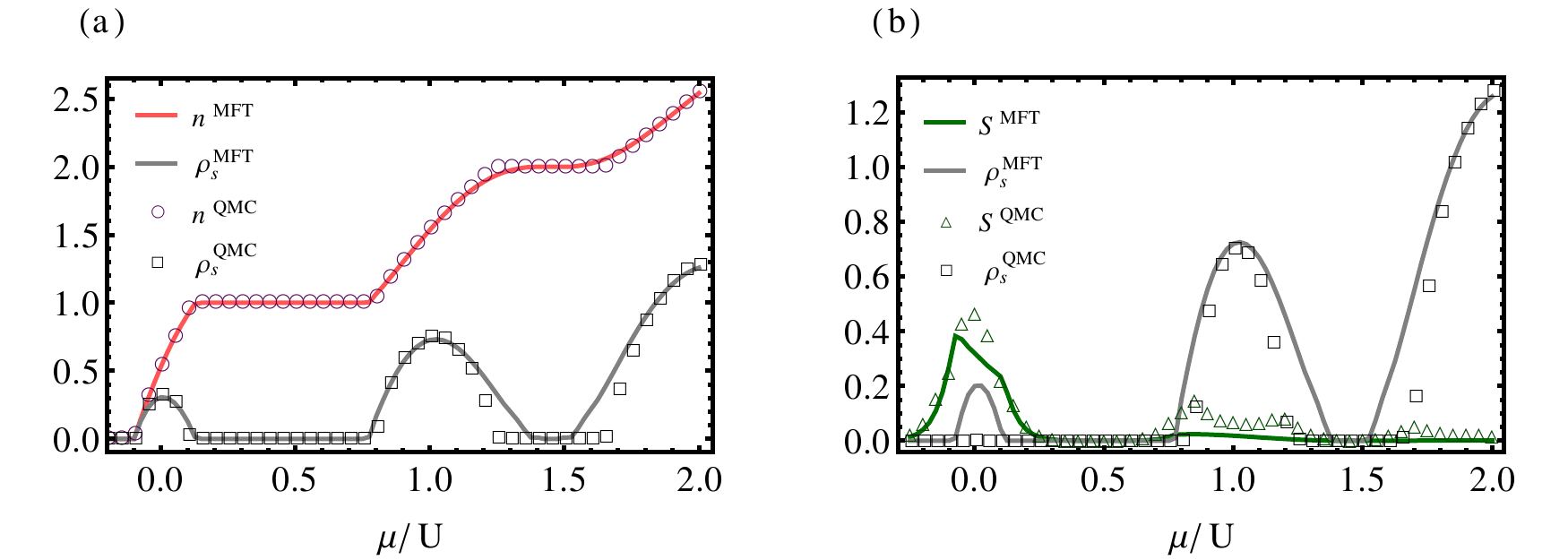}
  \caption{Comparison of MFT  and QMC in two dimensions at $t/U=0.1$ and (a) $T/t=0.1$, (b) $T/t=0.4$. (a) Quantum fluctuations in 2D significantly suppress the superfluid relative to MFT, but away from the QCPs the agreement is remarkable. (b) At finite temperatures, MFT calculations of thermodynamic quantities like the entropy $S$ only agree qualitatively with QMC.}
  \label{fig:MF_QMC_comp}
\end{figure*}

\begin{figure*}
  \centering
  \includegraphics[width=0.4\textwidth]{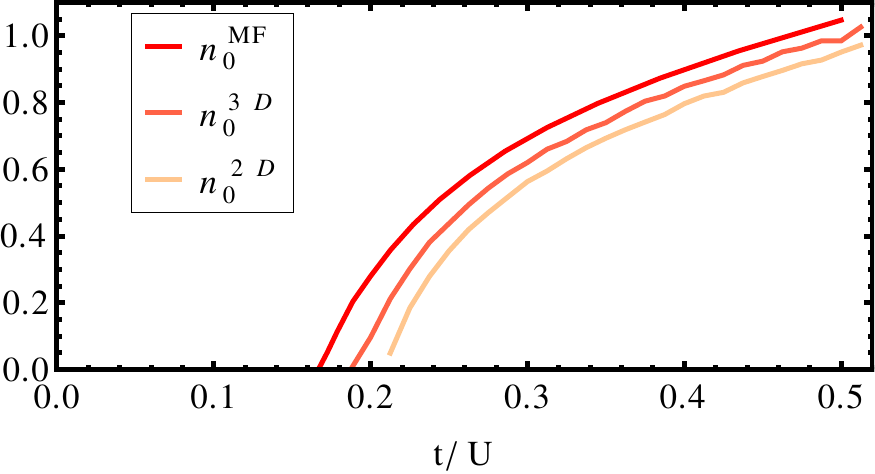}
  \caption{Comparison of the condensate fraction $n_0$ calculated within MFT and QMC in two and three dimensions at very low temperature. The phase fluctuations are included in QMC and are enhanced in lower dimensions, and suppress $n_0$ and the superfluid phase.}
  \label{fig:MFTQMCn0comp}
\end{figure*}

QMC results confirm that MFT is qualitatively correct in two and three dimensions, as shown in Fig.~\ref{fig:MF_QMC_comp}.
 Away from the QCP separating the Mott insulator and the superfluid, the agreement is quite good. The difference between QMC and MFT at the transition reflects the 
 effect of thermal and quantum phase fluctuations not included in MFT, which suppress the superfluid phase. The agreement is best in 3D and becomes progressively worse in 2D and 1D as phase fluctuations play an increasingly large role as shown in Figure~\ref{fig:MFTQMCn0comp}.

\begin{figure*}
  \centering
  \includegraphics[width=0.9\textwidth]{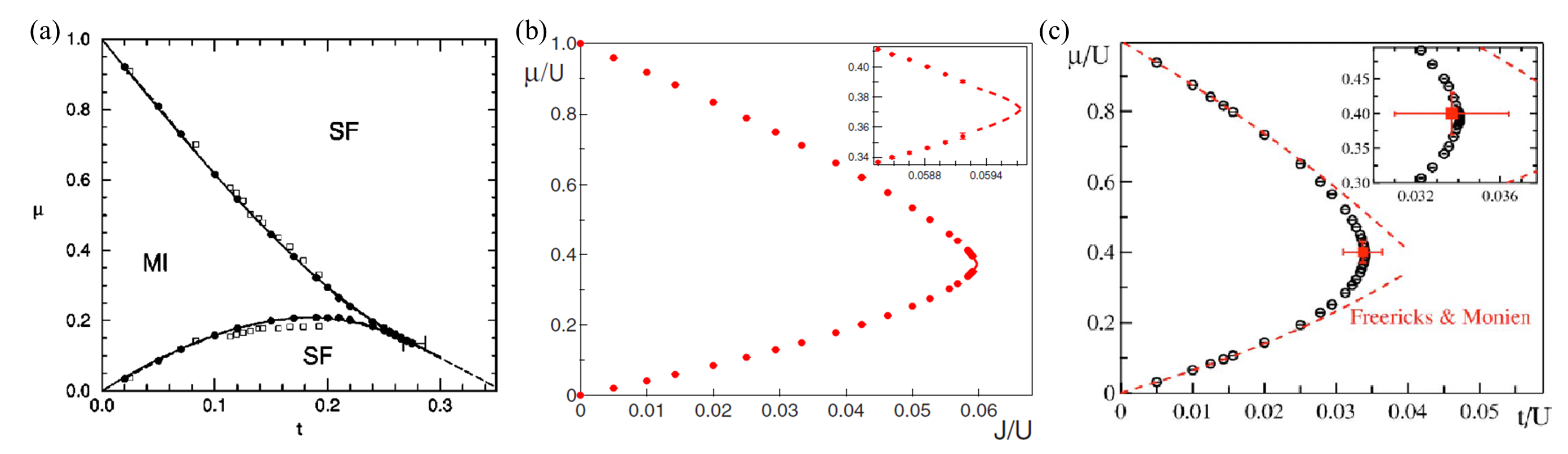}
  \caption{This is the ground state phase diagram in $d=1,2,3$ calculated from (a) DMRG and (b,c) QMC for the $n=1$ Mott insulator to superfluid transition. The definition of the hopping energy is $J=t/z$ in (b) and (c). Adapted from \cite{kuhner1998,capogrosso2008,capogrosso2007}.}
  \label{fig:QMCphasediagram}
\end{figure*}

\subsubsection{Critical Points and Phase Transitions}
 The zero temperature phase diagram has been exhaustively mapped numerically in $d=1,2,3$ spatial dimensions (see Fig.~\ref{fig:QMCphasediagram}). In every dimension, as the tunneling $t/U$ increases, the Mott insulator-superfluid phase boundary tapers to a particle-hole symmetric quantum critical point for each of the $n=1,2,\ldots$ Mott states\cite{fisher1989}. This characteristic shape is called a Mott lobe and the quantum critical point at the tip of each Mott lobe is in the $d+1$ dimensional $XY$ universality class because the system is particle-hole symmetric. 
In two and three dimensions, all the other quantum critical points along the MI-SF boundary are in the mean field universality class.

The quantum critical points (QCPs) in the BHM occur for $t/U$ of order unity and are extremely challenging to precisely locate within the phase diagram. 
Similar to classical critical points, large fluctuations arise near QCPs from new degrees of freedom that must form as the system transits from one phase to the other. The QCPs leave a definite footprint in the thermodynamic and response functions for a large range of temperatures and parameter values near the QCP. 
These fluctuations affect quantities like the density, superfluid density, compressibility and energy and cause them to scale with the temperature and the distance from the QCP. 
The special symmetry at $d+1$ $XY$ QCPs enables a mapping onto a $d+1$ dimensional classical spin system as well as onto quantum rotor models \cite{sachdev2011}. This is useful for several reasons, including that the quantum critical region may be easier to investigate in other models.

Near a quantum critical point, spatial fluctuations develop over a diverging length scale $\xi_x\sim \delta^{-\nu}$
and temporal fluctuations over a diverging time scale $\xi_\tau\sim \xi^z \sim \delta^{-z\nu}$, where $\nu$ and $z$ are critical exponents and 
$\delta=\left|t-t_c\right|/t_c$ or $\delta=\left|\mu-\mu_c\right|/\mu_c$ is the distance from a QCP at $(t_c,\mu_c)$.
The divergent time scales correspond to vanishing energy scales $\Omega\sim \delta^{z\nu}$ at the QCP.
At the quantum phase transition, the thermodynamic and response functions show characteristic singularities depending on the universality class of the transition.
Furthermore, the singular contributions to thermodynamic and response functions show scaling behavior and are independent of the details of the Hamiltonian.
This aspect can be exploited to
determine the location of QCPs in both simulations and experiments.

\begin{figure}[h]
\centering
\includegraphics[scale=0.60]{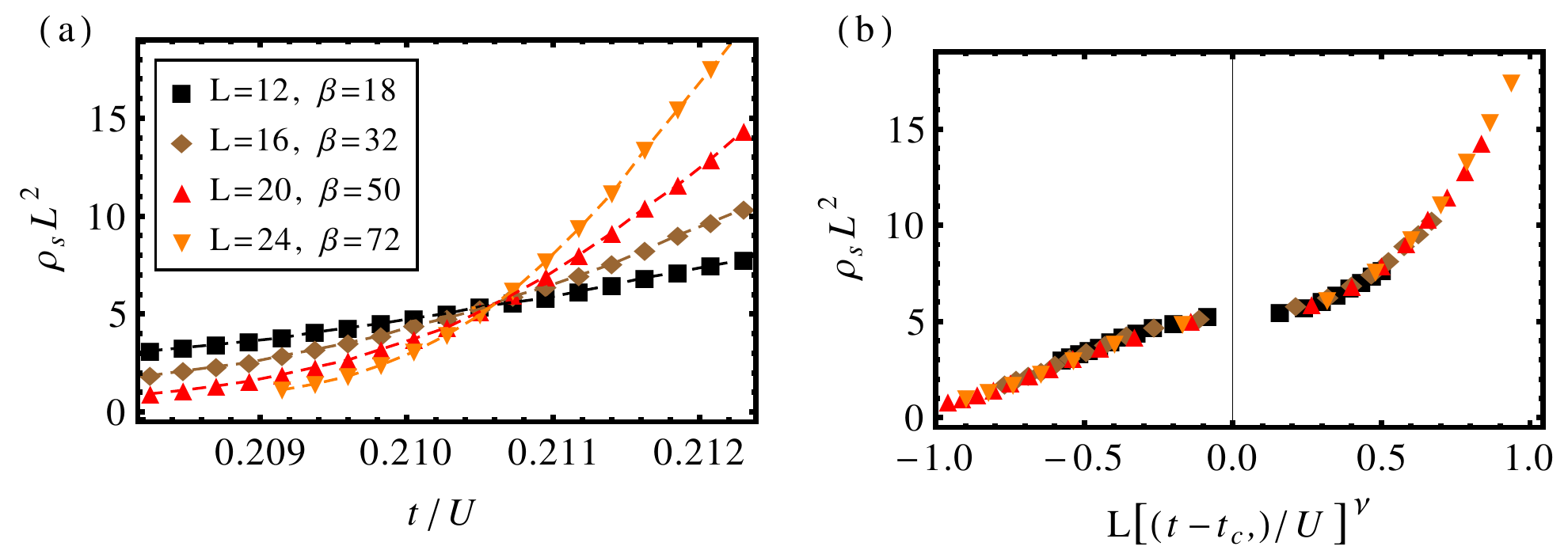}
\caption{Determination of the mean field type QCP in $d=2$ dimensions at $\mu/U=0.5$. (a) The superfluid density $\rho_s$ scaled by $L^z$ as a function of $t/U$ for different lattice sizes $L$ crosses at the QCP $t_c/U=0.21055$. For each $L$ the temperature $T=1/\beta$ is chosen so as to keep $\beta/L^z=0.125$ fixed. (b) With $t_c$ and the appropriate choice of the critical exponent $\nu=0.5$, $\rho_s$ collapses onto a universal curve [see Eq. \ref{eq:rhosSCALING}]. The error of these QMC results is smaller than the point size.
}
\label{fig:QCPscaling}
\end{figure}

The scaling form of $\rho_s$, the superfluid order parameter, is particularly important in numerical investigations. We begin with the long-wavelength and low-energy effective action of the BHM at temperature $T=1/\beta$ in imaginary time $\tau$,
\be
S_\text{eff}\sim \int_0^{\beta} d\tau \int_0^L d^d x\left(2 i \rho \partial_{\tau}\phi+\kappa\left|\partial_{\tau}\phi\right|^2+\rho_s\left|\nabla\phi\right|^2\right).
\label{eq:BHMeffectiveaction}
\ee
Simple dimensional analysis shows that in the thermodynamic limit of $\beta\rightarrow\infty$ and $L\rightarrow\infty$
\begin{flalign}
\rho_s\sim\xi_{\tau}^{-1}\xi_x^{2-d}&=\delta^{\nu(z+d-2)}\\
\kappa\sim\xi_{\tau}\xi_x^{-d}&=\delta^{\nu(d-z)}.
\end{flalign}
But near a QCP, $T_c$ must vanish according to $T_c\sim \delta^{z\nu}$. Substituting for $\delta$ in terms of $\rho_s$ according to
$\delta\sim \rho_s^{1/\nu(z+d-2)}$ we obtain the following relation between $\rho_s$ at zero temperature and $T_c$\cite{kopp2005,franz2006}
\be
T_c\propto \rho_s^{z/(z+d-2)}.
\label{eq:n0Tc}
\ee

For finite $\beta$ and sample size $L$, the diverging correlation lengths are cut off and $\rho_s\sim \xi^{2-d-z}$ is replaced by
\begin{flalign}
\rho_s&\sim L^{2-d-z} X\left(L/\xi_x,\beta/L^z\right)=L^{2-d-z} X\left(L\;g^{\nu},\beta/L^z\right)\label{eq:rhosSCALING}\\
\kappa &\sim L^{z-d} Y\left(L/\xi_x,\beta/L^z\right)=L^{z-d} Y\left(L\;g^{\nu},\beta/L^z\right)
\end{flalign}
where $X$ and $Y$ are universal scaling functions for $\rho_s$ and the full compressibility $\kappa$, respectively. Note that $\delta=0$ at the QCP, 
so if the ratio $\beta/L^z$ is fixed and $L^{d+z-2}\rho_s(\mu)$ or $L^{d-z}\kappa(\mu)$ are plotted for several values of $L$ and $\beta$, all the curves must intersect at the 
$(t_c,\mu_c)$ of the QCP. 
The mean field universality class is characterized by exponents $z=2$ and $\nu=0.5$ while the 3D XY universality class has $z=1$ and $\nu=0.6717$ \cite{fisher1989,hasenbusch1999}. Once $\mu_c$ is determined, the curves collapse onto the universal scaling function $X$ or $Y$ for the proper choice of the critical exponent $\nu$ and demonstrated in Figure~\ref{fig:QCPscaling}. This type of finite-size scaling analysis for $\rho_s$ and the Mott gap is the ideal method for determining the location of a critical point from numerical simulations\cite{fisher1989,capogrosso2007,capogrosso2008}.

A similar dimensional analysis of the singular part of the free energy density $f_s$ leads to \cite{fisher1989,campostrini2010,zhou2010,fang2011,hazzard2011}
\begin{flalign}
n_{s}(\mu,T)=n(\mu,T)-n_r(\mu,T)=T^{(d+z-1/\nu)/z}\mathcal{X}\left(\frac{\mu-\mu_c}{T^{1/z\nu}}\right)\label{eq:denSCALE}\\
\kappa_{s}(\mu,T)=\kappa(\mu,T)-\kappa_r(\mu,T)=T^{(d+z-2/\nu)/z}\mathcal{Y}\left(\frac{\mu-\mu_c}{T^{1/z\nu}}\right)\label{eq:compSCALE},
\end{flalign}
where the singular part of the density $n_s=-\partial f_s/\partial \mu$ and the singular part of the compressibility $\kappa_s=-\partial^2 f_s/\partial \mu^2$. 
The regular parts of the density $n_r$ and compressibility $\kappa_r$ can be estimated analytically \cite{hazzard2011}. This general scheme has recently been implemented in a 2D experiment and the critical exponents roughly estimated \cite{zhang2012}. It is important to note that at a QCP above the upper critical dimension (i.e. mean field QCP in $d=3$), a modified finite size scaling must be employed. This incorporates a dangerous irrelevant variable which scales like $L^{4-d}$ and is carefully justified in \cite{binder1985,chen1998,katoTHESIS}.

At finite temperatures far from QCPs, the superfluid order is destroyed by thermal fluctuations. Phase fluctuations excited by thermal energy eventually destroy the superfluid as the temperature rises above the critical temperature, $T_c$. This classical critical point is in the $d$-dimensional $XY$ universality class. For $d=2$, this transition is in the special Berezinskii-Kosterlitz-Thouless universality class where the phase fluctuations are vortex-antivortex pairs that proliferate and become unbound above $T_c$. Vortices cause correlations to decay algebraically and impose quasi-long-range order. In contrast, the Mott state experiences a smooth crossover to a normal state as temperature increases since it 
does not spontaneously break any symmetry. Particle-hole pair thermal fluctuations in the Mott insulator multiply as the temperature increases and give the state a finite compressibility.

\subsubsection{Illuminating QMC Results}
%

\begin{figure*}[t]
  \centering
  \includegraphics[width=0.95\textwidth]{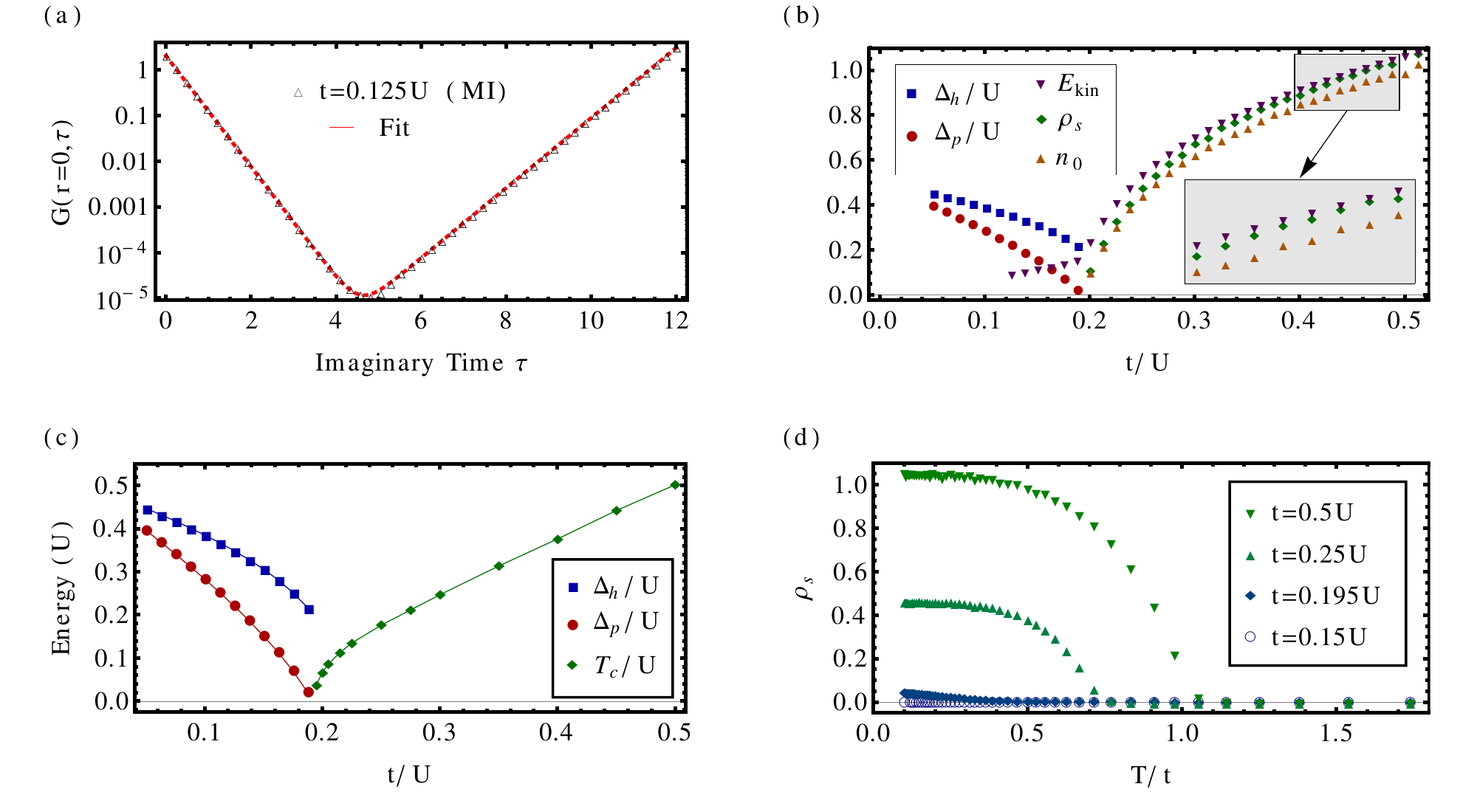}
  \caption{The characteristic energy scales within the BHM for $\mu/U=0.5$ in a uniform system with at least $12^3$ sites. (a) Imaginary time Green Function in the Mott insulator used to estimate the gap energy scales fit by Eq. \ref{eq:MIgap}. (b) Energy scales calculated with QMC. The inset shows $E_{kin}/t\geq\rho_s/t\geq n_0$ within the superfluid. (c) Both the particle gap scale and the superfluid critical temperature $T_c$ vanish at the QCP near $t/U=0.192$. (d) Interactions suppress $\rho_s$ even at low temperature as $t/U$ is tuned across the QCP point.}
  \label{fig:BHMenergyScales}
\end{figure*}

\begin{figure*}[h]
  \centering
  \includegraphics[width=0.95\textwidth]{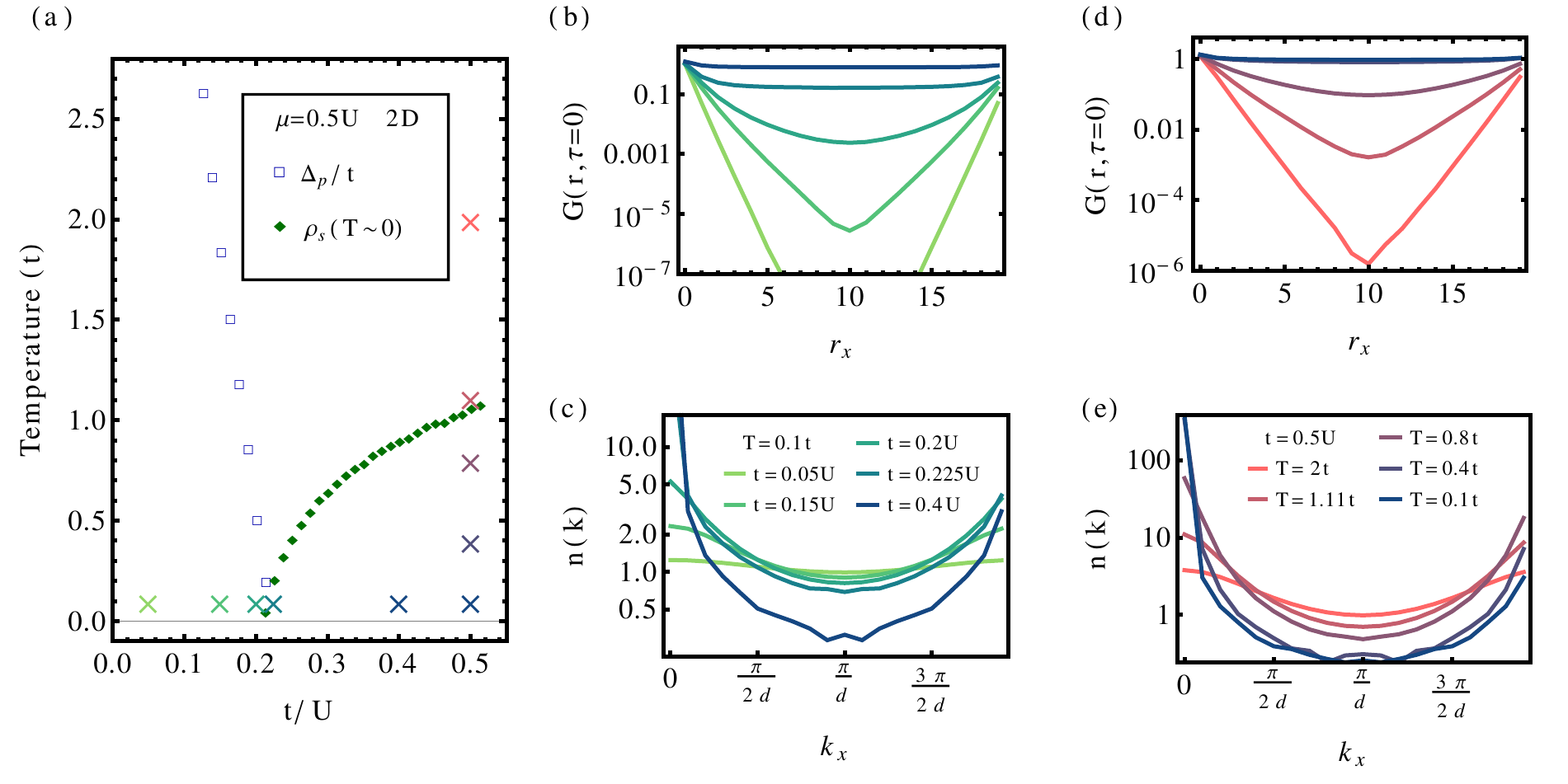}
  \caption{The real space Green function and the momentum distribution in a uniform, periodic, two dimensional, $20^2$ site lattice at $\mu/U=0.5$. (a) The particle gap energy and $\rho_s(T=0)$ indicate each phase's energy scale and reveal the quantum critical point near $t/U=0.21$. (b) The Green function $G(r,\tau=0)$ and (c) the momentum distribution $n(k)$ are plotted at a low temperature for several $t/U$. The temperature dependence of the same quantities are plotted in (d) and (e). Interactions and thermal excitations both deplete the condensate $n(k=0)$, but each momentum is roughly equally occupied in the Mott insulator while $n(k)$ follows the Maxwell-Boltzmann distribution of velocities at high temperature.}
  \label{fig:BHMgrnk}
\end{figure*}

Many properties in the superfluid and Mott insulator phases can be captured exactly within QMC simulations. The energy scales within each ground state are particularly important for estimating critical or crossover temperatures and identifying the quantum critical points. 

In the Mott state, the energy cost to add $(\Delta_p(\vc{q}))$ and remove $(\Delta_h(\vc{q}))$ a particle at temperature $T=0$ and momentum $\vc{q}$ are the crucial energy scales. Away from the QCP, these two modes define the entire spectral function in the Mott insulator
\be
A(\vc{q},\omega)=-S_{h,\vc{q}}\delta(\Delta_h(\vc{q})-\omega)+S_{p,\vc{q}}\delta(\Delta_p(\vc{q})-\omega).
\ee
These gap scales are easily extracted from QMC simulations by the fit to the imaginary time Green function (Eq.~\ref{eq:gf}) in momentum space
\be
\lr{\hatad(\vc{q},0)\hat{a}(\vc{q},\tau)}\approx S_{\vc{q}} e^{-\Delta_h(\vc{q})\tau}+(S_{\vc{q}}+1) e^{-\Delta_p(\vc{q})(\beta-\tau)} \;\;\;\;\mbox{in the Mott state}\label{eq:MIgap}
\ee
where the coefficient $S_{\vc{q}}$ is the spectral weight for the hole excitations shown in Figure~\ref{fig:BHMenergyScales}(a). The spectral weight for the particle excitation must be $S_{\vc{q}}+1$ to satisfy the sum rule $\int_{-\infty}^{+\infty}d\omega A(\vc{q},\omega)/2\pi=1$ for each $\vc{q}$. These gap scales are suppressed and the spectral lines are broadened as the QCP is approached. At the tip of the Mott lobe, both $\Delta_h(0)$ and $\Delta_p(0)$ disappear, while at a QCP at $\mu/U$ above (below) the tip, only $\Delta_p(0)$ $(\Delta_h(0))$ disappears. This is illustrated for $\mu/U=0.5$ in $d=2,3$ in Figs.~\ref{fig:BHMenergyScales} and \ref{fig:BHMgrnk}.

Several related energy scales can be used to characterize the superfluid. The condensate fraction $n_0$ measures the occupation of the zero momentum state and is strictly bounded from above by $\rho_s$. Within linear response theory, it can be shown that the kinetic energy $E_{kin}=t\sum_{\lr{i,j}}\lr{\hatad_i\hataa_j}/L^d$ is a strict upper bound on $\rho_s$, shown in Figure~\ref{fig:BHMenergyScales}(b) \cite{paramekanti1998}. Near the QCP and also as the dimension decreases from infinity in MFT to $d=3$ and $d=2$, interactions and quantum fluctuations increase in strength and suppress $\rho_s$ and $n_0$, shown in Figure~\ref{fig:BHMenergyScales}(b,c).

The finite temperature phase diagram for $d=3$ is shown in Figure~\ref{fig:BHMenergyScales}(c). The smaller of the two gap scales in the MI indicates a crossover energy scale and the superfluid $T_c$ is estimated by the finite-size scaling method outlined above. As $t/U$ increases, $T_c\rightarrow t$. As the QCP is approached within the superfluid, the condensate is destroyed by interactions and both $T_c$ and $\rho_s$ are suppressed as in Figure~\ref{fig:BHMenergyScales}(d). At finite temperatures and near the QCP at $t/U\approx0.192$, observables scale with temperature according to the mean field universality class appropriate for this QCP.

The momentum distribution defined in Eq.~\eqref{eq:nk} is a very important quantity in both theory and experiment. In these remarks we confine ourselves to $d=2$ with $T_c\propto \rho_s(T\rightarrow 0)$ (Eq.~\eqref{eq:n0Tc}) with a numerically determined coefficient somewhat less than 1. The Green function $G(r_x,\tau=0)$ decays to $n_0\sim 1$ in the superfluid and decays exponentially in both the Mott and the high-temperature normal state in Figs.~\ref{fig:BHMgrnk}(b,d). $n(\vc{k})$ shows that in the deep Mott state, $t/U\rightarrow 0$, the momentum states become uniformly occupied while at high temperatures $n(\vc{k})$ is well approximated by the Maxwell-Boltzmann distribution, Figs.~\ref{fig:BHMgrnk}(c,e). It is also clear that as $T$ decreases toward $T_c$, there is a significant occupation of low momentum states even before the system condenses.



\subsubsection{The Local Density Approximation for Trapped Systems}
\begin{figure*}[t!]
  \centering
  \includegraphics[width=0.4\textwidth]{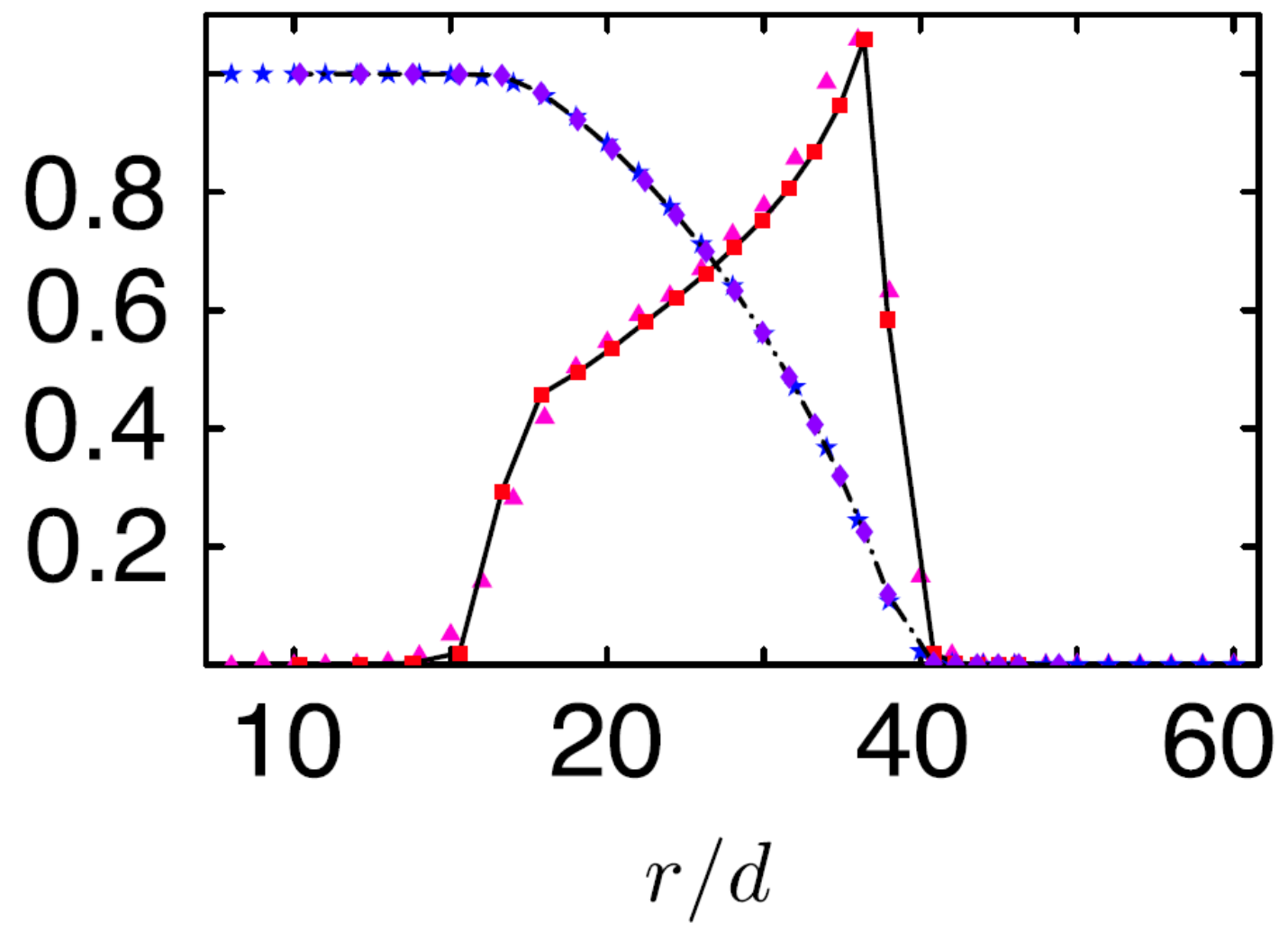}
  \caption{The local density approximation using the mapping (Eq. \ref{eq:LDA}) from a uniform system at chemical potential $\mu$ onto a trapped system at position $r(\mu)$ in 3D. As a function of lattice site within the trap $r/a_{\text{latt}}$, the density (blue stars) and local compressibility (red boxes, defined in Eq. \ref{eq:localcompfirst}) obtained from QMC with the trap compare well with the density (purple diamonds) and local compressiblity (pink triangles) calculated in the equivalent uniform systems. For all the QMC simulations here, $T/t=0.1$ and $t/U=0.15$. The trap frequency was $\omega=2\pi\times 30$Hz and contained $10^5$ bosons. From \cite{zhou2009a}.}
  \label{fig:LDA}
\end{figure*}

A complication specific to optical lattice emulation of the BHM is the presence of an overall confining or trapping potential, which is usually approximated as harmonic, $V_T\propto r^2$. 
The effect of a slowly varying confining potential can be included in the Hamiltonian by introducing a local chemical potential
\be
-\mu\sum_i \hat{n}_i\rightarrow \sum_i\mu_i \hat{n}_i=\sum_i\left(\mu_0-\Omega \left|\vc{r}_i\right|^2\right)\hatn_i,
\label{eq:LDA}
\ee
 where $\mu_0$ is the chemical potential at the center of the trap $\vc{r}_0$, and $\vc{r}_i$ is the distance from $\vc{r}_0$ measured in units of lattice spacing $a_{\text{latt}}$. The trap curvature is parameterized by $\Omega=m a_{\text{latt}}^2\omega^2/2$, where $m$ is the boson mass and $\omega$ is the trap frequency.

The modified chemical potential gives rise to a non-uniform density profile within the trap. 
As the distance from the trap center increases, both the local chemical potential and the local density decrease. In the strong-interaction regime, this gives rise to a density profile often called a ``wedding cake'' since a dense core is surrounded by alternating shells of superfluid and Mott insulating states with decreasing density.

The confining potential can be explicitly modeled within QMC, but that precludes estimation of the superfluid density and is numerically expensive. The local density approximation (LDA) is frequently used to map the properties of a uniform system with $\mu=\mu_i$ onto the those at site $i$ and effective chemical potential $\mu_i$ in the trapped system. This procedure is very accurate in three dimensional systems and remains reasonably good in lower dimensions, especially if $\mu_i$ varies slowly on the scale of $a_{\text{latt}}$, see Figure \ref{fig:LDA}. This approximation is exploited to estimate the superfluid density and to create an accurate picture of a trapped system without resorting to computationally expensive QMC of the entire trapped system. The density profile of any given trap can be easily visualized using the calculated properties of a uniform system. At a fixed $t/U$, the trap center $\mu_0$ is located on a diagram like Figure \ref{fig:MFTrhorhos}. Then $n(r(\mu))$ or $\rho_s(r(\mu))$ can be estimated by lowering $\mu$ at fixed $t/U$ until the vacuum is reached using \cite{zhou2009a}
\be
\mu_{LDA}(r)\approx\mu_0-\Omega r^2
\label{eq:muLDA}
\ee 

The LDA does break down in predictable situations. In regions where $\mu(r)$ changes very quickly, the bosons experience a lower effective dimensionality that cannot be captured by mapping to the uniform system. Similarly, at the boundary between the superfluid and the normal or Mott state, the length scales of correlations diverge\cite{wessel2004,mahmud2011}. Then, even for slowly varying $\mu(r)$, local properties like the critical temperature are affected by the confining potential and are not identical to those of the equivalent uniform system.

\subsection{Experimental Probes}

\renewcommand{\arraystretch}{1.5}
\begin{table}[h]
\begin{centering}
\begin{tabular}{| p{6cm} | p{6.5cm} | l |}
\hline
Experiment Technique & Properties Observed & Scope \\
\hline \hline
Time-of-Flight \cite{greiner2002,spielman2008,folling2005,trotzky2010} & Momentum Distribution, Coherence, Noise Correlations & Global \\
\hline
Potential Gradient and Bragg Spectroscopy \cite{greiner2002, du2010,grimm2006} & The Mott and Superfluid Excitation Spectrum & Global \\
\hline
Magnetic Resonance Imaging Across a Quench \cite{gerbier2006} & Global Number Fluctuations and Number Squeezing, Temperature & Global \\
\hline
{\it In situ} Measurements \cite{gemelke2009,bakr2009,sherson2010} & Single- or Few-site Density, Onsite Number Fluctuations, Temperature & Local \\
\hline
Mass Transport Measurements \cite{hung2010, sherson2010,bakr2010, natu2011} & Local and Global Equilibration Timescales & Local \\
\hline
\end{tabular}
\caption{A brief summary of common techniques to probe atoms in optical lattices.}
\label{tab:exptclass}
\end{centering}
\end{table}
\renewcommand{\arraystretch}{1}

Two broad categories of experimental measurements, local and global, have been developed to probe these isolated optical lattice systems. They are reviewed in Table \ref{tab:exptclass}. The first to be developed were global measurements that average over the entire system. These result in large signals from coherent components of the system but at the cost of losing detailed information about exactly which phases are present and where they are within the system. As discussed above, the bosons are confined in a harmonic potential in addition to the lattice periodic potential, 
which allows multiple phases to coexist within a single system. 
The second class are {\it in situ} measurements that address single or small numbers of sites. The Mott state is easily characterized using these local density measurements, but the 
presence of the superfluid order is difficult to deduce. Another caveat is that in contrast to global-type measurements, many identical runs of the experiment must be averaged over to estimate the local average density and its fluctuations.

These optical lattices are isolated systems, so attaching leads to measure transport or constructing a torsional oscillator to measure $\rho_s$ is difficult;
nevertheless there have been some creative attempts in these directions.

\subsubsection{Mapping Momentum from the Lattice to the Continuum}
Time-of-flight (TOF) experiments were some of the first to show evidence of superfluidity in optical lattices\cite{greiner2002,spielman2008,folling2005}. 
By switching off all potentials, allowing the cloud of bosons to freely expand and observing the bosons' positions using resonant absorption imaging,
these experiments probe phase coherence within the system. The interpretation proved more subtle than first expected, but the experiments were eventually well understood. Direct comparisons between experiments and similar-sized QMC simulations of traps provided definitive proof that the BHM correctly describes these systems.

\begin{figure*}[h]
  \centering
  \includegraphics[width=0.6\textwidth]{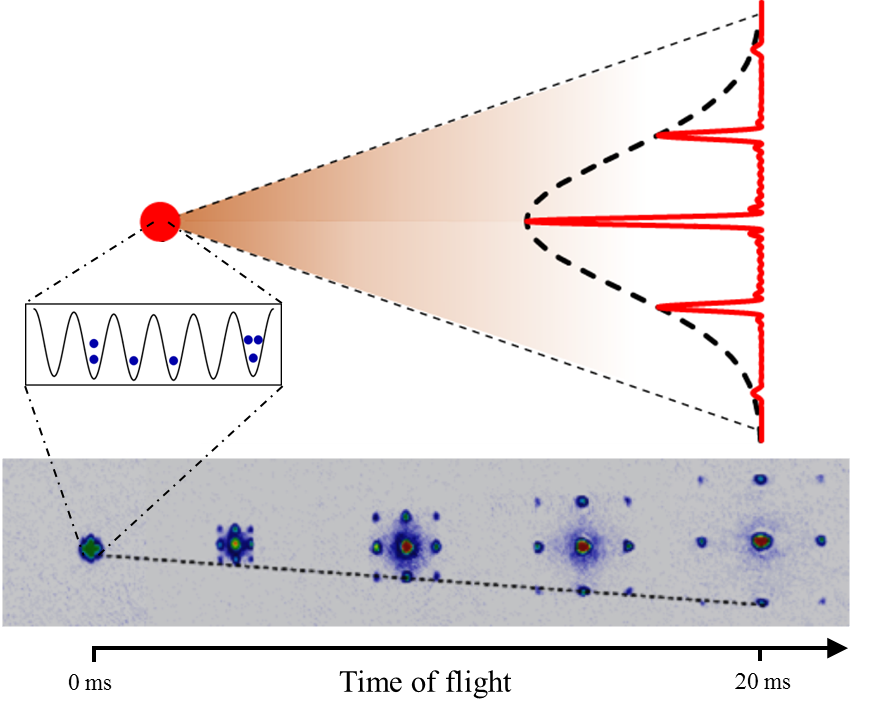}
  \caption{The microscopic optical lattice system is most easily characterized through a time-of-flight expansion. Upon the removal of all optical potentials, atoms occupying quasi-crystal momenta are projected onto continuum momenta and expand from $~25\mu$m to $~500\mu$m. This matter field interferes according to each atom's original phase $\theta_i$, captured in $n(\vc{k})$. In the far-field (long expansion time), the interference pattern is a reliable indicator of the initial state in the optical lattice. }
  \label{fig:TOFschematic}
\end{figure*}

How the atom cloud expands is the crux of time-of-flight measurements and is sketched in Fig.~\ref{fig:TOFschematic}. If the cloud is imaged after a very short time $t_{\rm tof}$, the bosons have not had time to move and the small clouds are difficult to resolve using optical techniques. 
After slightly more time, the bosons move from their initial positions in the lattice but have not had time to expand according to their momenta into resolvable features. Optimally, the cloud is imaged when it is much larger than the optical lattice and the interference pattern is far-field, but before it becomes very dilute. In this regime, the bosons' positions are directly proportional to the initial momentum by $\vc{r}=\hbar t_{\rm tof}\vc{k}/m$. The strong confinement energy necessary to maintain the atoms in the optical lattice is released when the lattice is switched off.
Interaction effects are only a small perturbation during the ballistic expansion.

 There are two essential components in the analysis of time-of-flight experiments\cite{pedri2001}. The first is the momentum distribution $n(\vc{k})$ discussed above (Eq.~\eqref{eq:nk}), which has a singular contribution at $\vc{k}=\vc{0}$ in the superfluid and is a smooth function for all other momenta. 
The second aspect is the mapping from the momentum distribution within the optical lattice to real-space. This requires the Fourier transform of the Wannier function whose form is fixed by the lattice potential and the recoil energy, as discussed above. The resulting interference pattern is
\be
\tilde{n}(\vc{r})=\left(\frac{m}{\hbar t_{\rm tof}}\right)
^d\left|W\left(\vc{k}=\frac{m\vc{r}}{\hbar t_{\rm tof}}\right)\right|^2n\left(\vc{k}=\frac{m\vc{r}}{\hbar t_{\rm tof}}\right).
\ee
where it is important to remember that $\vc{r}$ in the TOF image is related to the initial $\vc{k}$. In three dimensions, the time-of-flight image records the column-integrated density 
$\tilde{n}_{\rm col}\left(k_x,k_y\right)=\int dk_z \tilde{n}(\vc{r}=\hbar t_{\rm tof} \vc{k}/m)$.

\subsubsection{Interpreting Time-Of-Flight Images}
\begin{figure*}[ht]
  \centering
  \includegraphics[width=0.9\textwidth]{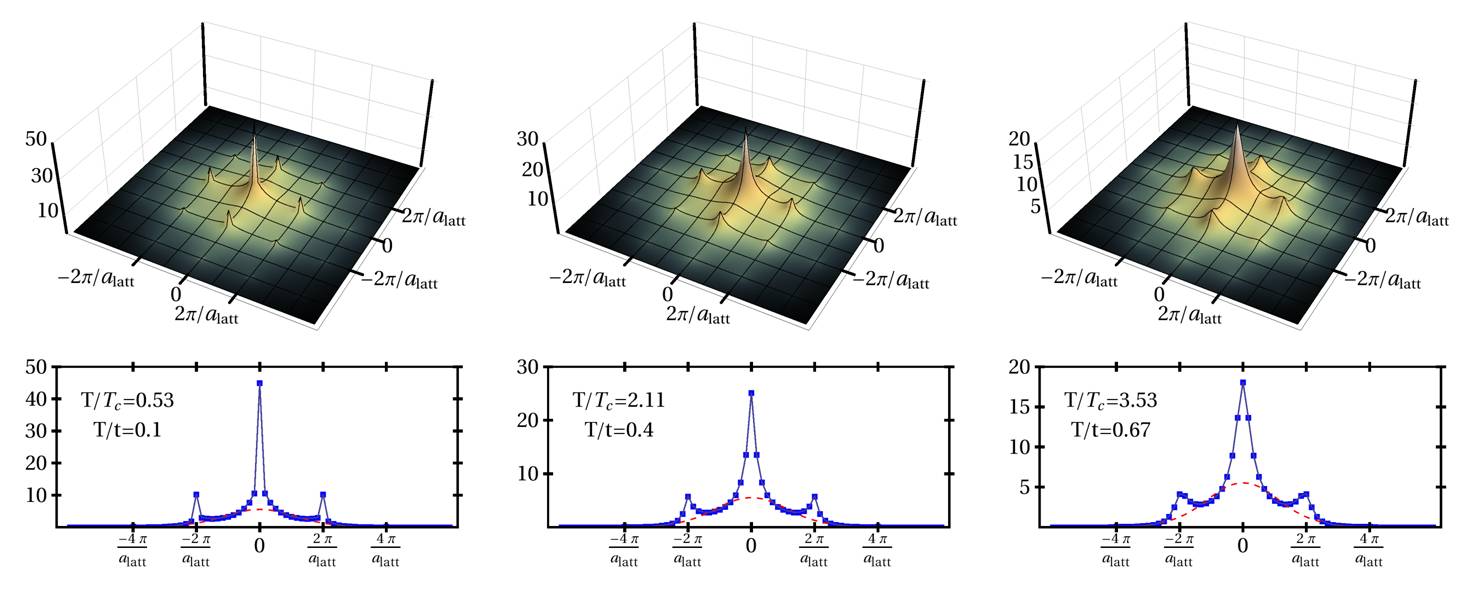}
  \caption{Column-integrated time-of-flight $\tilde{n}_{\rm col}(k_x,k_y)$ images from QMC simulations at $t/U=0.195$, $\mu/U=0.5$, the indicated temperatures and with $12^3$ sites. The top row is the full $\tilde{n}_{\rm col}(k_x,k_y)$ and the bottom row is the cut $\tilde{n}_{\rm col}(k_x,0)$. Above the transition temperature $T_c$,
quantum fluctuations contribute to peaks above the Gaussian background arising from the Wannier function envelope (dashed red curve) 
and below $T_c$ the peaks are sharpened by the singular contribution of $n_0$. These parameters place the simulation very close to the Mott insulator-superfluid boundary, which suppresses $n_0\sim 0.04$ at $T=0$ \cite{kato2008}.}
  \label{fig:peaks}
\end{figure*}

The results are relatively easy to interpret when the system is far from a critical point. Both at high temperature and in the Mott insulator at $T=0$, $\tilde{n}(\vc{r})$ is a broad Gaussian determined by the Wannier function envelope with width $2\pi/a_{latt}$. This observation is trivial deep in the Mott where $n(\vc{k})$ is constant for all $\vc{k}$. At high temperature $n(\vc{k})$ is the Boltzmann distribution with a width proportional to $1/\lambda_{db}\gg 1/a_{latt}$ so the convolution is dominated by the Wannier function envelope. In contrast, in the superfluid most of the particles populate the condensate fraction $n_0=n(\vc{k}=\vc{0})$, which maps onto momenta at reciprocal lattice vectors in the continuum and survive the time-of-flight expansion as sharp peaks due to phase coherence. 

The situation is murkier near a critical point. In addition to singular peaks from $n_0$ within the superfluid and the broad Wannier function envelope, another distinct feature emerges\cite{diener2007,kato2008}. The long-range critical fluctuations strongly affect $n(\vc{k})$ by depleting $n_0$ while still resulting in a peaked momentum distribution
because of enhanced occupancy at low momenta. As shown in Fig.~\ref{fig:peaks}, $\tilde{n}(\vc{r})$ remains sharply peaked even when critical fluctuations have destroyed the condensate. The effect is particularly visible in Fig.~\ref{fig:peaks} where the small $\rho_s\sim0.04$ generates a small singular peak, the low temperature means little weight is associated with the Wannier function envelope, and a significant proportion of $\tilde{n}(\vc{r})$ is determined by the fluctuations.

\subsubsection{Theory-Experiment TOF Comparison}
\begin{figure*}[ht]
  \centering
  \includegraphics[width=0.9\textwidth]{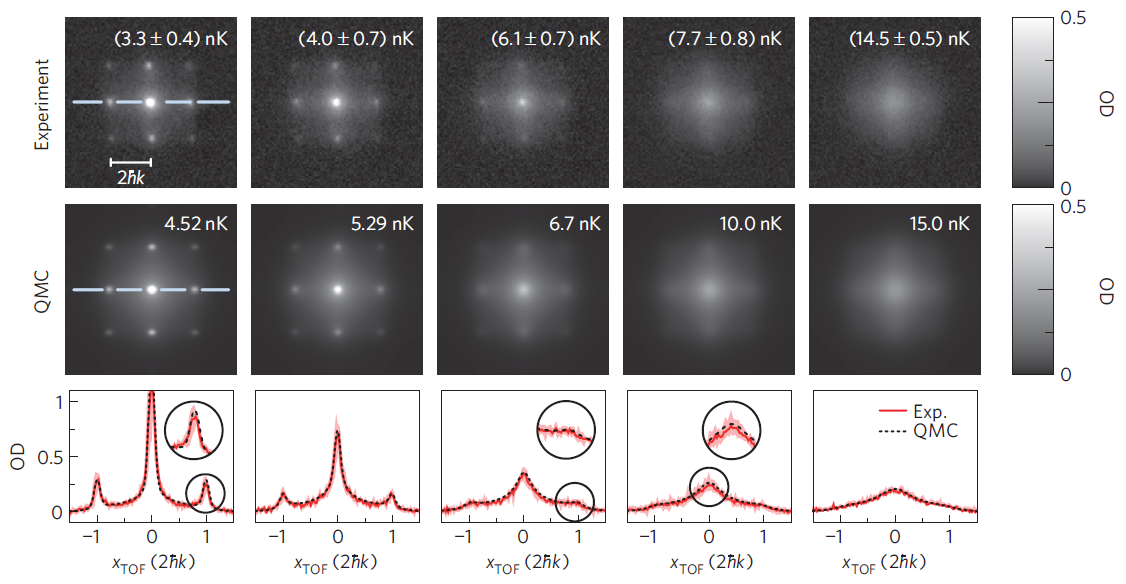}
  \caption{Integrated time-of-flight images $\tilde{n}_{\rm col}(k_x,k_y)$ in experiment (top row) and from QMC simulations (middle row), for $t/U=0.22$ and $T_c=5.3$nK in the uniform system. The results are directly compared in the bottom row for $\tilde{n}_{\rm col}(k_x,0)$. The experimental temperature is estimated by estimating the temperature and entropy before application of the optical lattice potential, assuming adiabaticity during the lattice turn-on, and extracting the bosons' temperature from a calculated entropy per particle in the lattice. The temperature shift for this coupling is on the order of $30\%$ \cite{trotzky2010}.}
  \label{fig:troyerComp}
\end{figure*}

A tour de force experiment-simulation combination confirmed that bosons in optical lattices do emulate the BHM and validated the previous discussion with stunning accuracy\cite{trotzky2010}. Shown in Fig.~\ref{fig:troyerComp}, the time-of-flight images from experiment and from QMC simulations with similar particle number and trapping potential agree extremely well. This conclusively showed that analyses of TOF images can reveal the presence of a superfluid within a trapped system. However, the mismatch between the estimated temperature in the experiment and the QMC simulation's temperature demonstrated a need for better thermometry of these extraordinarily isolated systems, since direct QMC-experiment comparisons are computationally expensive. 

Many proposals concerning the extraction of detailed phase information without direct comparison with QMC have been put forward. 
Within TOF imaging, the quantities that correlate with the condensate fraction include the height or width of the central peak and the visibility or the intensity of the satellite peaks relative to the background \cite{diener2007,trotzky2010}. Another set of proposals remap the phase diagram from the $t/U-\mu/U$ to the $t/U-\tilde{\rho}$ parameter space, where $\tilde{\rho}$ is a characteristic density that rescales the total boson number by the trap length scale $\left(\Omega/t\right)^{-d}$ \cite{rigol2009,garcia2010,mahmud2011}. Trap-size scaling to extract critical properties is another approach \cite{campostrini2010}. These approaches have merits but are unable to match the detailed information about each individual phase within the trap available from {\it in situ} imaging.

\subsubsection{{\it In Situ} Experiments}
\begin{figure*}[h]
  \centering
  \includegraphics[width=0.5\textwidth]{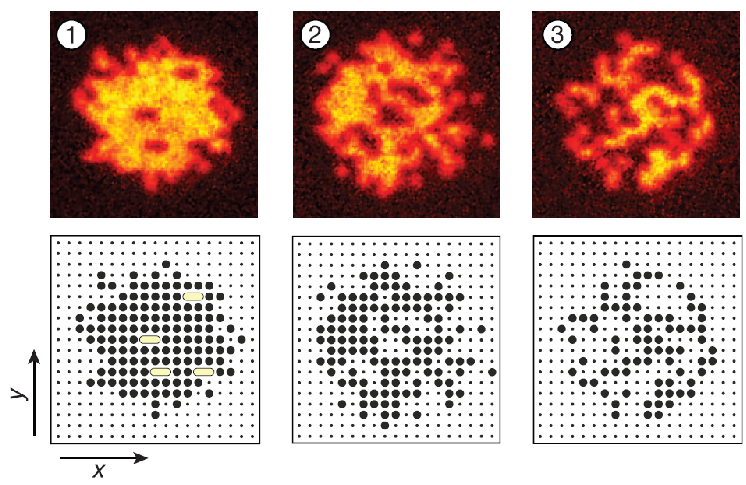}
  \caption{{\it In situ} density measurements (top column) and reconstructed site occupation modulo 2 (bottom row). Deep in the Mott state, hole and doubly occupied states are recorded as zero (modulo 2) and are tightly bound (left) while in the superfluid (right) the on-site occupation number has a much broader distribution. Adapted from \cite{endres2011}.}
  \label{fig:insituimage}
\end{figure*}

Several experimental groups achieved single- or few-site {\it in situ} imaging of optical lattice systems nearly simultaneously using multiple techniques \cite{bakr2009,sherson2010,gemelke2009}. This breakthough enabled detailed studies of the local density distribution within each phase in a trap, although the highest resolution techniques can only measure whether the onsite density is odd or even (Fig.~\ref{fig:insituimage}). Almost immediately, it became clear that similar systems could have radically different and unexpected timescales for global equilibration. In these experiments, the lattice potential was ramped to a relatively large potential $(t/U\sim0.05-0.1)$ and the equilibration timescales estimated by measuring $n(r)$ at a series of times after the ramp and finding how long it took to stabilize. Under some conditions, equilibration occurred in a fraction of the tunneling time $\sim 0.1 t^{-1}$ \cite{bakr2010} but it could also take longer than the experimental timescales $\sim 10 t^{-1}$ \cite{hung2010}.

Results from time-dependent MFT and DMRG indicated how to resolve this discrepancy\cite{natu2011,bernier2011,bernier2012}. Instead of the tunneling timescale, it is the particle-hole timescale $U^{-1}$ that sets the equilibration time in the superfluid. After a ramp, the amplitude of each occupation number $(0,1,2,\ldots)$ on each site fluctuates before being damped by the interaction $U$. This result, valid in a superfluid in either a trap or a uniform system, is consistent with the former set of experiments. 
The longer timescale is directly related to the formation of Mott insulator shells within the trapped system. Since a Mott state has a very small compressiblity at low temperatures, the mass transport of bosons across a Mott shell occurs very slowly and can hinder global equilibration if, to reach the equilibrium state, many bosons must be transported across a Mott shell. It is this trap-dependent distinction between fast equilibration within a superfluid shell but slow equilibration between superfluid shells that separated the experiments' timescales so widely. At low temperatures, superfliud shells are isolated by the Mott shells and each can exhibit a distinct temperature and effective chemical potential.

\subsubsection{Adapting the Fluctuation-Dissipation Theorem}
\begin{figure*}[t!]
  \centering
  \includegraphics[width=0.9\textwidth]{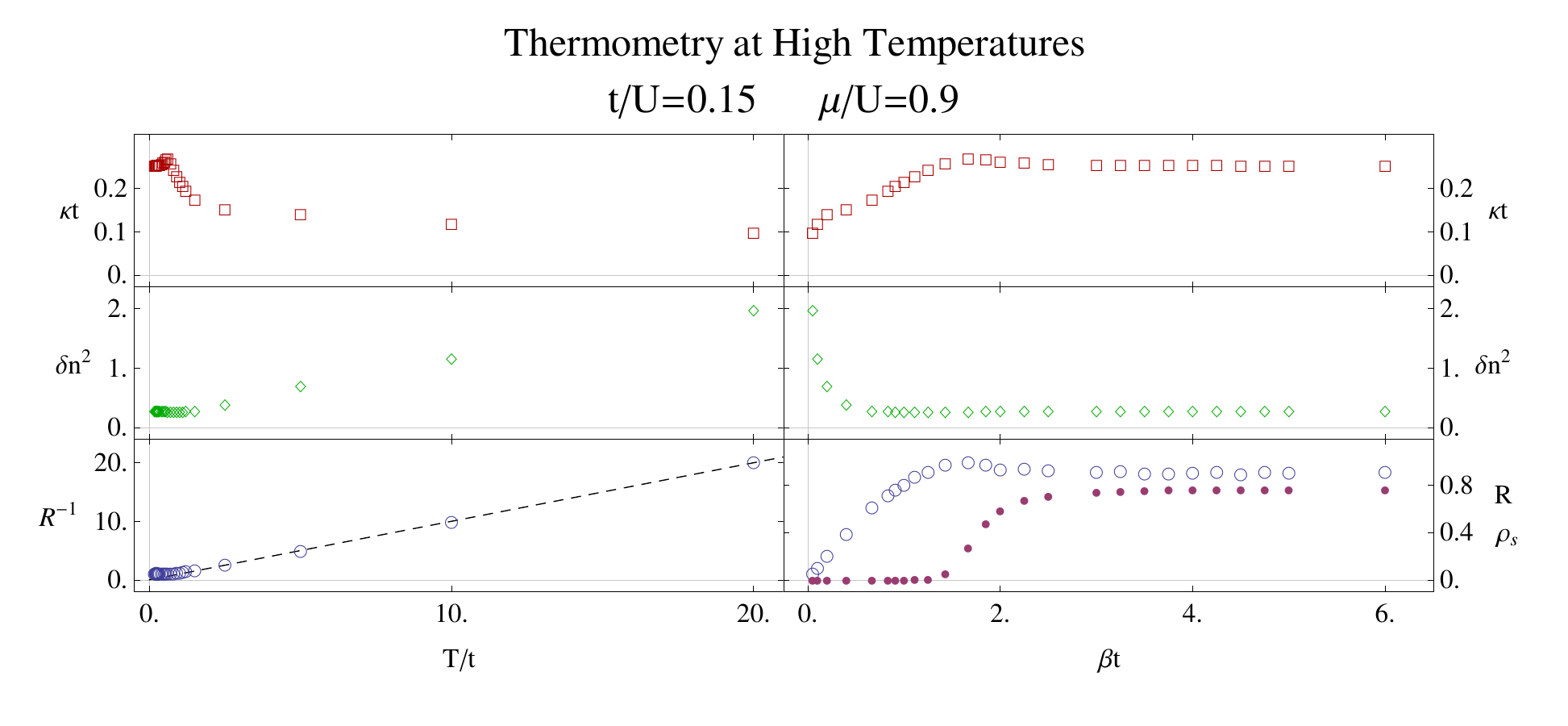}
  \caption{The application of the fluctuation-dissipation theorem to thermometry. At high temperatures only local number fluctuations contribute to Eq.~\ref{eq:FDT} so that $R(\mu)=\kappa(\mu)/\delta n_i^2=1/T$ (shown as the dashed line in the lower left figure). This is primarily due to the onsite number fluctuations $\delta n_i^2\sim T$ since $\kappa$ is approximately independent of temperature in this regime. At low temperature, $R^{-1}$ deviates from $T$ due to quantum effects which, in this case, result in a transition into the superfluid state (see $\rho_s$ filled disks in lower right figure).
}
  \label{fig:thermometry}
\end{figure*}

\begin{figure*}[t!]
  \centering
  \includegraphics[width=0.65\textwidth]{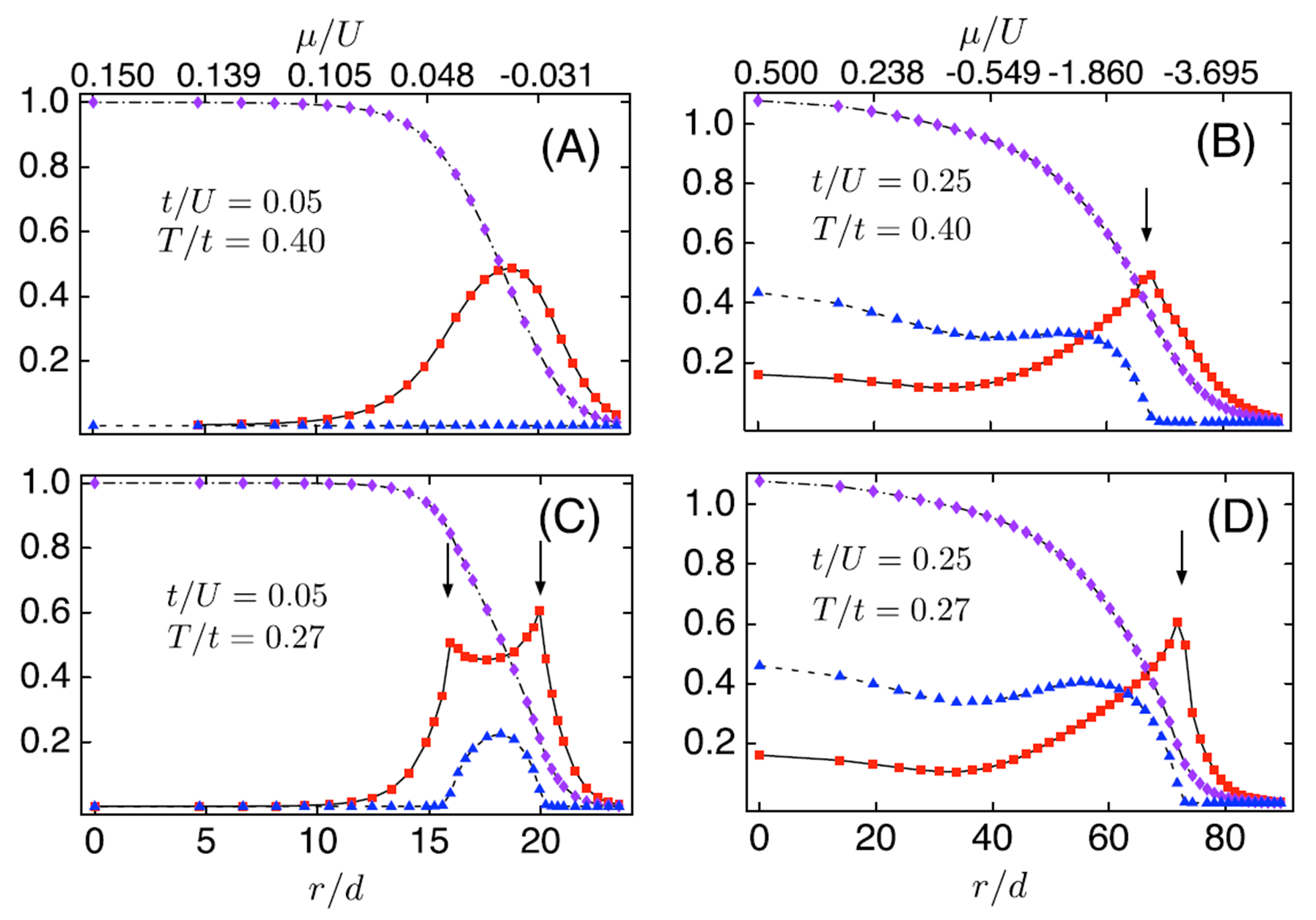}
  \caption{Kinks in the local compressibility appear where the superfluid vanishes. The density (purple diamonds), local compressibility (red boxes) and superfluid density (blue triangles) are shown at (a,b) low and (c,d) high temperatures. In (a) no portion of the system is superfluid, while in (b,c,d) the arrows indicate the boundaries of the superfluid region. These are QMC results of a uniform system mapped onto a trap using LDA \cite{zhou2009a}.}
  \label{fig:kinks}
\end{figure*}

The development of {\it in situ} imaging spurred theoretical work into the connections between experimentally measurable local properties like density $\lr{\hatn_i(\mu)}$ and 
density fluctuations $\langle \delta \hatn_i^2(\mu)\rangle=\langle (\hatn_i-\lr{\hatn_i})^2\rangle$ at site $i$ and desired information like the temperature, characteristic energy scales
and the phase of the system. The natural starting place is the fluctuation-dissipation theorem (FDT) which we apply to a large patch of the system with a volume $\Omega_{\rm patch}$
\be
\kappa(\mu)={{\partial \lr{\hatn(\mu)}}\over{\partial\mu}}=\frac{1}{T}\frac{\lr{\delta \hat{N}^2(\mu)}}{V}
\label{eq:FDT}
\ee
that exactly relates the compressibility to the {\em total} number fluctuations $\lr{\delta \hat{N}^2(\mu)}=\sum_{ij}\lr{\delta \hatn_i \delta \hatn_j}$ 
within the patch at temperature $T$ in a uniform system. Here $\hat {N}=\sum_i \hatn_i$, where the sum runs over sites within the patch and $\delta\hatn_i=\hatn_i-\lr{\hatn_i}$.
In the cold atom experiments, it is possible to independently measure two quantities: (a) $\kappa(\mu)$ the change in the density due to a small
change in the chemical potential around a reference value achieved, for example, by tightening or softening the trapping potential,
and (b) $\lr{\delta N^2(\mu)}$ the boson number fluctuation within the patch, with the statistics generated through multiple experiment runs with the trapping potential fixed at the reference value. In an equilibrated patch, the ratio $\kappa(\mu)/\lr{\delta\hat{N}^2(\mu)}$ provides an accurate estimate of the temperature. More detailed analyses of these observables reveal significant local properties, including the superfluid phase boundary, energy scales, and the local state (superfluid, normal, Mott or quantum critical). The Mott insulating state can be identified using $\kappa(\mu)=0$ since $\kappa$ is finite in the other states. Careful analysis of $\kappa(T)$ or $\kappa(\mu)$ shows it also encodes the superfluid critical temperature $T_c$ or critical filling $\mu_c$. 

The FDT is easily adapted to local thermometry within the local density approximation. 
QMC simulations show that density-density correlations $\lr{\delta \hatn_i\delta\hatn_j}$ quickly decay to zero with a characteristic length $\xi_{nn}\lesssim 5$ lattice spacings, except very close to critical points. In a uniform system the FDT sum over number fluctuations can be restricted to a patch of sites near site $i$, as mentioned above, of linear dimension $\xi_P$. We then obtain a ``local" definition of the compressibility 
\be
\kappa_i(\mu)=\frac{1}{T}\lr{\delta \hatn_i(\mu) \sum_j \delta \hatn_j(\mu)}\approx\frac{1}{T}\lr{\delta \hatn_i(\mu) \sum_{|r_i-r_j|\leq\xi_P} \delta \hatn_j(\mu)}
\label{local-kappa}
\ee
without a significant loss of accuracy, where $\kappa_i(\mu)=\partial\lr{\hatn_i(\mu)}/\partial\mu$. It is important to notice that the variation 
of the density at site $i$ is taken with respect to a global chemical potential $\mu$; as a result $\kappa_i$ includes correlations between site $i$ and others within the patch.
In a trapped system by using LDA we next relate position in the trap $r_i$ to chemical potential (equation \ref{eq:muLDA}) and obtain
\be
\kappa_i=-\frac{1}{2\Omega r_i}\left. \frac{\partial\lr{\hatn(r)}}{\partial r}\right|_{r\rightarrow r_i}.\label{eq:localcompsecond}
\ee
As in the uniform system, QMC simulations have shown $\xi_{nn}$ is on the order of several lattice spacings in the trapped system. In fact, at very high temperature $\xi_{nn}\sim a_{latt}$ so the temperature can be estimated directly from the ratio $\kappa(\mu)/\delta n_i^2$ (see Fig.~\ref{fig:thermometry}). It is important to remember that this method is less reliable wherever $n(r)$ varies rapidly or near a transition between a superfluid and a normal or Mott insulating state.

It is useful to contrast $\kappa_i$ with the truly local compressibility $\kappa^L_i(\mu_i)=\partial\lr{n_i(\mu_i)}/\partial\mu_i=\lr{\hatn_i^2}-\lr{\hatn_i}^2$. The latter quantity $\kappa^L_i(\mu_i)$  
measures the {\it local} change in density in response to a {\it local} change in the chemical potential at site $i$. It is impossible for $\kappa^L_i$ to pick up the singular behavior near a phase transition that is encoded in the behavior of long range correlation functions. In contrast, the ``local'' compressibility $\kappa_i(\mu)$ in Eq.~\eqref{eq:localcompsecond} is a variation with respect to a {\it global} change in $\mu$ and is clearly sensitive to the long-wavelength excitations associated with a phase transition, as indicated in Eq.~\eqref{eq:FDT}.

QMC simulations show that kinks appear in $\kappa(\mu)$ where singular fluctuations cause $\rho_s$ to vanish in both trapped and uniform systems (Figs.~\ref{fig:LDA} and \ref{fig:kinks}). The kinks are rounded in the trap due to finite size and reduced dimensionality.

\begin{figure*}[t!]
  \centering
  \includegraphics[width=0.9\textwidth]{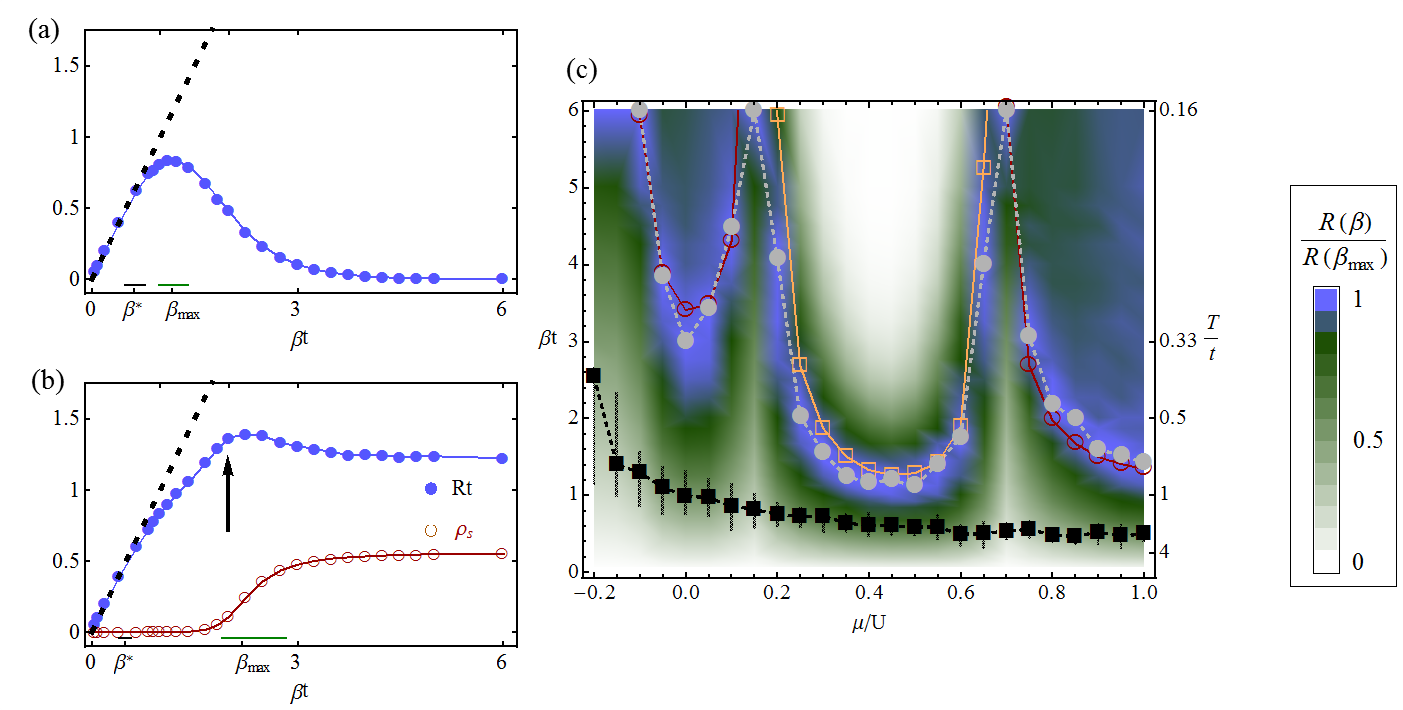}
  \caption{Using a single observable to map the finite temperature phase diagram. Panels (a) and (b) show characteristic $R(T)$ (Eq. \ref{eq:FDTratio}) in a Mott insulator $(\mu/U=0.4)$ and a superfluid $(\mu/U=0.8)$, respectively, and $t/U=0.15$. The dashed black line is the high temperature case $R(T)\sim\beta$. Panel (c) A finite temperature phase diagram constructed from the temperature $T^\ast$ where $R(T)$ deviates from $\beta$ (black boxes) and the temperature $T_{max}$ of the peak in $R(T)$ (gray discs) agrees well well with the superfluid $T_c$ estimated from $\rho_s\rightarrow 0$ (red circles) and from vanishing $\kappa(\mu)$ (orange squares). The background is a density plot of $R(\beta,\mu)$ normalized so $R(\beta_{max})=1$ for each value of $\mu$. From \cite{duchon2012}.}
  \label{fig:FDTpic}
\end{figure*}

With thermometry and phase identification under control, attention has turned to locating QCPs and identifying their universality classes. Several numerical studies successfully scale the local density [Eq.~\eqref{eq:denSCALE}] and local compressibility [Eq.~\eqref{eq:compSCALE}] near QCPs \cite{zhou2010,hazzard2011,fang2011} but the sole experiment to date has had difficulty in obtaining critical exponents\cite{zhang2012}. One of the complications is the identification of what data should be scalable, i.e. in the quantum critical region. 
The insights gained from using the FDT for thermometry and from identifying the superfluid transition by kinks in $\kappa(\vc{r}_i)$ can be combined into a proposal that maps the quantum critical region \cite{duchon2012}.

Our proposal is to construct a ratio 
\be
R(\mu)=\frac{\kappa(\mu)}{\delta n^2(\mu)}
\label{eq:FDTratio}
\ee
of the compressibility as defined in Eq.\eqref{eq:localcompsecond} and the local number fluctuations $\delta n^2(\mu)=\lr{\delta \hatn_i\delta\hatn_i}$.
At high temperatures $\xi_{nn}$ is negligble and $\kappa(\mu)\approx\beta \delta n^2(\mu)$ giving a ratio $R\sim \beta$ or $1/R\sim T$.
However, as the temperature is lowered, the ratio deviates from $\beta$ at a characteristic temperature $T^\ast$, which is identified as an upper bound to the quantum critical region and is shown in Fig.~\ref{fig:FDTpic}(a,b). 
If a superfluid emerges at low temperature, the ratio $R(T)$ peaks due to the kink in $\kappa(\mu)$ at the critical temperature for a system with chemical potential $\mu$ and asymptotes to a constant at low temperatures since $\kappa(\mu),\delta n^2(\mu)>0$ at $T=0$. In a Mott state $R(T)\rightarrow 0$ as $T\rightarrow 0$ since $\kappa(\mu)$ vanishes, while quantum fluctuations maintain $\delta n^2(\mu)$ at a finite value. At intermediate temperatures a peak emerges that is correlated with the formation of a Mott state. The temperature $T_{max}$ of the peak in $R(T)$ and the temperature $T^\ast$ when it diverges from $1/T$ can be combined into a finite temperature phase diagram obtained from the single quantity $R$, shown in Fig.~\ref{fig:FDTpic}(c). Note that QCPs can be located using this ratio because there $R(T)$ behaves like neither phase. As discussed above, this method is easily extended to trapped systems using LDA. Indeed the presence of several coexisting phases makes this phase-diagram-mapping method more efficient because a single equilibrated experiment will provide information about the phases of a wide range of chemical potentials at a single temperature.

\subsection{ Future Directions for Bosons in Optical Lattices}
The fundamental groundwork to identify and understand the phases and phase transitions of the BHM is in place both theoretically and experimentally, but the BHM has several other exciting directions. 
The dynamics and excitation spectra near QCPs remain an active field of study. Particularly near the particle-hole-symmetric QCP in two dimensions, the structure of the spectral density and whether the Higgs mode can be observed as a well-defined excitation at zero momentum are important questions being addressed through analytic methods and analytic continuation of QMC correlations to real frequencies \cite{sachdev2011,pollet2012,podolsky2012}. 
Experimentally, one of the first observations of the Higgs mode in a condensed matter system has been made through periodic modulations of the optical lattice \cite{endres2012}. 
The phases of the disordered BHM are also becoming better understood using the strong disorder real-space renormalization group and large-scale QMC \cite{iyer2012,soyler2011,lin2011,meier2012}. These studies investigated the fate of the Mott insulator and the superfluid, the emergence of the Mott and Bose glass phases, estimated critical exponents and examined the universal DC conductivity value. These could shed light on systems like $^4$He in Vycor, superconductor-insulator transitions, and disordered optical lattice experiments \cite{white2008,pasienski2010}. 

An even greater variety of systems require only minor modifications of the basic BHM discussed here and explorations of them have begun both theoretically and in optical lattice experiments. The square or cubic lattice geometry can be altered to mimic the honeycomb structure of graphene or to triangular and other geometries  \cite{soltan2011a}. Further manipulation can yield spin-dependent optical lattices so that with multiple species of bosons, novel phases emerge such as spin-ordered insulators and superfluids with a complex order parameter and non-trivial orbital symmetry \cite{altman2003,soltan2011b}. With multiple species comes the possibility to study spin textures and topological states in systems with an artificial gauge field \cite{huber2011,cole2012}. Frustration in an antiferromagnetic system is yet another branch of the BHM in an artificial gauge field \cite{eckardt2010}. As outlined above for the basic BHM, experimental and theoretical advances are continuing almost simultaneously in the study, detection and characterization of these novel systems.

\newpage

\section{Fermions in Optical Lattices: Fermi Hubbard Model}

\subsection{Motivation from high-$T_c$ superconductors} 
The Fermi Hubbard model is the simplest lattice model that incorporates interactions between fermions. It was proposed in the 1960s by Hubbard~\cite{Hubbard1963} and 
investigated extensively by Anderson~\cite{Anderson1950} and Gutzwiller~\cite{Gutzwiller1965} as a way to understand the origin of superexchange interactions driving antiferromagnetism and of electron correlations in narrow band systems. More recently, interest in this model peaked because of the discovery of high temperature superconductivity 
in cuprates in 1987 and the suggestion, once again by Anderson, that the single-band Hubbard model could encapsulate the complex and unusual phenomenology of the cuprates. 

\begin{figure*}[t!]
  \centering
  \includegraphics[width=0.45	\textwidth]{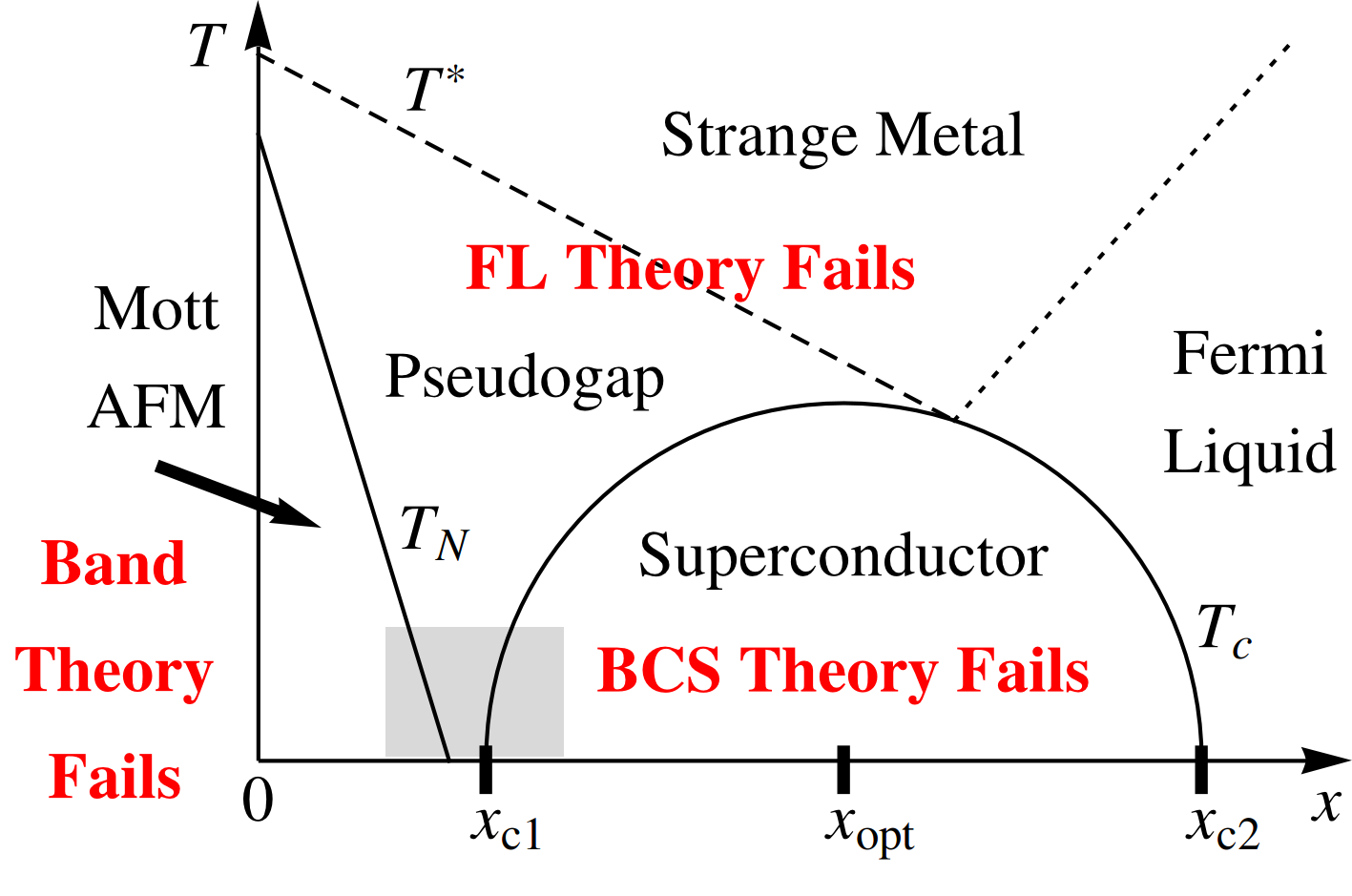}
  \caption{Schematic phase diagram for the high temperature cuprates showing the breakdown of three major paradigms 
of condensed matter physics: the failure of band theory at $x=0$ where a half-filled system is Mott insulating due to strong correlation effects; the failure of the BCS paradigm at finite doping yielding not one, but two energy scales, the higher $T^\ast$ a crossover scale describing pair formation and a lower $T_c$ describing a phase transition to a coherent
d-wave superconducting state; and the failure of Fermi liquid theory in the normal state which is anything but normal showing a 
suppression of density of states in the pseudogap region, and the absence of well-defined quasi-particles in the strange metal region.
}
  \label{fig:hightc}
\end{figure*}

One of the reasons research in high-$T_c$ superconductivity (HTSC) was pursued with such fervor for over two decades is because three major paradigms of condensed mater physics break down 
in different portions of the phase diagram, as illustrated in Fig.~\ref{fig:hightc}. Starting with a square plane of CuO$_2$ with one electron in the $d_{x^2-y^2}$ orbital 
of the parent compound La$_2$CuO$_4$, it is possible to remove electrons by substituting $x\%$ of trivalent La with divalent Sr thereby generating $x\%$ of holes in the lattice.
A rich phase diagram with multiple surprises emerges as a function of increasing hole concentration that breaks band theory, Fermi liquid theory and BCS theory paradigms.

Already at $x=0$ the system shows its first surprise: the formation of a Mott insulator with a large gap to single charge excitations even though a half-filled
system should be metallic according to band theory. The presence of a Mott insulator is indicative of the importance of strong Coulomb correlations. Once the charge excitations are gapped out, the low-lying excitations are 
dominated by the spin degrees of freedom. The ground state at half-filling is antiferromagnetic with linearly dispersing magnons. 

As the system is ``doped'' with holes, Mott insulating behavior and antiferromagnetism are weakened and the system becomes an unconventional superconductor with a d-wave pairing symmetry whose transition temperature $T_c$ 
shows a non-monotonic dome shape. This superconductor does not fit the standard BCS paradigm which had successfully described all superconductors prior to the discovery of HTSCs.  
At least on the underdoped side, cuprates show the existence of two distinct temperature scales: a scale $T^\ast$ where pairing correlations develop and a scale $T_c$ where long-range phase coherence develops.  
Note that $T^\ast$ could be influenced by other factors such as competing orders, disorder, and Coulomb effects.
The temperature range between $T^\ast$ and $T_c$ shows a pseudogap with a suppressed density of states.
This is in stark contrast with the BCS paradigm where a single temperature scale $T_c$ is determined by the superconducting gap, the mean field description works extremely well, and phase fluctuations can be ignored. 

The third paradigm uprooted by HTSCs is Landau Fermi liquid theory, which describes a typical normal state above $T_c$ with well-defined quasiparticles and a resistivity $\rho\sim T^2$.
In the cuprates the resistivity is distinctly non-Fermi liquid-like with $\rho\sim T$, known as a ``strange metal".

Turning now to the Hubbard model that is purported to describe high Tc superconductors, including the Mott-antiferromagnetic insulator, the pseudogap and the strange metal phases,
it is humbling to note than an exact solution can only be found in 1D by resorting to Bethe ansatz methods. Even then it is difficult to obtain correlation functions and spectral functions.
This is where the utility of optical lattice emulators containing fermions with two hyperfine species becomes useful. We hope that a direct emulation of the fermions in the optical lattice
will give us an understanding of just how much of the phenomenology 
of the cuprates can be captured by the Hubbard model.

\subsection{The Fermi Hubbard Model}

For the sake of concreteness, we will focus the discussion on the case of two species of fermions, $\up$ and $\dn$, in a cubic optical lattice.  We will assume that the lattice is sufficiently deep that the physics of the lowest Bloch band can be approximated by a single-band Hubbard model, and we will assume that the ``action'' takes place only in the lowest Bloch band.  Then the system is described by the Hubbard Hamiltonian
	\begin{align}
	\hat{H}^\text{trad}
	&=
	-	\sum_{\rrr\rrr'\sigma} t_{\rrr\rrr'} \cdag_{\rrr\sigma} \cccc_{\rrr'\sigma}
	-	\sum_{\rrr\sigma} \mu^\text{trad}_\sigma \hatn_{\rrr\sigma}
	+	\sum_{\rrr}		U \hatn_{\rrr\up} \hatn_{\rrr\dn}
	,
	\label{HubbardHamiltonianTraditional}
	\end{align}
where $\rrr$ and $\rrr'$ are site indices, $\sigma=\up,\dn$ distinguishes the two species, $t_{\rrr\rrr'}$ are tunneling amplitudes such that $t_{\rrr\rrr'}=t$ for neighboring sites and $0$ otherwise, $\cdag_{\rrr\sigma}$ and $\cccc_{\rrr\sigma}$ are fermion creation and annihilation operators, $\mu^\text{trad}_\sigma$ is the chemical potential for each species defined in the ``traditional'' way, and $\hatn_{\rrr\sigma} = \cdag_{\rrr\sigma} \cccc_{\rrr\sigma}$ are number operators.  For a unified theoretical discussion it is very convenient to rewrite the Hamiltonian in the symmetrized form $\hat{H}=\hat{H}^\text{trad} + \mu^\text{trad} - \sum_\rrr \tfrac{U}{4}$,
	\begin{align}
	\hat{H}
	&=
	-	\sum_{\rrr\rrr'\sigma} t_{\rrr\rrr'} \cdag_{\rrr\sigma} \cccc_{\rrr'\sigma}
	-	\sum_{\rrr\sigma} \mu_\sigma \hatx_{\rrr\sigma}
	+	\sum_{\rrr}		U \hatx_{\rrr\up} \hatx_{\rrr\dn}
	\label{HubbardHamiltonian}
	\end{align}
where $\hatx_{i\sigma} = \hatn_{i\sigma} - \half$ are number operators measured from half-filling and $\mu_\sigma = \mu^\text{trad}_\sigma - \frac{U}{2}$ are chemical potentials with respect to the band center.

%

For nearest-neighbor hopping, the Hamiltonian is particle-hole symmetric so $F(n)=F(2-n)$, or the free energy at a density $n=N_{fermions}/N_{sites}$ is equal to that at a density $(2-n)$. Half-filling ($n=1$) is taken as the reference and densities
less than half-filling are considered ``hole" doping that decrease particle number to $n=0$, the empty lattice, whereas densities greater than half-filling denote ``particle" doping and can reach a maximum of $n=2$ at the band insulator.

\begin{figure}[htbp] \centering
		\includegraphics[width=0.95\textwidth]{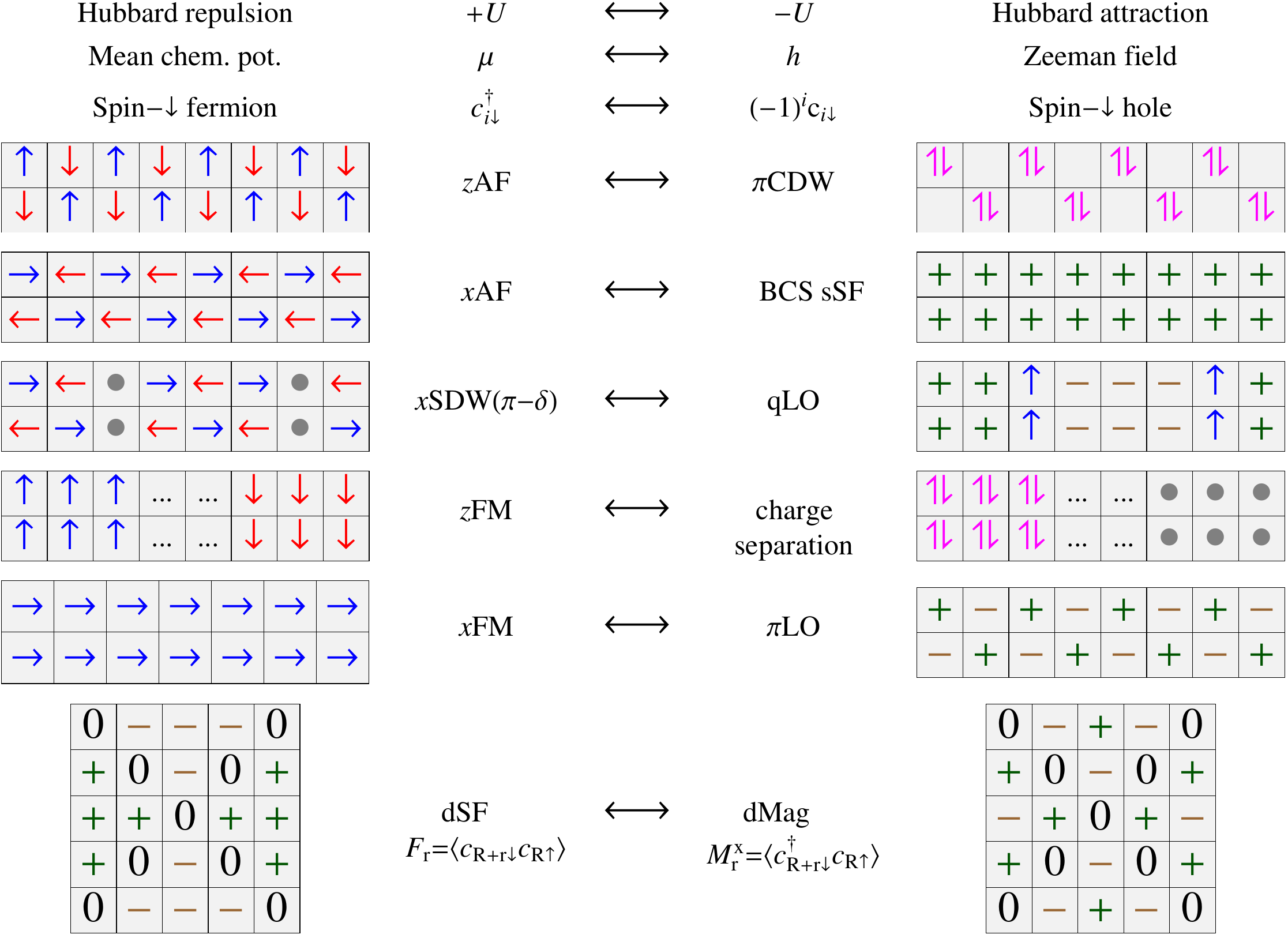} 
	\caption{
		\label{LMTCartoons}
		Simplified 2D depictions of various possible ground states of the Fermi Hubbard model.
		Repulsive interactions tend to favor magnetic order, 
		whereas attractive interactions tend to favor density order or paired superfluids.
		Arrows represent spin-half fermions, gray circles represent holes, 
		and other symbols ($0,+,-$) generally represent the sign of the pairing order parameter.
		The figures omit many subtleties. 
		For example, actual ground states typically contain admixtures of many other configurations 
			due to quantum fluctuations of charge and spin.
		An LO state is characterized by oscillations of the pairing amplitude $\Delta(\RRR,\rrr)$
			with the pair center-of-mass position $\RRR$ (as depicted), but the pairing symmetry is still s-wave.
		In contrast, $\Delta$ in a dSF state is independent of $\RRR$
			but has d-wave-like symmetry with respect to the relative coordinate $\rrr$
			 (as depicted).
	}
	\end{figure}

The form of the Hamiltonian is invariant under SU(2) spin rotations and U(1) phase rotations, and under the $\mathbb{Z}_2$ operations of parity, charge conjugation, time reversal, and the ``Lieb-Mattis'' transformation (LMT).  The LMT relates the repulsive and attractive Hubbard model on a bipartite lattice through a particle-hole transformation on the down spins with a $\pi$ phase shift for the B sublattice: $\cccc_{A\dn} \rightarrow \cdag_{A\dn}$, $\cccc_{B\dn} \rightarrow -\cdag_{B\dn}$.  
Figure~\ref{LMTCartoons} illustrates the mapping between magnetic phases of the repulsive Hubbard model to density-ordered or paired phases of the attractive Hubbard model.  These phases include ferromagnets (FM), antiferromagnets (AF), spin density waves (SDW) or stripes, charge density waves (CDW), Bardeen-Cooper-Schrieffer (BCS) s-wave superfluids, Larkin-Ovchinnikov (LO) phases, d-wave superfluids (dSF), and exotic ``d-wave magnets'' (dMag).

The tunneling ($t$) term is diagonal only in momentum space, whereas the interaction ($U$) term is diagonal only in real space.  Thus the Hubbard model cannot be solved exactly, and one must resort to some sort of approximation. Before discussing the different methods that have been applied to this model, the phase diagram is reviewed.

\subsection{Phase Diagram}
For a bipartite lattice at half-filling,
the Fermi surface is perfectly nested (such that it overlaps itself when translated by a nesting wavevector).
In two dimensions, it is diamond-shaped with nesting vector ${\bf q}=(\pi,\pi)$, while in three dimensions it has a complicated multiply connected shape with nesting vector ${\bf q}=(\pi,\pi,\pi)$ as illustrated in Fig.~\ref{FermiSurfaces}. 
Thus a small repulsive $U$ produces weak antiferromagnetic order, opening up an exponentially small gap in the single particle spectrum 
$E_g\sim t e^{-c\sqrt{t/U}}$ $(E_g\sim t e^{-ct/U})$ in two (three) dimensions where $c$ is a constant of order unity. 
This is the weak coupling 
regime of a spin-density wave insulator for which the antiferromagnetic Neel temperature $T_N\sim E_g$.
The gap in the density of states closes at $T_N$ and the system is an ordinary Fermi liquid above that temperature. 
As $U$ increases there is a crossover to a regime 
where a window in temperature opens up between $T^\ast\sim U$ where local moments form and $T_N\sim t^2/U=J$ where the preformed local 
moments order into a quantum antiferromagnetic state and is on the order of the exchange scale. In this regime the state above $T_N$ is no longer a Fermi 
liquid but a Mott-insulator-like state with a pseudogap.

We next consider the effects of doping. For $U\rightarrow \infty$ the problem of a single hole was considered 
by Nagaoka~\cite{Nagaoka1966} and shown to lead to a ferromagnetic ground state. Later variational~\cite{Shastry1990} and quantum Monte Carlo calculations~\cite{Becca2011} further confirmed that the ferromagnetic ground state persisted even for finite doping for large $U$. At large doping or low filling, 
the lattice dispersion can be replaced by a continuum dispersion and the Fermi liquid theory was developed as an expansion in $na^3\ll 1$ in 3D~\cite{Huang-Yang1957,Lee-Yang1957} or in terms of $1/{\rm log}(na^2)$ in 2D~\cite{Engelbrecht1992,Engelbrecht1991}. 

The open questions about this model pertain to the low-doping regime for large $U$. If this model is to describe the high-$T_c$ experiments, this is the regime where unconventional SC should be found. We discuss below the progress 
made using various approaches.

\subsection{Theoretical Methods}

\begin{figure}[tbh] \centering
	\subfigure[
		Density of states
		]{
		\includegraphics[width=0.33\textwidth]{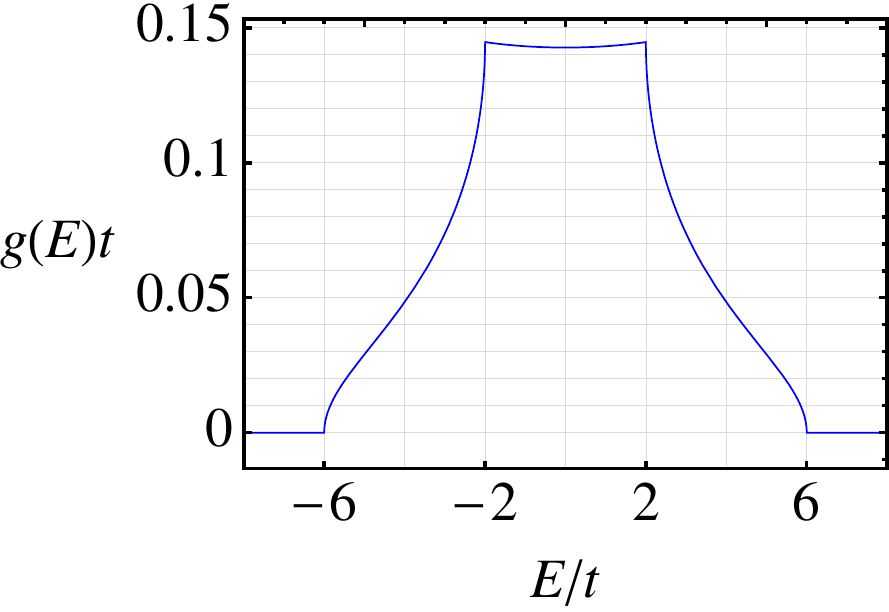} \label{CubicLatticeDoS}
	}
	\subfigure[
		$\mu=-3t$
		]{
		\includegraphics[height=0.23\textwidth,clip=true,trim=35mm 15mm 30mm 15mm]
		{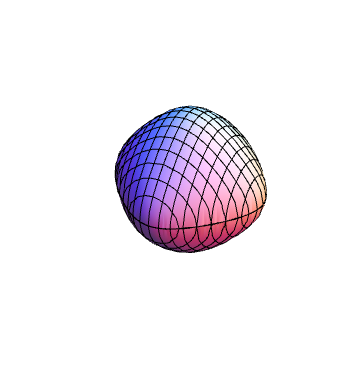} \label{FermiSurfaceMuMinus3}
	}
	\subfigure[
		$\mu=-2t$
		]{
		\includegraphics[height=0.23\textwidth,clip=true,trim=20mm 15mm 15mm 15mm]
		{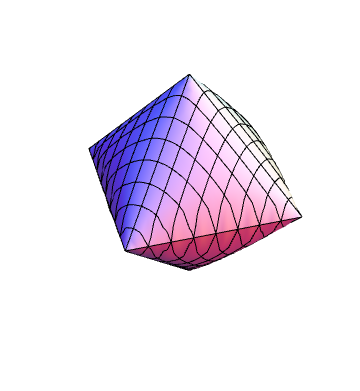} \label{FermiSurfaceMuMinus2}
	}
	\subfigure[
		$\mu=0$
		]{
		\includegraphics[height=0.23\textwidth,clip=true,trim=15mm 15mm 10mm 15mm]
		{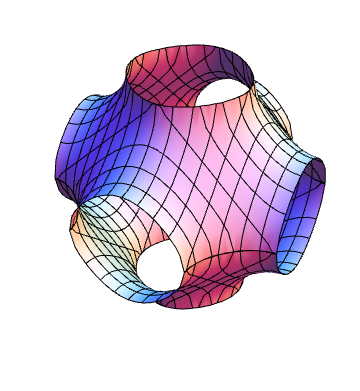} \label{FermiSurfaceMu0}
	}
	\caption{
		\label{FermiSurfaces}
		(a) Density of states of free particles in a cubic lattice.
		(b,c,d) Fermi surfaces at various fillings.
		In the continuum limit ($\mu \approx -6t$) the Fermi surface is spherical.
	}
	\end{figure}

It is instructive to write the Hamiltonian in momentum space as
	\begin{align}
	\hat{H}
	&=
	\sum_{\kkk} (\vare_\kkk-\mu_\sigma) \cdag_{\kkk\sigma} \cccc_{\kkk\sigma}
	+	\sum_{\kkk_1\kkk_2\kkk_3} U 
		\cdag_{\kkk_1+\kkk_2-\kkk_3,\up} \cdag_{\kkk_3\dn} \cccc_{\kkk_2\dn} \cccc_{\kkk_1\up}
	\end{align}
where the momenta $\kkk,\kkk_1,\kkk_2,\kkk_3$ run over the first Brillouin zone and the free-particle dispersion is
	\begin{align}
	\vare_\kkk &= 
	 	-2t(\cos k_x + \cos k_y + \cos k_z)
	.
	\end{align}
The density of states $g(E)$ and typical Fermi surfaces are illustrated in Figure~\ref{FermiSurfaces}.  
The cubic lattice density of states can be written analytically\cite{joyce1972,guttmann2010}
as 
$g(\vare) = -\frac{1}{\pi} \Im G(\vare + i0^+)$ where
	\begin{align}
	G(\vare) 
	&= 
		\frac{4}{\pi^2 \vare} ~
		\frac{1 - 9 x^4}{ (1 - x)^3	(1 + 3x) }
		\mathtt{EllipticK} \left[   \frac{16 x^3}{ (1 - x)^3 (1 + 3x) }  \right] ^2
				,\quad
	x = \sqrt{   \frac{ \vare - \sqrt{\vare - 2t} \sqrt{\vare + 2t}}{ \vare + \sqrt{\vare - 6t} \sqrt{\vare + 6t}}  }
	,
	\end{align}
where $0^+$ is a positive infinitesimal and $\mathtt{EllipticK}$ is the complete elliptic integral of the first kind as implemented in \emph{Mathematica}.  In particular, $g(0) \approx  0.143/t$.

\subsubsection{{A.  Perturbation theory\label{FermiPerturbationTheory}}}
\begin{figure}[htb] \centering
		\includegraphics[height=0.25\textwidth]
			{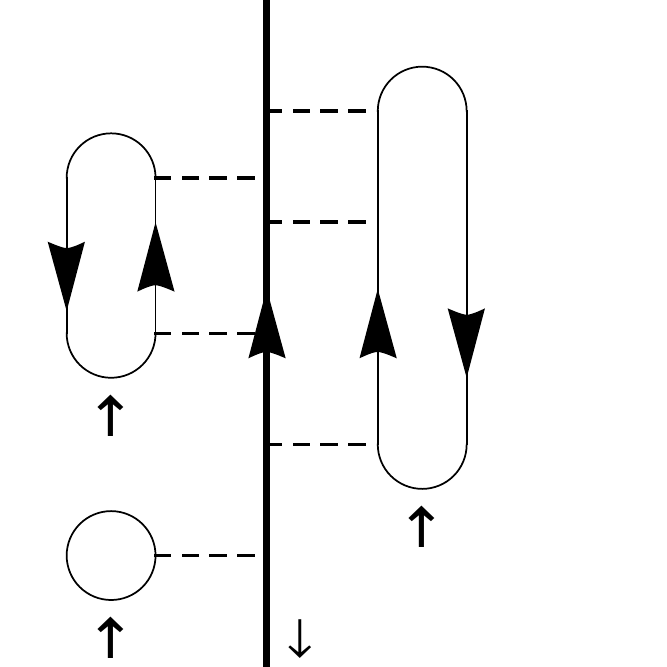} \label{DiagramSigma}
	\caption{
		\label{Polaron}
		This is an example of a diagram contributing to the self-energy $\Sigma(\kkk,E)$ of a single down-spin fermion 
			in an up-spin Fermi sea.
		Solid curves represent bare fermion Green functions, which may be in a continuum or a lattice.
		Dashed lines (``vertices'') represent interactions.
		Diagram techniques work well when the interaction or density are small parameters
			and the system is far from any order-disorder phase transition.
	}
	\end{figure}

Far from half-filling, or at sufficiently high temperatures, the Hubbard model is approximately a free Fermi gas with energy levels $\vare_\kkk$ occupied according to Fermi-Dirac distributions with chemical potentials $\mu_\sigma$.  Then, one can develop many-body diagrammatic calculations involving perturbative expansions\cite{AGD,FetterWalecka,Mahan} with  $U/t$ or density $n$ as small parameters \cite{lee1957,huang1957,galitski1958}.
Diagram techniques have been quantitatively successful in treating the Fermi polaron problem (one minority-species fermion in a sea of majority fermions)  \cite{combescot2007}, corresponding to the ``Lifshitz'' phase transition between a one-component Fermi liquid and a two-component Fermi liquid.  See Fig.~\ref{Polaron}.

	\begin{figure}[htb] \centering
	\subfigure[
		]{ 
		\includegraphics[height=0.25\textwidth]
			{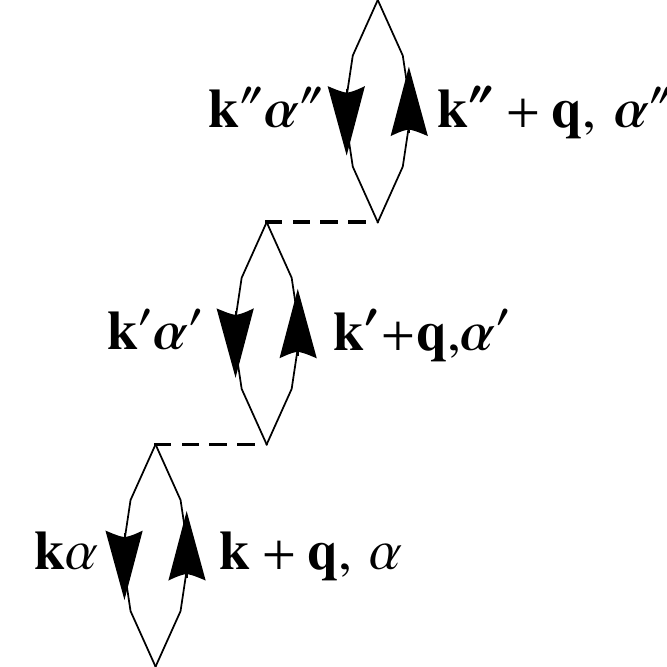} \label{DiagramKappa}
	}
	\subfigure[
		]{ 
		\includegraphics[height=0.25\textwidth]
			{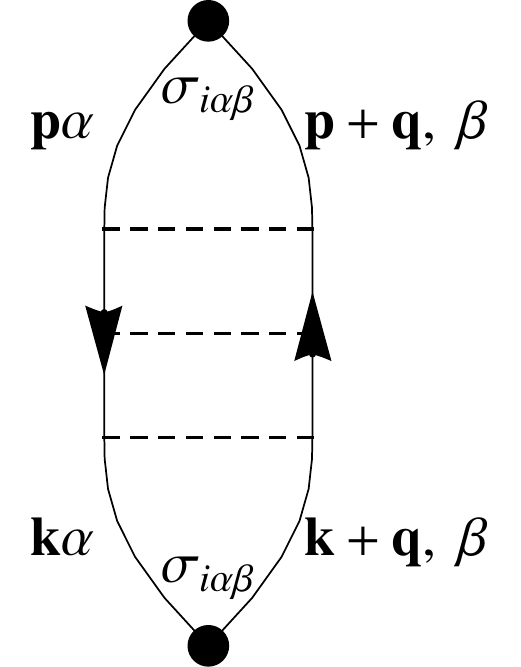} \label{DiagramChi}
	}
	\subfigure[
		]{ 
		\includegraphics[height=0.25\textwidth]
			{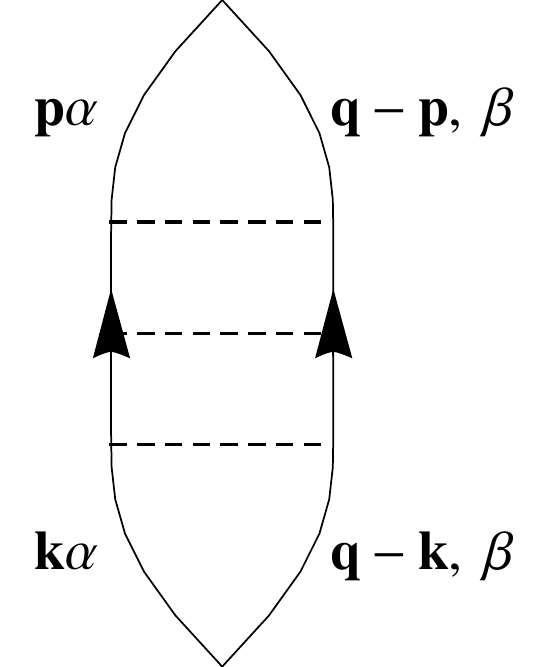} \label{DiagramPi}
	}	
	\subfigure[]{
		\label{DiagramGMB}
		\includegraphics[width=0.25\columnwidth]{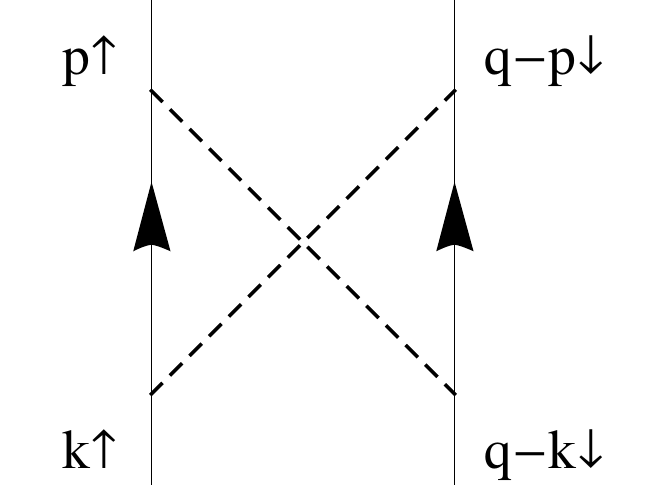}
	}
	\caption{
		\label{BubblesAndLadders}
		Examples of diagrams contributing to various Fermi liquid instabilities.  
		The Hubbard $U$ is written here as a density-density interaction $U\hat{n}_\rrr \hat{n}_\rrr'$.
		$\alpha$ and $\beta$ are spin indices.
		The first three figures are examples of diagrams contributing to divergences in the
		(a) compressibilitity $\kappa(\qqq) \sim \mean{\hatn(\qqq) \hatn(-\qqq)}$,
		(b) susceptibility $\chi (\qqq) \sim \mean{\hat{S}^i(\qqq) \hat{S}^i(-\qqq)}$, and
		(c) pairing susceptibility $\Pi(\qqq) \sim \mean{\hat{F}(\qqq) \hat{F}^\dag(\qqq)}$ where $F$ is the anomalous Green function defined in Table \ref{VarParTable}.
		Diagram (d) is the GMB vertex correction, 
		which reduces the strength of an attractive interaction
		from its bare value.
	}
	\end{figure}

	\begin{figure}[htb] \centering
		\includegraphics[height=0.25\textwidth]
			{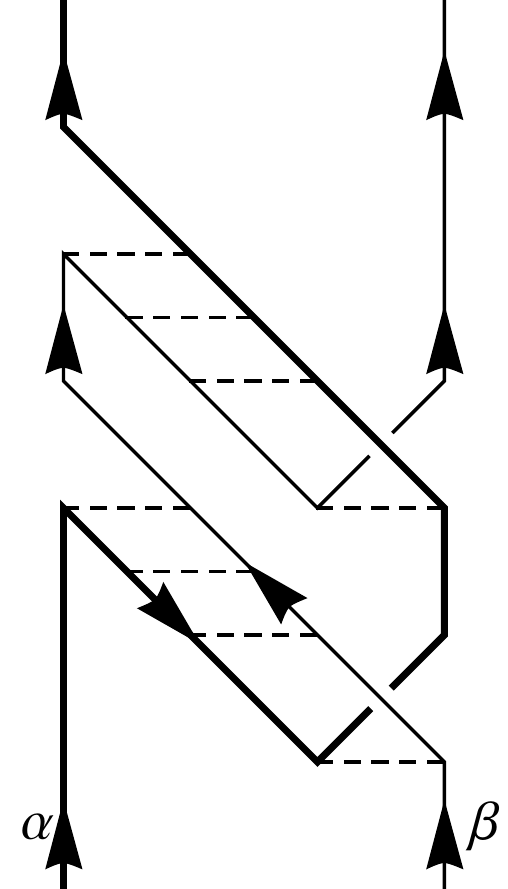}
	\caption{
		\label{DiagramFLEX}
		Example of a diagram contributing to spin-fluctuation-mediated pairing
			via the fluctuation exchange (FLEX) mechanism.
		Near a magnetic quantum critical point the magnetic susceptibility (Figure~\ref{DiagramChi})
			(particle-hole ladder diagrams) 	is large,
		and	may provide a pairing interaction 
			(i.e., it can be used as the ``rungs'' of particle-particle ladder-diagrams as above).
	}
	\end{figure}

\paragraph{\underline{Instability in spin channel:}}
Order-disorder phase transitions present a greater challenge.  We present a few examples to show why this is so.
Summing infinite diagram series (Figure~\ref{BubblesAndLadders}) shows that certain correlation functions diverge at low temperature, indicating instabilities towards ordered phases.  
Summing diagrams such as Figure~\ref{DiagramChi} gives the dressed spin susceptibility as 
$\chi(\qqq) = 1/[1/\chi^{(0)}(\qqq) - U]$, where $\chi^{(0)}$ is the susceptibility of the non-interacting system.  Thus $\chi(\qqq) \rightarrow \infty$ when $U = 1/\chi^{(0)}(\qqq)$.  This is the Stoner criterion for a magnetic instability at wavevector $\qqq$.  
For the cubic lattice at half-filling $(\mu=0)$ the Fermi surface, Figure~\ref{FermiSurfaceMu0}, is perfectly nested such that 
$\chi^{(0)}(\pi,\pi,\pi) \rightarrow \infty$ as $T \rightarrow 0$ and the situation is similar to the BCS pairing problem discussed later.  

However, for a generic Fermi surface such as that in Figure~\ref{FermiSurfaceMuMinus3},
$\chi^{(0)}  \sim g(0)  \sim  1/t$ even at $T=0$. In this case a magnetic instability requires a finite attraction $U \sim t$, 
so $U/t$ is no longer a small parameter and the conclusions drawn from this type of theory, such as Stoner ferromagnetism, should be treated with healthy skepticism.

\paragraph{\underline{Instability in pairing channel:}}
Summing diagrams such as Figure~\ref{DiagramPi} gives the dressed pairing susceptibility as $\Pi(\0) = 1/[1/\Pi^{(0)}(\0) + U]$.  
Given a generic Fermi surface, there are a large number of ways to add fermion pairs of opposite momenta and spin 
($\kkk\up$ and $-\kkk\dn$) close to the Fermi surface, so the zero-momentum bare pairing susceptibility $\Pi^{(0)} (\0)$ diverges logarithmically as $T\rightarrow 0$, when the Fermi surface becomes sharp.  Keeping the $O(\ln T)$ and $O(1)$ terms,
	\begin{align}
	\Pi^{(0)}
	&\approx  g(\mu) \ln \frac{2e^\gamma \omega_D}{\pi T}
			~~\text{where}\quad
	g(\mu) \ln \omega_D	
	=
	g(\mu) \ln \sqrt{36-\frac{\mu^2}{t^2}} + \int_{-6t}^{6t} d\xi~ \frac{g(\xi) - g(\mu)}{2 \abs{\xi - \mu} } 
			,
	\end{align}	
where $\omega_D(\mu)$ plays the role of the Debye frequency from the original BCS theory, $g(\mu)$ is the density of states at the Fermi level, and $\gamma \approx 0.577$ is Euler's constant. 
This means that an infinitesimal attraction ($U<0$) is sufficient to produce a pairing instability, albeit at an exponentially small critical temperature:
%
%
	\begin{align}
	T_c
	&\approx 
		\frac{2e^\gamma \omega_D}{\pi}  \exp \frac{-1}{\abs{U} g(\mu) }.
	\label{TcBCS}
	\end{align}	
However, this expression is not asymptotically correct even in the limit $U\rightarrow 0$.  Although $U/t$ is a small parameter, $T_c$ is also small, and it is affected by classes of diagrams that we have neglected.
To get the prefactor correct, one needs to take into account Gor'kov-Melik-Barkhudarov (GMB) vertex corrections
\cite{gmb1961,vandongen1994,tahvildarzadeh1997,kim2009}
as illustrated in Fig.~\ref{DiagramGMB}.   
For a spherical 3D Fermi surface \cite{gmb1961}, the GMB correction factor is $(4e)^{-1/3} = 0.451$.
For the half-filled cubic lattice Hubbard model, the correction factor is approximately 0.282. \cite{vandongen1994}
In certain situations, such as the 2D attractive Hubbard model near half-filling, one has to include even more sophisticated corrections (e.g., by going to the fluctuation exchange and pseudopotential parquet approximations) to capture even qualitative features in the behavior of $T_c$. \cite{bickers1989,moreo1991,luo1993,paiva2004}

\paragraph{\underline{Spin-fluctuation-mediated pairing:}}
Return to the repulsive ($U>0$) Hubbard model.
A very interesting and important situation occurs close to a magnetic quantum critical point.
Suppose the Fermi surface is nearly nested, so that 
$\chi^{(0)}(\qqq)$ is large near $(\pi,\pi,\pi)$,
and
$\chi(\qqq) = 1/[1/\chi^{(0)}(\qqq) - U]$ is very large but not infinite.
Then the system has nearly antiferromagnetic paramagnons -- damped collective excitations in the dynamic spin susceptibility $\chi(\qqq,\omega)$ near $(\pi,\pi,\pi)$ at small $\omega$.  
Whereas the bare vertex (the Hubbard $U$) is repulsive and independent of frequency and momentum,
the spin-fluctuation-dressed vertex $U_\text{eff} (\qqq,\omega)$ is a complicated function of $\qqq$ and $\omega$ with repulsive and attractive components in different channels.
Eliashberg calculations within the fluctuation-exchange (FLEX) approximation suggest that
spin fluctuations can indeed provide the ``glue'' 
\cite{miyake1986,bealmonod1986,monthoux1991,moriya2003,scalapinoRMP2012}
for unconventional d-wave pairing
(see Figure~\ref{DiagramFLEX}). 
The FLEX mechanism explains $d$-wave pairing in heavy-fermion superconductors, and it may also apply to high-temperature superconductivity in cuprates, although the latter topic has always been controversial; there are other mechanisms for explaining $d$-wave pairing coming from other directions on the phase diagram.

%

%
%

Ultimately, to describe critical phenomena and the many possible ordered phases that can arise in the Hubbard model, non-perturbative methods are required.  We describe these in the next few sections.

\subsubsection{B. Mean-field theories}
As discussed in Sec.~\ref{BoseMFT}, for the Bose-Hubbard model one can construct a mean-field theory by working in real space, where the $U$ term is diagonal, and decoupling the $t$ term.  An alternative approach is to work in momentum space, where the $t$ term is diagonal, and decouple the $U$ term.  For the Fermi Hubbard model the latter approach is generally more useful.

In the simplest mean-field decoupling scheme, one approximates number operators by their averages as follows,
	\begin{align}
	U \hatn_{\rrr\up} \hatn_{\rrr\dn}
	 &\approx 
		U \mean{\hatn_{\rrr\dn}} \hatn_{\rrr\up} + U \mean{\hatn_{\rrr\up}}\hatn_{\rrr\dn}
	- U \mean{\hatn_{\rrr\up}} \mean{\hatn_{\rrr\dn}}
		\nonumber\\
	 &\approx 
		V^H_{\rrr\up} \hatn_{\rrr\up} + V^H_{\rrr\dn}  \hatn_{\rrr\dn}
	- U \mean{\hatn_{\rrr\up}} \mean{\hatn_{\rrr\dn}}
	\end{align}
where $V^H_{\rrr\sigma}$ is the Hartree potential on site $\rrr$ for fermions of spin $\sigma$.  The bilinear mean-field Hamiltonian can then be solved and one can find $\mean{\hatn_{\rrr\up}}$ 
and $\mean{\hatn_{\rrr\dn}}$ self-consistently.

If $\mean{\hatn_{\rrr\up}}$ and $\mean{\hatn_{\rrr\dn}}$ are restricted to be spatially uniform, ferromagnetism is the only type of order that can be described.
However, the word ``mean'' in ``mean-field theory'' refers to quantum and thermal averaging, not spatial averaging. In general the mean field can be spatially inhomogeneous.  Furthermore, there are various non-trivial ways to decouple the $U$ term.  This gives rise to a family of inhomogeneous mean-field theories -- Hartree-Fock (HF), unrestricted Hartree-Fock or generalized Hartree-Fock, and Bogoliubov-de Gennes (BdG) -- that allow various types of superconducting and magnetic order as shown in Fig.~\ref{LMTCartoons}.

There are several ways to develop mean-field theory, such as the approximation method above or the Hubbard-Stratonovich transformation, but these suffer from over-counting problems when one tries to decouple the interaction in more than one channel.  Thus we shall describe the variational approach, which is the most rigorous and extensible one.
In principle one could use the variational method in its original form, where one writes down a trial wavefunction (such as a Slater determinant) and minimizes its Rayleigh quotient to obtain an upper bound to the true ground state energy.  We will find it more convenient to use the $\Tr \rho \ln \rho$ variational mean-field formalism \cite{chaikin}, as described below. 


Let $\mu_\rrr$ and $h_\rrr$ be the average chemical potential and Zeeman field in the $z$-direction at site $\rrr$.
Write the Hubbard Hamiltonian (Eq.~\ref{HubbardHamiltonian}) in the form
	\begin{align}
	\hat{H} &= \hat{H}_t + \hat{H}_U + \hat{H}_\mu
		,\\	
	\hat{H}_t &=	-	\sum_{\rrr\rrr'\sigma} t_{\rrr\rrr'} \cdag_{\rrr\sigma} \cccc_{\rrr'\sigma}
		,\\	
	\hat{H}_U	&=	\sum_{\rrr}		U \hatx_{\rrr\up} \hatx_{\rrr\dn}
		,\\	
	\hat{H}_\mu 	&=	-	\half \sum_{\rrr}	
	\pmat{
		\cdag_{\rrr\up} \\
		\cdag_{\rrr\dn} \\ 
		\cccc_{\rrr\up} \\
		\cccc_{\rrr\dn} \\
	}
	\pmat{
		\mu_\rrr + h_\rrr & 0 & 0 & 0 \\
		0 & \mu_\rrr - h_\rrr & 0 & 0 \\
		0	& 0 & -\mu_\rrr - h_\rrr & 0 \\
	  0 & 0 & 0 & -\mu_\rrr + h_\rrr \\
	}
	\pmat{
		\cccc_{\rrr\up} \\
		\cccc_{\rrr\dn} \\ 
		\cdag_{\rrr\up} \\
		\cdag_{\rrr\dn} \\
	}
	,
	\label{NambuHamiltonian}
	\end{align}
where the factor of $\half$ compensates for the particle-hole doubling.

	\begin{table}
	\begin{tabular}{cccc}
	\hline \hline
		Hartree field  
	& $\Sigma^1_\rrr = h^\text{X(Hartree)}_\rrr$
	& Magnetization	 
	&	$G^1_\rrr = m^X_{\rrr} 
			= \half \mean{\cdag_{\rrr\up}\cccc_{\rrr\dn} + \cdag_{\rrr\dn}\cccc_{\rrr\up}}$
		\\
		{}
	& $\Sigma^2_\rrr = h^\text{Y(Hartree)}_\rrr$
	&
	& $G^2_\rrr = m^Y_{\rrr} 
			= \tfrac{1}{2i} \mean{\cdag_{\rrr\up}\cccc_{\rrr\dn} - \cdag_{\rrr\dn}\cccc_{\rrr\up}}$
		\\
		{}
	& $\Sigma^3_\rrr = h^\text{Z(Hartree)}_\rrr$
	&
	& $G^3_\rrr = m^Z_{\rrr} 
			= \tfrac{1}{2} \mean{\cdag_{\rrr\up}\cccc_{\rrr\up} - \cdag_{\rrr\dn}\cccc_{\rrr\dn}}$
		\\
		Anomalous potential
	& $\Sigma^4_\rrr = \Re \Delta_\rrr$
	& Anomalous density
	&	$G^4_\rrr = \Re F_{\rrr} 
			= \half \mean{\cdag_{\rrr\up}\cdag_{\rrr\dn} + \cccc_{\rrr\dn}\cccc_{\rrr\up}}$
		\\
		{}
	& $\Sigma^5_\rrr = \Im \Delta_\rrr$
	&
	& $G^5_\rrr = \Im F_{\rrr} 
			= \tfrac{1}{2i}  \mean{\cccc_{\rrr\dn}\cccc_{\rrr\up} - \cdag_{\rrr\up}\cdag_{\rrr\dn}}$
		\\
	Hartree chem. pot.
	& $\Sigma^6_\rrr = \mu^\text{Hartree}_\rrr$
	& Density
	&	$G^6_\rrr = x_{\rrr} 
			= \half \mean{\cdag_{\rrr\up}\cccc_{\rrr\up} + \cdag_{\rrr\dn}\cccc_{\rrr\dn}} - 1$
		\\
		\hline \hline
	\end{tabular}
	\caption{
		\label{VarParTable}
		Physical meanings of the $6N$ variational parameters (self-energies) and the $6N$ densities.
		Note that $-\infty < \Sigma^l_\rrr < \infty$ and $-\half \leq   G^l_\rrr \leq  +\half$.
	}
	\end{table}
Let $N$ be the number of sites.
Define $6N$ variational parameters $\Sigma^l_\rrr$ ($l=1,2,3,4,5,6$) whose physical interpretations are shown in Table \ref{VarParTable}.
Define a trial Hamiltonian in which the original interaction term has been decoupled in terms of the variational parameters:
	\begin{align}
	H^\text{trial} &=  H_t + H_\mu
	-	\half \sum_{\rrr}	
	\pmat{
		\cdag_{\rrr\up} \\
		\cdag_{\rrr\dn} \\ 
		\cccc_{\rrr\up} \\
		\cccc_{\rrr\dn} \\
	}^T
	\pmat{
		\Sigma^6_\rrr + \Sigma^3_\rrr & -\Sigma^1_\rrr+i\Sigma^2_\rrr & 0 & \Sigma^4_\rrr+i\Sigma^5_\rrr \\
		-\Sigma^1_\rrr-i\Sigma^2_\rrr & \Sigma^6_\rrr-\Sigma^3_\rrr & -\Sigma^4_\rrr-i\Sigma^5_\rrr & 0 \\
		0	& -\Sigma^4_\rrr+i\Sigma^5_\rrr  & -\Sigma^6_\rrr-\Sigma^3_\rrr & \Sigma^1_\rrr-i\Sigma^2_\rrr \\
		\Sigma^4_\rrr-i\Sigma^5_\rrr & 0 & \Sigma^1_\rrr+i\Sigma^2_\rrr & -\Sigma^6_\rrr+\Sigma^3_\rrr \\
	}
	\pmat{
		\cccc_{\rrr\up} \\
		\cccc_{\rrr\dn} \\ 
		\cdag_{\rrr\up} \\
		\cdag_{\rrr\dn} \\
	}
	.
	\end{align}

The trial Hamiltonian $\hat{H}^\text{trial}$ corresponds to a partition function
$Z^\text{trial} = \Tr \exp (-\beta \hat{H}^\text{trial})$ 
and a density matrix
$\hat{\rho}^\text{trial} = Z^{-1} \exp (-\beta \hat{H}^\text{trial})$.
Since $\hat{H}^\text{trial}$ is bilinear in the fermion operators, it can be ``solved'' exactly as follows.
Construct the $4N\times 4N$ matrix kernel, $\HHH$, and diagonalize it to find the $4N$ eigenvalues $E_m$ and eigenvectors $Y^{(m)}_{\rrr s}$, where $m=1,\dotsc,4N$ is the eigenmode index and $s=1,2,3,4$.~
\footnote{
Computing all eigenvectors of $\HHH$ takes $O(N^3)$ time.  Alternative approaches are possible.  For example, one can write $\Omega$ and $\GGG_\rrr$ in terms of $\cosh \tfrac{\beta \HHH}{2}$ and $\tanh \tfrac{\beta \HHH}{2}$ and expand these functions in Chebyshev polynomials $T_n(c_1 + c_2 \HHH)$, which can be computed by recursion.
} 
Find the free energy of the ``trial'' system
	\begin{align}
	\Omega &= - \tfrac{T}{2} \sum_m  \ln \big(   2 \cosh \tfrac{\beta E_m}{2}  \big)
	\end{align}
and the equal-time on-site matrix Green function
	\begin{align}
	\GGG_\rrr &=
	\mean{
	\pmat{
		\cdag_{\rrr\up} \\
		\cdag_{\rrr\dn} \\ 
		\cccc_{\rrr\up} \\
		\cccc_{\rrr\dn} \\
	}
	\pmat{
		\cccc_{\rrr\up} \\
		\cccc_{\rrr\dn} \\ 
		\cdag_{\rrr\up} \\
		\cdag_{\rrr\dn} \\
	}^T
	}
	\equiv
	\pmat{
		 G_\rrr^6 +  G_\rrr^3 & -G_\rrr^1 + iG_\rrr^2 & 0                     &  G_\rrr^4 + iG_\rrr^5 \\
		-G_\rrr^1 - iG_\rrr^2 &  G_\rrr^6 -  G_\rrr^3 & -	G_\rrr^4 - iG_\rrr^5 &  0                    \\
		0                     & -G_\rrr^4 + iG_\rrr^5 & -	G_\rrr^6 -  G_\rrr^3 &  G_\rrr^1 - iG_\rrr^2 \\
		 G_\rrr^4 - iG_\rrr^5 &0                      & G_\rrr^1 + iG_\rrr^2  & -G_\rrr^6 +  G_\rrr^3 \\
	},
			\nonumber\\
	(\GGG_\rrr)_{ss'}
	&= - \half \sum_m 
			\tanh \tfrac{\beta E_m}{2}   
			\big[  Y^{(m)}_{\rrr s} \big]^* 
			Y^{(m)}_{\rrr s'}
							\qquad(s=1,2,3,4)
	.
	\end{align}
Extract the $6N$ densities $G^l_\rrr$, whose physical interpretations are listed in Table~\ref{VarParTable}.  The variational free energy, $\Omega_\text{var} = -T \Tr (\hat{\rho}^\text{trial} \ln \hat{\rho}^\text{trial})$, is then given by
	\begin{align}
								&
	\Omega_\text{var}
	= \Omega
		+ \sum_\rrr
			U\Big[
					(G^4_\rrr)^2 + (G^5_\rrr)^2 + (G^6_\rrr)^2 - (G^1_\rrr)^2 - (G^2_\rrr)^2 - (G^3_\rrr)^2
			\Big]
						\nonumber\\&~~~~~~~~~~{}
		+ 2\sum_\rrr
			\Big( 
				\Sigma^1_\rrr G^1_\rrr 
			+	\Sigma^2_\rrr G^2_\rrr 
			+	\Sigma^3_\rrr G^3_\rrr 
			+	\Sigma^4_\rrr G^4_\rrr 
			+	\Sigma^5_\rrr G^5_\rrr 
			+	\Sigma^6_\rrr G^6_\rrr 
		\Big) .
	\end{align}
Numerically minimizing $\Omega_\text{var} (\{\Sigma\})$ with respect to the $6N$ variational parameters gives a least upper bound to the free energy of the Hubbard model, together with approximate values for the densities, Green functions, and any other quantities of interest.

One should incorporate gradient information into the minimization procedure.  In fact, one may convert the $6N$-dimensional minimization problem into a $6N$-dimensional root-finding problem by writing
$\frac{\dd \Omega_\text{var}}{\Sigma^l_\rrr} = 0$, which leads to the criteria
	\begin{align}
	\Sigma^1_\rrr &= +U G^1_\rrr		\qquad~~ 
	\Sigma^2_\rrr = +U G^2_\rrr		\qquad~~ 
	\Sigma^3_\rrr = +U G^3_\rrr		\qquad~~ \\
	\Sigma^4_\rrr &= -U G^4_\rrr		\qquad~~ 
	\Sigma^5_\rrr = -U G^5_\rrr		\qquad~~ 
	\Sigma^6_\rrr = -U G^6_\rrr.		\qquad~~ 
	\label{SelfConsistencyConditions}
	\end{align}
Equations~\eqref{SelfConsistencyConditions} may be solved by fixed-point iteration (repeatedly computing $G$ and $\Sigma$ from each other until they stop changing).  There are various ways to accelerate convergence, such as ``linear mixing'' and the Broyden method.  It is nevertheless a good idea to keep track of $\Omega_\text{var}$ to verify that the procedure has converged to a minimum (and not a maximum or saddle-point), and to discriminate between local minima.
	
Equations~\eqref{SelfConsistencyConditions} may be recognized as the BdG and HF self-consistency conditions $\hhh^\text{Hartree}_\rrr=+U\mathbf{m}_\rrr$, $\Delta_\rrr=-UF_\rrr$, and $\mu^\text{Hartree}_\rrr=-U x_\rrr$.  The physical interpretation is obvious.  In the repulsive Hubbard model ($U>0$), a positive local magnetization $\mathbf{m}_\rrr$ (an excess of $\up$ fermions) produces a Hartree field $\hhh^\text{Hartree}_\rrr$ that repels $\dn$ fermions and maintains the magnetization self-consistently.  In the attractive Hubbard model ($U<0$), a positive local density $x_\rrr$ produces a positive Hartree chemical potential $\mu^\text{Hartree}$ that maintains the density self-consistently.  Similarly, the presence of a ``condensate wavefunction'' or ``anomalous density'' $F_\rrr \sim \mean{\cccc_{\rrr\dn} \cccc_{\rrr\up}}$ can produce a self-consistent ``pairing amplitude'' or ``anomalous potential'' $\Delta_\rrr$.  On top of this competition between spin, pairing, and density channels, the external potentials $\mu_\rrr$ and $h_\rrr$ enter via Eq.~\eqref{NambuHamiltonian} to bias the system one way or another.

The above description of variational mean-field theory is intended to give a unified, overarching viewpoint.  In practice it is rarely necessary to implement self-consistency in all six channels.  For example, for the attractive Hubbard model with equal populations ($U<0, h=0, \mu\neq 0$) it is sufficient to decouple in channels 4 and 6 ($\Re \Delta$ and $\mu^\text{Hartree}$), because all other self-energies converge to zero; for unequal populations ($U<0, h\neq 0, \mu\neq 0$) it is usually sufficient to decouple in channels 3, 4, and 6.  This decoupling allows one to use real $2N\times 2N$ matrices instead of complex $4N\times 4N$ matrices, giving a significant speedup.  If one makes a specific ansatz for the spatial dependence (e.g., FM, AF, BCS) one can solve the mean-field equations analytically to recover known results, e.g., the BCS gap equation.

	\begin{figure}[htb] \centering
	\subfigure[2FL
		]{
		\includegraphics[width=0.31\textwidth]{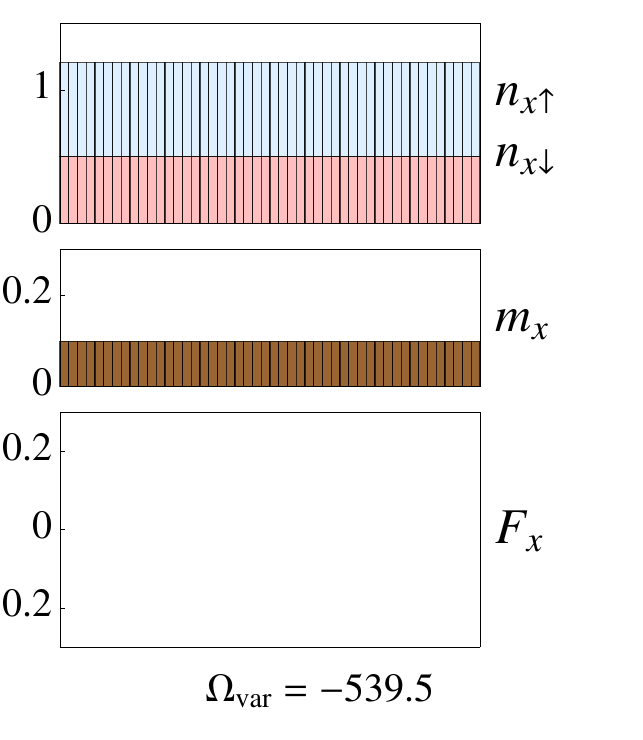} \label{BdGFL}
	}
	\subfigure[LO 
		]{
		\includegraphics[width=0.31\textwidth]{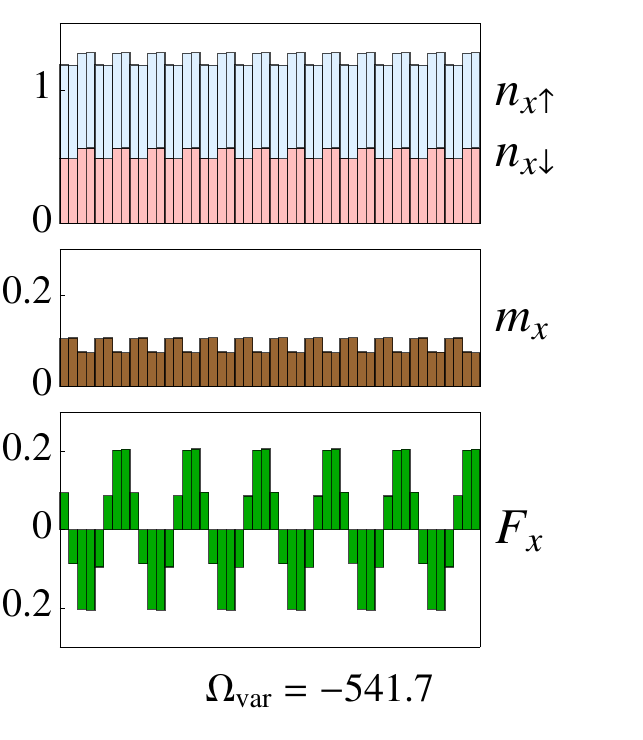} \label{BdGLO}
	}
	\subfigure[BCS
		]{
		\includegraphics[width=0.31\textwidth]{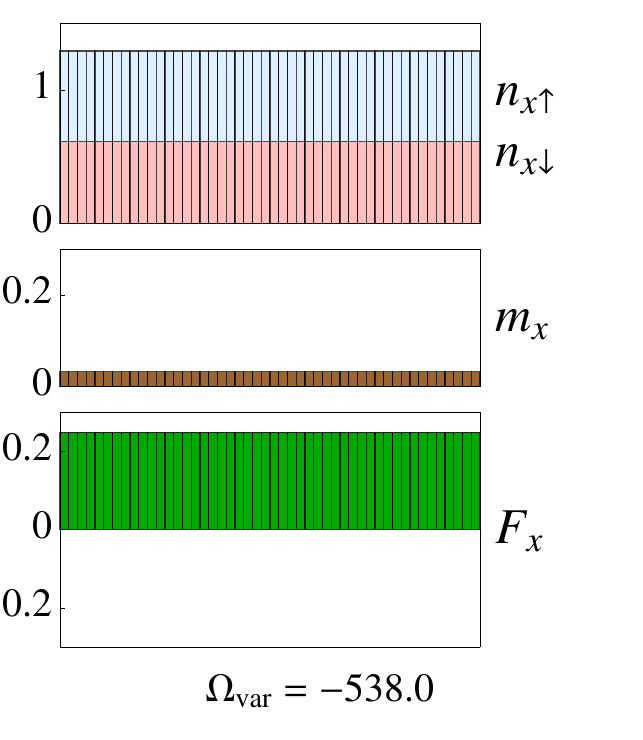} \label{BdGBCS}
	}
	\caption{
		\label{BdGConfigs}
		Pedagogical illustration of solutions of the BdG equations
		for the Hubbard model on a $48\times 6$ square lattice with 
		$U=-4, t=1, T=0.1, \mu=-0.4, h=1.1$.
		Shown are
			the number density $n_{x\sigma} \in [0,1]$,
			magnetization $m_x=n_{x\up} \in [-\half, \half]$,
			and anomalous density $F_{x\sigma} \in [-\half,\half]$
			as functions of $x$ (averaged over $y$).		
		$\Omega_\text{var}$ is the variational free energy.
		(One often quotes $\Delta=-U F$ instead.)
		In an LO state the magnetization 
			is concentrated near the sign changes of the pairing amplitude.
		The BCS s-wave superfluid has a small magnetization 
			due to thermally excited quasiparticles.
	}
	\end{figure}

Figure~\ref{BdGConfigs} illustrates three solutions of the BdG equations for a particular parameter set.  Although the three solutions are all self-consistent, the Larkin-Ovchinnikov (LO) state has the lowest variational free energy.  See Sec.~\ref{FHMPhaseDiagrams} for more discussion of LO states.

Where Fermi liquid instabilities are concerned, mean-field theory gives the same phase boundaries as resummed perturbation theory; we emphasize again that GMB corrections reduce critical temperatures by a factor of 2--3 \emph{even in the weak-coupling limit}.

There are many ways to develop more sophisticated mean-field theories (e.g., slave boson approaches) to capture physics that one hopes to see, but it appears that most interesting phases of the Hubbard model require fluctuations beyond MFT for an accurate description.  Hence we focus the rest of the discussion on numerical methods.

%


\subsubsection{C. Variational quantum Monte Carlo}

In the variational formulation of mean-field theory we worked with a bilinear trial density matrix that could be solved relatively easily.  The variational principle can also be applied with a more complicated trial wavefunction (or density matrix).  For example, one can take a simple many-body wavefunction, such as
	\begin{align}
	\ket{\Psi_\text{FL}} 
	&= \Big(  \prod_\kkk \cdag_{\kkk\up} \cdag_{\kkk\dn}   \Big)  \ket{\Psi_\text{vac}} 
		\\ \text{or~}
	\ket{\Psi_\text{BCS}} 
	&= \Big(  \prod_\kkk a_\kkk \cdag_{\kkk\up} + b_\kkk \cccc_{\kkk\dn}   \Big) \ket{\Psi_\text{FL}} 
	= \Big(  \prod_\kkk A_\kkk \cdag_{\kkk\up} \cdag_{\kkk\dn}  + B_\kkk   \Big) \ket{\Psi_\text{vac}} 
	\end{align}
and build in two-particle correlations by applying a Gutzwiller projection operator or multiplying by a Jastrow factor:
	\begin{align}
	\hat{P}_\text{G} 
	&= \prod_\rrr \exp \big( - g n_{\rrr\up} n_{\rrr\dn}   \big) 
		\\ \text{or~}
	\hat{P}_\text{J} 
	&= \prod_\rrr \exp \big(  - \half v_{\rrr-\rrr'} n_{\rrr} n_{\rrr'}   \big) 
	.
	\end{align}
One could take any other type of wavefunction, such as a valence bond solid (VBS) \cite{affleck1988}, resonating valence bond (RVB) spin liquid \cite{andersonScience1987}, or flux phase \cite{affleck1988,kishine2001}.
Even though the trial wavefunction may be simple to write down, it is usually too difficult to find an exact expression for the trial energy (or for any other expectation value).  However, one can estimate the trial energy by sampling the wavefunction using a Metropolis-type method.  This is known as variational quantum Monte Carlo (VQMC)
\cite{ceperleyScience1986,paramekanti2001,paramekanti2004}.  Compared with other QMC methods, VQMC has the disadvantage that it is biased by the choice of form of the trial wavefunction, however it has the advantage that it does not suffer from the sign problem.  

	\begin{figure}[htb] \centering
	\includegraphics[width=0.5\textwidth]{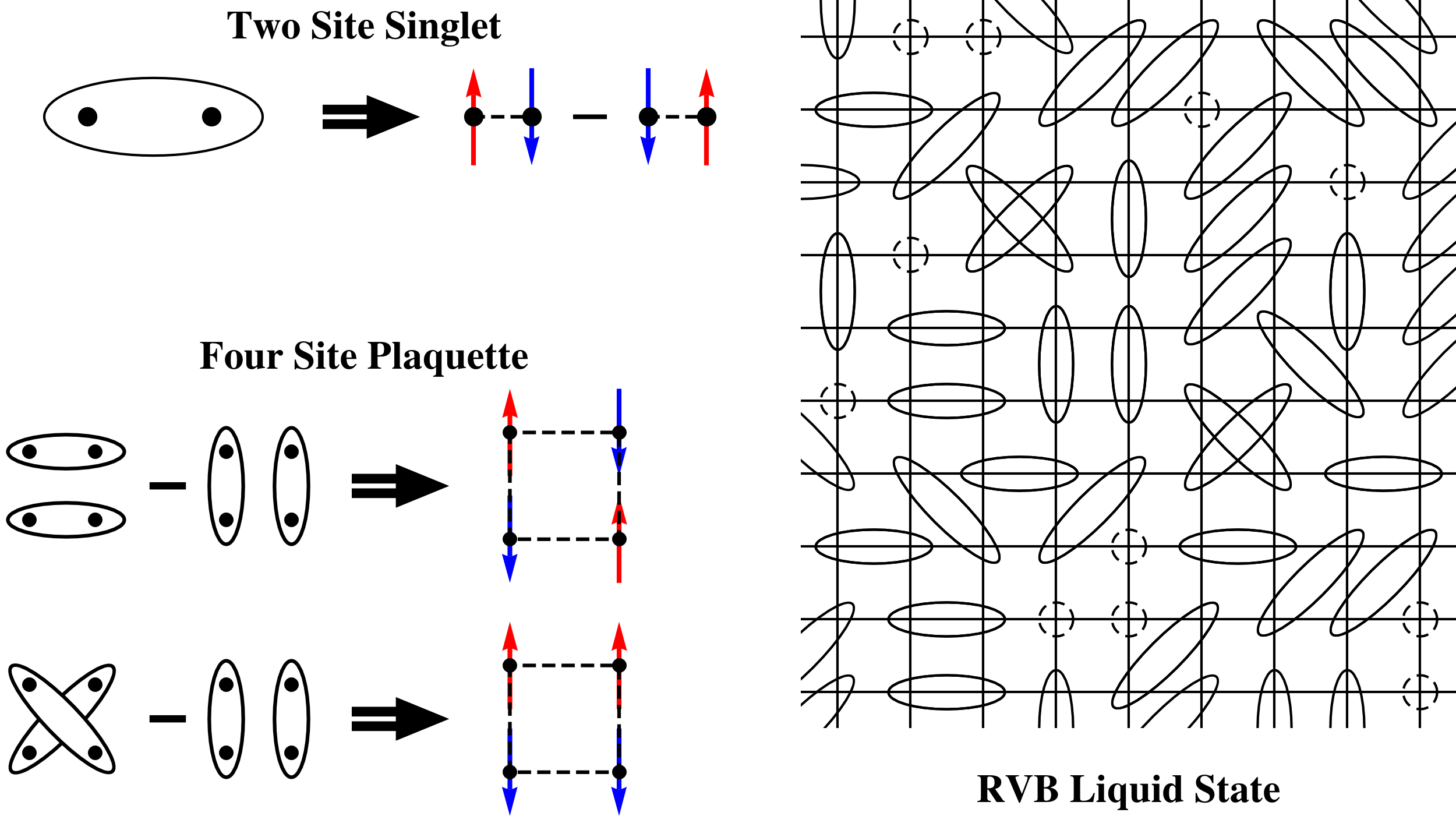}
	\caption{
		\label{fig:rvbSCH}
                Schematic illustration of the RVB trial wavefunction.
	}
	\end{figure}

We next demonstrate the insights from the resonating valence bond state that have 
been extremely successful in describing the phenomenology of the cuprates. Given the ``sign problem"  in an unbiased simulation of the fermion Hubbard model for general fillings, the variational approach appears to be particularly attractive.
Instead of applying the RVB state to the Hubbard model, we turn to the effective Hamiltonian for $U\gg t$, where to second order in $t/U$ we obtain the t-J model~\cite{Gros1987,Yoshioka1988} described by

\begin{equation}
H_{\rm tJ}=-t\sum_{\langle i,j\rangle\sigma} \left( \hat{{\cal P}}^{-1}  \hat{c}^\dagger_{i\sigma} \hat{c}_{j\sigma} \hat{{\cal P}} + h.c. \right) + J\sum_{\langle i,j\rangle}
\hat{{\bf S}}_i \cdot \hat{{\bf S}}_j
\label {tJ}
\end{equation}
where $J=t^2/U$ is the antiferromagnetic exchange scale and
$\hat{{\cal P}}=\prod_{\bf r} (1-n_\uparrow (\bf r) n_\downarrow (\bf r))$ is the fully projected Gutzwiller projection operator 
in which all double occupied sites are projected out.

 To understand the structure of the RVB wave function which we propose as a trial wave function for the t-J Hamiltonian,
 consider two {\em distinct} sites and two electrons of opposite spin. The exchange term 
couples these spins into a singlet ground state as shown in Fig.~\ref{fig:rvbSCH}. For four distinct sites with two up and two down spins,
there are two possible configurations. Note the effect of the projection operator is implicitly accounted for by taking distinct sites.
These configurations can be obtained by expanding the Slater determinant 
\begin{equation}
\Psi_{RVB}={\cal P} \left\|\begin{array}{cc}
\phi( {\bf r}_{1\uparrow}-{\bf r}_{1\downarrow}) & \phi({\bf r}_{1\uparrow}-{\bf r}_{2\downarrow} )\\
 \phi({\bf r}_{2\uparrow}-{\bf r}_{1\downarrow}) & \phi({\bf r}_{2\uparrow}-{\bf r}_{2\downarrow})\\
\end{array}\right\|
\label{4sitervb}
\end{equation}
Here $\phi({\bf r}-{\bf r}^\prime)=\sum_{\bf k} \exp(i{\bf k} \cdot ({\bf r}-{\bf r}^\prime)) \phi({\bf k})$ is 
the paired wave function and, by analogy with BCS theory, $\phi({\bf k})=v_{\bf k}/u_{\bf k}=\Delta_{\bf k}/
\left[\xi_{\bf k}+\sqrt{\xi_{\bf k}^2 + \Delta_{\bf k}^2}\right]$. The two variational parameters $\mu_{\rm var}$ and $\Delta_{\rm var}$ 
determine $\phi({\bf k})$ through $\xi_{\bf k}=\epsilon({\bf k}) - \mu_{\rm var} $ and $\Delta_{\bf k}=\Delta_{\rm var}
({\rm cos} k_x-{\rm cos}k_y)/2$ for a d-wave gap on a square lattice.

Such a singlet wavefunction, known as the resonating valence bond (RVB) state because it involves a linear superposition
of singlets or valence bonds, can be generalized for $N_\uparrow=N_\downarrow$ spins in order to describe an entangled state of spin singlets as depicted in Fig.~\ref{fig:rvbSCH}.

\begin{figure}[htb] \centering
		\includegraphics[width=0.9\textwidth]{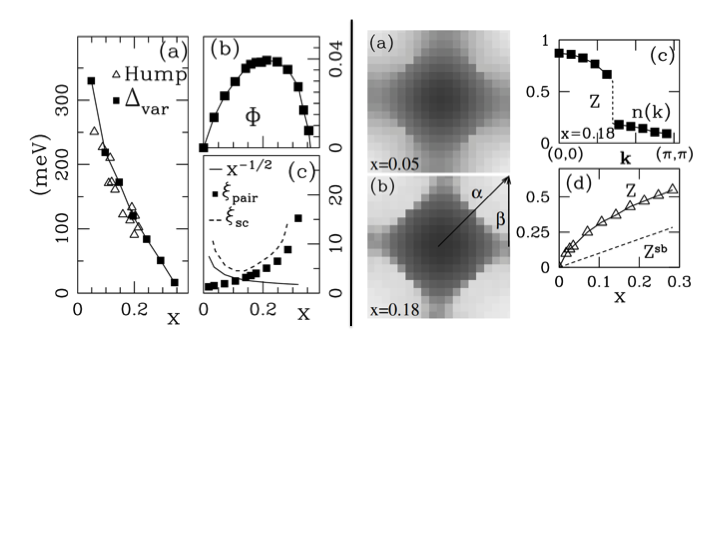} \label{rvb1}
\vspace{-3.5cm}
	\caption{Left Figure: Panel (a) The variational parameter $\Delta_{\rm {var}} $ (filled squares) compared to the $(\pi,0)$ hump scale (open triangles)
		in ARPES as a function of doping. $\Delta_{\rm {var}}\propto T^\ast$ and defines a temperature scale where pairing correlations start to develop.  Panel (b) 
		Doping dependence of the d-wave superconducting order parameter $\Phi$. This scale is qualitatively related to the transition temperature $T_c$ 
		where the preformed pairs condense into a phase coherent state. Panel (c) The coherence length $\xi_{SC}\ge {\rm max} (\xi_{\rm pair},1/\sqrt{x})$ 
		is the larger of the two scales, the pair size $\xi_{\rm pair} = v_F/\Delta_{\rm var}$ and interhole spacing.
		Right Figure: Panels (a) and (b) The momentum distribution $n({\bf k})$ plotted on a gray scale between 1 (black) and 0 (white) centered at ${\bf k}=(0,0)$ for $x=0.05$ and $x=0.18$ on a tilted $19\times 19 + 1$ lattice showing very little doping dependence of the large ``Fermi surface". Panel (c) $n({\bf k})$ plotted along the diagonal direction indicated as $\alpha$ in (b) showing the jump at $k_F$ which implies a gapless nodal quasiparticle of weight $Z$. Panel (d) Nodal quasiparticle weight $Z(x)$ compared with the simple slave boson mean field theory result $Z^{\rm sb}(x)=x$.}
\label{fig:rvb}
\end{figure}

Using RVB wavefunctions it has been possible to describe in quite some detail the phenomenology of the high-$T_c$ cuprates~\cite{paramekanti2001,paramekanti2004} as shown in Fig.~\ref{fig:rvb}. The variational parameter $\Delta_{var}$ is a measure of the maximum energy gap and the order parameter a measure of long range phase coherence. The energy scales constructed out of these quantities determine the behavior of $T^\ast$ and $T_c$ respectively showing very different doping dependencies.
What is interesting is that these bosonic tendencies toward pair formation above $T_c$ are found in a highly degenerate regime with a large Fermi sea, as seen from the momentum distribution function $n({\bf k})$. 
In the d-wave superconducting regime, an
anisotropic gap opens up around the Fermi surface which vanishes at the nodal point along the 
$(0,0)\rightarrow (\pi,\pi)$ direction. The finite quasiparticle weight $Z$ encoded in the jump in $n({\bf k})$ shows a strong doping-dependence and vanishes in the non-superconducting limit.
We also find that the Fermi velocity of these nodal quasiparticles is more or less independent of doping. This result indicates that superconductivity is lost not because the effective mass of the quasiparticles diverges at $x=0$,
as is known to occur in several metal-insulator transitions, 
but because the quasiparticles become increasingly incoherent as half filling is approached.

\subsubsection{D. Projective quantum Monte Carlo}
The basic idea of projective quantum Monte Carlo schemes is to take a starting wavefunction $\ket{
\Psi}$ and evolve it over a large imaginary time interval $\tau$, so that the coefficients of excited states decay, leaving only the ground state.  Observables may then be computed using
	\begin{align}
	\mean{\hat{A}}
	&=\lim_{\tau\rightarrow \infty} 
		\frac{  
			\bra{\Psi} e^{-\tau\hat{H}}    \hat{A}  e^{-\tau\hat{H}}  \ket{\Psi}
		}{
			\bra{\Psi} e^{-\tau\hat{H}}      e^{-\tau\hat{H}}  \ket{\Psi}
		}
	.
	\end{align}
This class of methods includes diffusion Monte Carlo \cite{ceperleyScience1986} and Green function Monte Carlo.  This allows one to refine a VQMC estimate of the ground state by using the best variational wavefunction as the starting wavefunction in a projective scheme.  However, for systems with a sign problem, the error grows exponentially during time evolution, which makes it difficult to access low temperatures.

\subsubsection{E. Auxiliary field quantum Monte Carlo}
For boson models (see Sec.~\ref{BoseQMC}) worldline QMC techniques are very useful.  However, for fermions in more than one dimension, worldline QMC techniques usually suffer from a sign problem.  
Fortunately, there exist auxiliary field QMC (AFQMC) methods that are free of the sign problem for an important class of models.
The general idea of AFQMC is to take the quantum Fermi Hubbard model in $d$ dimensions 
and decouple the interaction using an auxiliary Hubbard-Stratonovich field in $d+1$ dimensions.
The resulting Hamiltonian (or action or partition function) is then bilinear in the fermions, and one can integrate out the fermions to obtain an action in terms of the auxiliary fields alone, which can be sampled using Monte Carlo.  

\paragraph{\underline{Derivation:}}
Here we give a bare-bones description of the Hirsch-Fye/Blankenbecler-Sugar-Scalapino \onlinecite{hirsch1986,blankenbecler1981,bai2005} AFQMC algorithm, which is often referred to as determinant quantum Monte Carlo (DQMC).
Take the partition function of the Fermi Hubbard model.  Perform a Suzuki-Trotter discretization with $L_\tau$ time slices.  Use the identity
$
	e^{  -\tau_1 U \hat{x}_\up \hat{x}_\dn  }
	= \tfrac{1}{2} e^{-\tau_1 U/4} \sum_{s=\pm 1} 
		e	^{ \tau_1 \lambda s (\hat{x}_\up - \hat{x}_\dn)  }
$
to decouple the quartic terms using auxiliary Ising variables $s_{\rrr\tau}=\pm 1$, so that the exponents are now all bilinear in the form $\sum_{ij} h_{ij} \cdag_i\cccc_j$.  This allows us to employ an identity that turns the trace over a $4^N$-dimensional Hilbert space into a determinant of an $N\times N$ matrix:
	\begin{align}
	Z
	&= \Tr e^{-\beta \hat{H}}
			\nonumber\\
	&= \Tr \prod_\tau  
		\Big[
			\prod_{\rrr}
			e^{  - \tau_1                 U \hat{x}_{\rrr\up} \hat{x}_{\rrr\dn}  }
		\Big]
		\Big[
			\prod_{\rrr\rrr'\sigma}
			e^{    \tau_1    t_{\rrr\rrr'} \cdag_{\rrr\sigma} \cccc_{\rrr'\sigma}  }
		\Big]
			\qquad (\tau=0, \tau_1, 2\tau_1, \dotsc, \beta-\tau_1)
			\nonumber\\
	&= \Tr \prod_\tau  
		\Big\{
			\Big[
			\prod_\rrr
			\sum_{s_{\rrr\tau}}
			z e^{  \lambda s_{\rrr\tau}      (\hat{x}_{\rrr\up} - \hat{x}_{\rrr\dn} ) }
			\Big]
			e^{  ~\sum_{\rrr\rrr'\sigma}  \tau_1  t_{\rrr\rrr'} \cdag_{\rrr\sigma} \cccc_{\rrr'\sigma}  }
		\Big\}
			\qquad (z=\tfrac{e^{-\tau_1 U/4}}{2},
						\cosh \lambda= e^{U\tau_1 /2})
			\nonumber\\
	&= 
		Z_0
		\sum_{\{s\}}
		\Tr \prod_\tau  
			e^{ \Big[\textstyle
					\sum_\rrr\limits 
			  	\sum_\sigma\limits 
					\lambda \sigma  s_{\rrr\tau} \hat{x}_{\rrr\sigma} 
			  \Big]}
			e^{ \Big[\textstyle
			  	\sum_{\rrr\rrr'\sigma}\limits 
				  \tau_1  t_{\rrr\rrr'} \cdag_{\rrr\sigma} \cccc_{\rrr'\sigma}  
			  \Big]}
			\qquad (Z_0=z^{NL_\tau})
			\nonumber\\
	&= 
		Z_0
		\sum_{\{s\}}
		\Tr  \prod_\sigma \prod_\tau
			e^{ \Big[\textstyle
						\sum_{\rrr\rrr'}\limits
						 \lambda \sigma  s_{\rrr\tau}  \delta_{\rrr\rrr'}  
						\cdag_{\rrr\sigma} \cccc_{\rrr'\sigma} 
						\Big]}
			e^{ \Big[\textstyle
			  	\sum_{\rrr\rrr'}\limits 
				  \tau_1  t_{\rrr\rrr'} \cdag_{\rrr\sigma} \cccc_{\rrr'\sigma}  
			  \Big]}
			\nonumber\\
	&= 
		Z_0
		\sum_{\{s\}}
		\Tr  \prod_\sigma \prod_\tau
			e^{ \Big[\textstyle
						\sum_{\rrr\rrr'}\limits (\vvv_{\tau\sigma})_{\rrr\rrr'}
						\cdag_{\rrr\sigma} \cccc_{\rrr'\sigma} 
						\Big]}
			e^{ \Big[\textstyle
						\sum_{\rrr\rrr'}\limits (\kkk)_{\rrr\rrr'}
						\cdag_{\rrr\sigma} \cccc_{\rrr'\sigma}  
			  \Big]}
			\nonumber\\
	&= 
		Z_0
		\sum_{\{s\}}
		\prod_\sigma 
		\det
		\bigg[
			\mathbf{1}
		+	
			\prod_\tau
			e^{ \vvv_{\tau\sigma} }
			e^{ \kkk }
	  \bigg]
	  \qquad\text{(using an identity)}
			\nonumber\\
	&= 
	\sum_{\{s\}} W[s] .
	\end{align}
This is the partition function of a Ising model where the weight of each configuration is
	\begin{align}
		W(\{s_{\rrr\tau}\})
		&=
		Z_0
		\prod_\sigma
		\det
		\bigg[
			\mathbf{1}
		+	
			\prod_\tau
			e^{ \vvv_{\tau\sigma} }
			e^{ \kkk }
	  \bigg]
	  		,\quad
		(\vvv_{\tau\sigma})_{\rrr\rrr'}
		 = \lambda \sigma  s_{\rrr\tau}  \delta_{\rrr\rrr'}  
	  		,\quad
 		(\kkk)_{\rrr\rrr'}
		 = \tau_1  t_{\rrr\rrr'}
		 .
	\end{align}
Similarly, expectations of observables can usually be written as weighted averages,
	\begin{align}
	\langle \hat{A} \rangle
	&= Z^{-1} \Tr \hat{A} e^{-\beta \hat{H}}
	=
		\sum_{\{s\}} W[s] A[s]  \bigg/   		\sum_{\{s\}} W[s]
		 .
	\end{align}
Thus the Hubbard model in $d$ dimensions maps to an Ising model in $d+1$ dimensions.

	\begin{figure}[htb] \centering
	\subfigure[Fermi-liquid-like configuration
		]{
		\includegraphics[width=0.31\textwidth]{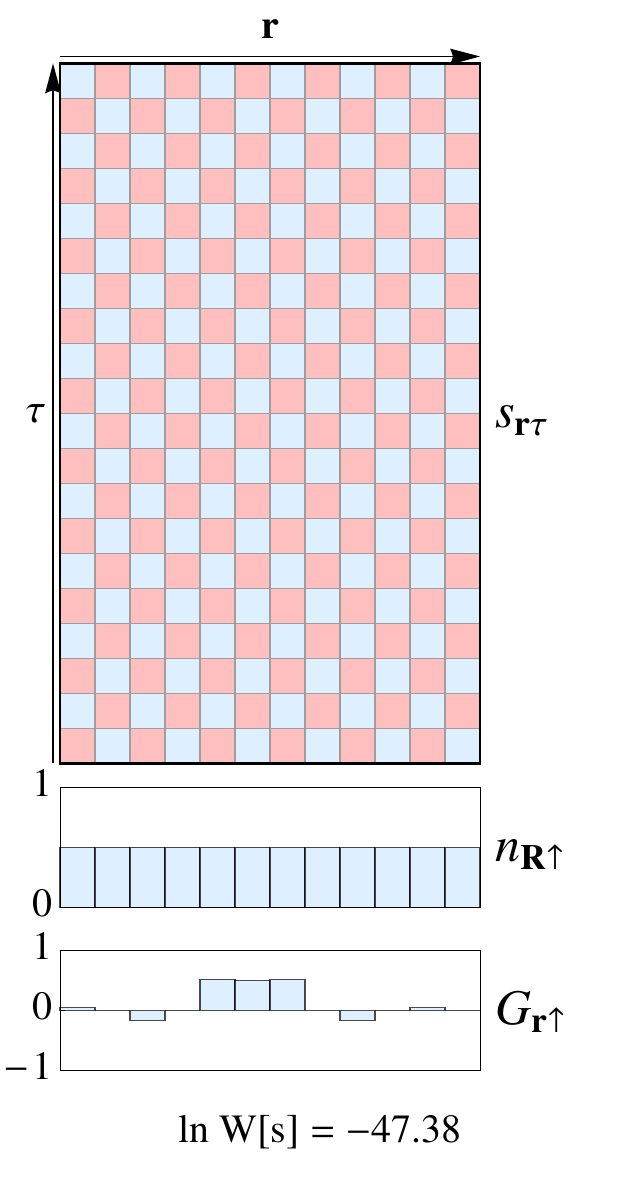} \label{DQMCFL}
	}
	\subfigure[Ferromagnetic
		]{
		\includegraphics[width=0.31\textwidth]{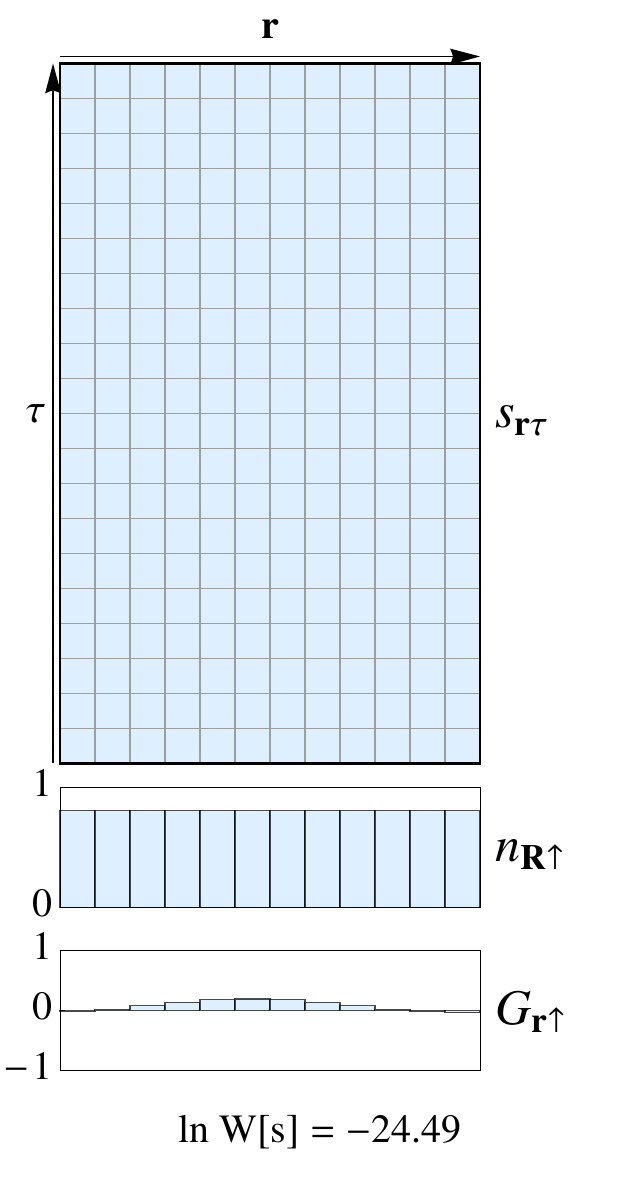} \label{DQMCFM}
	}
	\subfigure[Antiferromagnetic
		]{
		\includegraphics[width=0.31\textwidth]{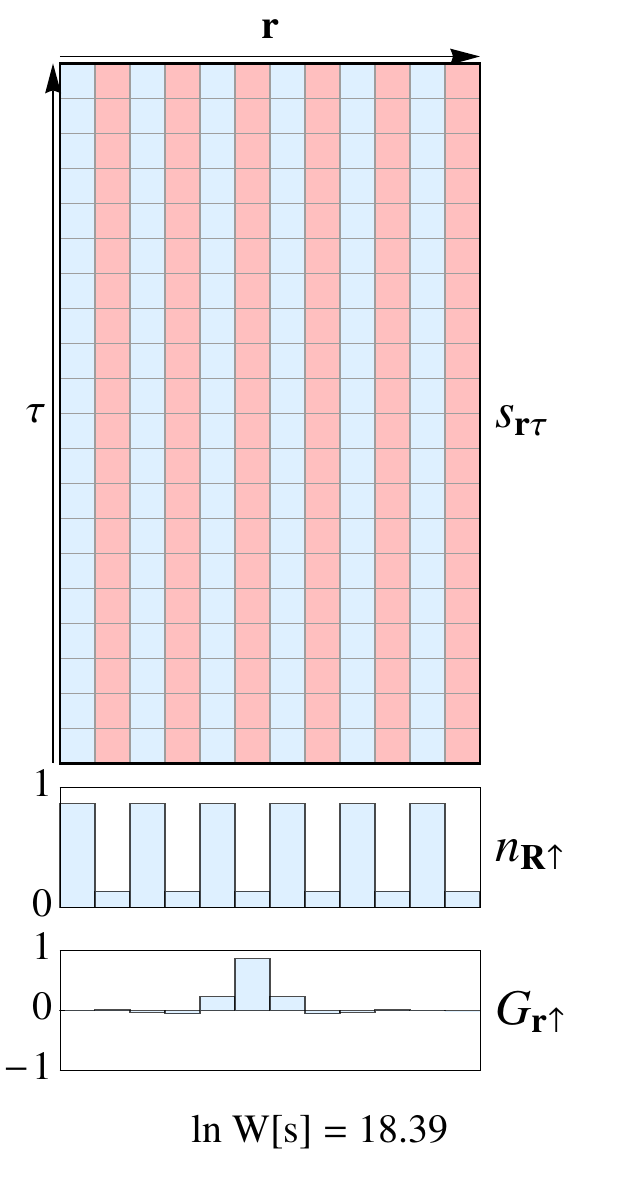} \label{DQMCAF}
	}
	\caption{
		\label{DQMCConfigs}
		Pedagogical illustration of quantities in DQMC.
		In each case, 
		the top panel shows examples of configurations 
			of the Hubbard-Stratonovich Ising field $s_{\rrr\tau}=\pm 1$,
			where blue represents $+1$ and light red presents $-1$,
			site index $\rrr=0,1,2,\dotsc,N-1$ runs horizontally,
			and imaginary time $\tau/\tau_1=0,1,2,\dotsc,L_\tau-1$ runs vertically.
		The lower panels show the spin-up fermion density $n_{\rrr\up}$,
			one row $G_{\rrr\up}$ of the Green function matrix (see text),
			and the weight $W[s]$ of the configuration.
		Parameters were
			$N=12$, $L_\tau=20$, $t=1$, $\mu=0$, $U=1$, $\beta=8$.
		A second-order Trotter decomposition was used
			instead of the first-order one in the text.
	}
	\end{figure}

\paragraph{\underline{Basic algorithm:}}
Let us describe the algorithm explicitly in a simple case.  Consider a Hubbard model with $N=4$ sites and $L_\tau$ slices in imaginary time
with tunneling $t$, chemical potential $\mu$, repulsion $U$, and temperature $T=1/\beta$.
Define $\tau_1=\beta/L_\tau$ and $\lambda=\mathop{\mathrm{arccosh}} \exp (U\tau_1 /2)$.
Set up a $N \times L_\tau$ array of random Ising spins, $s_{\rrr\tau}=\pm 1$, 
where $\rrr=1,2,\dotsc,N$ and $\tau/\tau_1=1,2,\dotsc,L_\tau$.  Define the $N\times N$ kinetic and potential energy matrices 
	\begin{align}
	\kkk
	= -\tau_1 
		\psmat{
			\mu & t & 0 & 0 \\
			t & \mu & t & 0 \\
			0 & t & \mu & t \\
			0 & 0 & t & \mu			
		},
		\qquad
	\vvv_{\sigma \tau}
	= \sigma \lambda
		\psmat{
			s_{1\tau} & 0 & 0 \\
			0 & s_{2\tau} & 0 & 0 \\
			0 & 0 & s_{3\tau} & 0 \\
			0 & 0 & 0 & s_{1\tau}
		},
	\end{align}	
where $\sigma=\pm 1$.  Calculate the $N\times N$ matrices
$
	\MMM_\sigma
	= \III + 
			e^\kkk e^{\vvv_{\sigma 1}}
			e^\kkk e^{\vvv_{\sigma 2}}
			\dots
			e^\kkk e^{\vvv_{\sigma \beta}}
$,
where $\III$ is the $N\times N$ identity matrix.
The statistical weight for the Ising configuration $\{s_{\rrr\tau}\}$ is
$
	W[s]
	=\det \MMM_\up \det \MMM_\dn
$.
Sample the distribution $W[s]$ using the Metropolis algorithm.  That is, propose a new Ising configuration $\{s_\text{new}\}$ and accept the move with probability $\min(1, W[s_\text{new}] / W[s])$.  
After a burn-in period, accumulate statistics of estimators for the desired quantities, such as energy, density, fermion Green functions, and correlation functions.

Figure~\ref{DQMCConfigs} gives a pedagogical illustration for a $12\times 1\times 1$ lattice comparing various Ising configurations $s_{\rrr\tau}$ with the corresponding densities 
$n_{\rrr\up} = \mean{\cdag_{\rrr\up} \cccc_{\rrr\up}}$,
Green functions
$G_{\rrr\up} = \mean{\cdag_{\rrr\up} \cccc_{\rrr_0\up}}$ (the amplitude for inserting a spin-up fermion at $\rrr_0=N/2$ and removing it at $\rrr$),
and weights $W[s]$.
This exposition shows how mean-field theory is in some sense a subset of DQMC, or how DQMC extends MFT to include all fluctuations.
If the Ising variables fluctuate rapidly on a short $\tau$ scale (Figure~\ref{DQMCFL}),
they average out to zero; if the DQMC simulation is dominated by this type of configuration, this suggests that the system is in a normal Fermi liquid state.
The ``ferromagnetic'' configuration $s_{\rrr\tau}=+1$ (Figure~\ref{DQMCFM}) is equivalent to a mean-field decoupling of the $U$ term as a uniform Zeeman field $h=+\lambda$,
whereas a spatially antiferromagnetic configuration of Ising variables (Figure~\ref{DQMCAF}) corresponds to an alternating Zeeman field.  
For the given parameters the AF Ising configuration has a much larger weight than the FL or FM configurations.
Nevertheless, a real DQMC simulation would contain fluctuations away from the pure AF that suppress or destroy the staggered moment.
Note that DQMC, like any other kind of QMC, does not spontaneously break symmetry on a finite lattice, so phase transitions must be located through scaling analyses of correlation functions rather than with non-vanishing order parameters.

The basic algorithm described above is incredibly inefficient.  A brute-force implementation (using $L_\tau$ operations on dense matrices of size $N\times N$) would take $O(N^3 L_\tau)$ time to evaluate $W[s]$ for any given $s$.  In reality there are a huge number of optimizations that make DQMC an attractive algorithm. 
If one proposes to flip one spin at a time, 
this produces only a small change in $\{\vvv_{\sigma\tau}\}$ and $\{\MMM_\sigma\}$.  
Then, $\MMM_\sigma$ (or rather, the Green function matrix $\GGG_\sigma = \MMM_\sigma{}^{-1}$) can be updated in $O(N^2)$ time using the Sherman-Morrison formula.
Other details include handling of sparse matrices, stabilization of matrix multiplications, multi-spin updates, higher-order Trotter decompositions, and so on.  Nevertheless, even in the absence of a sign problem, DQMC for fermions on an $N$-site lattice is much more expensive than worldline QMC for bosons on the same lattice; it may be that the Fermi problem is inherently more difficult.

\paragraph{\underline{Sign problem and statistical error:}}
The most important consideration for DQMC is the infamous fermion sign problem.  In certain situations the determinants $\MMM_\sigma$ can become negative, so that $W[s]$ is negative, and cannot be treated as a probability distribution to be simulated using Markov chain Monte Carlo.  This does not preclude the use of DQMC, as one can sample the probability distribution $\abs{ W[s] }$ instead and write
	\begin{align}
	\langle \hat{A} \rangle
	&= 
		\frac{ \sum_{\{s\}} W[s] A[s] }{   		\sum_{\{s\}} W[s]   }
	=
		\frac{ \sum_{\{s\}} \abs{W} A \sgn W }{   		\sum_{\{s\}} \abs{W}   }
		\Bigg/
		\frac{ \sum_{\{s\}} \abs{W} \sgn W       }{   		\sum_{\{s\}} \abs{W}   }
	=
		\frac{		\mean{A \sgn W}   }{ 		\mean{\sgn W}   }
		.
	\end{align}
However, in systems that suffer from the sign problem, the average sign (the denominator) typically goes to zero exponentially in the system size, so that exponentially more Monte Carlo samples are required to attain the same statistical accuracy.

There is no sign problem in the repulsive Hubbard model at half-filling on a bipartite lattice ($U\geq 0$ and $\nu=0$, i.e., right half of Figure~\ref{HubbardPDUh}).
Equivalently, there is no sign problem in the attractive Hubbard model with equal populations ($U\leq 0$ and $h=0$, i.e., left half of Figure~\ref{HubbardPDUmu}).
Empirically, it is found that the sign problem is most severe in the regions of the phase diagram where LO and qSDW phases are expected to occur.
In general, the existence of the sign problem is related to time-reversal symmetry of the fermion matrix \cite{wu2005SignProblem}.

In DQMC, as with other Markov chain Monte Carlo methods, statistical errors decrease with $1/\sqrt{N_\text{samples}}$, provided that the simulation is long enough to be ergodic; one encounters the usual problems with critical slowing down.
Systematic errors arise from the finite lattice size $N$ and the finite number of imaginary time slices $L_\tau$.  These can be reduced (or even eliminated) by scaling and extrapolation to the limit $N\rightarrow \infty, L_\tau \rightarrow \infty$.

In DQMC one keeps track of equal-time fermion Green function $G_{\rrr\rrr'}$ as part of the algorithm.  Thermodynamic quantities (kinetic energy, double occupancy, etc.) and equal-time correlation functions can be calculated cheaply.  If one wishes, the imaginary-time-dependent correlators can be estimated and then analytically continued to obtain dynamical response functions, but this is expensive and often controversial \cite{gubernatis1991}.  Also, for a simulation at temperature $T$, the spacing of the Matsubara frequencies is $2\pi T$, so it is not possible to extract information about response functions in the hydrodynamic regime $\omega \lesssim 2\pi T$.


\subsubsection{F. Other numerical methods}
There are a large number of other methods that have been developed for studying Hubbard-like models. We briefly mention some here.  

There are auxiliary field QMC methods using other Hubbard-Stratonovich transformations, e.g., with real or complex variables instead of Ising spins. 
There are ``ground state'' determinant simulations that work at $T=0$ and in the canonical ensemble.
There are also continuous-time QMC (CTQMC) methods \cite{werner2006} based on a stochastic sampling of a perturbation expansion in the impurity-bath hybridization parameter, similar in spirit to continuous-time worldline methods for bosonic systems \cite{prokofev1998} and stochastic series expansion methods for spin systems \cite{sandvik1991,syljuasen2002}.
Alternatively, instead of sampling the diagram expansion of the partition function, one can sample the connected diagrams for the free energy; this is the idea behind diagrammatic Monte Carlo methods \cite{vanHoucke2011}.

Dynamical mean-field theory (DMFT) can be viewed as an extension of MFT to include all local quantum fluctuations on a single site (i.e., fluctuations between the different quantum states of a single site as a function of imaginary time).  
The essential idea is to replace a lattice model by a single-site quantum impurity problem embedded in an effective medium that is determined self-consistently.  The impurity model offers an intuitive picture of the local dynamics of a quantum many-body system \cite{metzner1989,georgesRMP1996}.
The self-consistent functional equations for the self-energy can be interpreted as an Anderson impurity model in a bath; they can be treated using iterated perturbation theory or using quantum Monte Carlo techniques such as AFQMC and CTQMC.
DMFT is exact in the limit of infinite spatial dimensions $d\rightarrow \infty$, and has in fact been used to calculate detailed properties of the infinite-dimensional Hubbard model at half-filling \cite{georges1993}.
Cellular DMFT \cite{kotliar2001}
and the Dynamical Cluster Approximation (DCA) 
\cite{gullRMP2011}  
are generalizations of DMFT to include fluctuations in real space or momentum space; in some sense, they interpolate between single-site DMFT and brute-force QMC calculations on an infinite lattice.

Other methods include variational cluster perturbation theory \cite{senechal2005} and
functional renormalization group methods\cite{halboth2000,salmhofer2001}.

\subsubsection{G. Strong-coupling limit}
At half-filling and strong repulsion ($\mu=0$, $U\rightarrow\infty$) the repulsive Hubbard model maps to the antiferromagnetic Heisenberg model.  Charge degrees of freedom are frozen out, leaving only the spin degrees of freedom.  The lowering of kinetic energy by virtual hopping leads to an effective antiferromagnetic coupling $J=4t^2/U$ due to the superexchange mechanism, and hence to an antiferromagnetic ground state.
Equivalently, at zero field and strong attraction ($h=0$, $U\rightarrow -\infty$) the attractive Hubbard model maps to the hard-core boson model: pairs of fermions on the same site are bound into bosonic molecules, and boson tunneling is a two-step process with an effective amplitude $t_{b}=4t^2/\abs{U}$.  This leads to a Bose-Einstein condensate (s-wave superfluid), which is degenerate with a checkerboard CDW state.
These states are crudely illustrated in Fig.~\ref{LMTCartoons} as $z$AF, $x$AF, $\pi$CDW, and sSF.

The hard-core boson model (or the equivalent Heisenberg model) is still a non-trivial quantum many-body problem, but, being bosonic, it can be simulated efficiently using worldline Monte Carlo methods (Sec.~\ref{BoseQMC}).  For the cubic lattice Heisenberg model, $T_c = 0.946 J$ according to SSE simulations \cite{sandvik1998}.   Thus, for the cubic lattice Hubbard model at $U\gg t$ and $\mu=h=0$, $T_c = 0.946 J = 3.784 t^2/U$.

At strong repulsion away from half-filling, the Hubbard model maps approximately to the $t$--$J$ model, which is a difficult unsolved problem.  
In the limit $U\rightarrow \infty$ (or $J\rightarrow 0$) the ground state is massively degenerate, which causes traditional perturbative methods to fail.  At the time of writing it appears that a new type of formalism may be required to tackle this regime, such as extremely correlated Fermi liquid theory \cite{shastry2011} or hidden Fermi liquid theory \cite{casey2011}.

\subsection{Phase Diagrams\label{FHMPhaseDiagrams}}
In this section we will consider the phase diagram of the Hubbard model as a function of the parameters $(\mu,h,T,U)$.  It is impossible to completely describe the phase diagram in this four-dimensional space, but we can at least get an idea of the general shape by taking various two-parameter cross-sections.

%

	\begin{figure}[htb] \centering
	\subfigure[
		]{
		\includegraphics[width=0.33\textwidth]{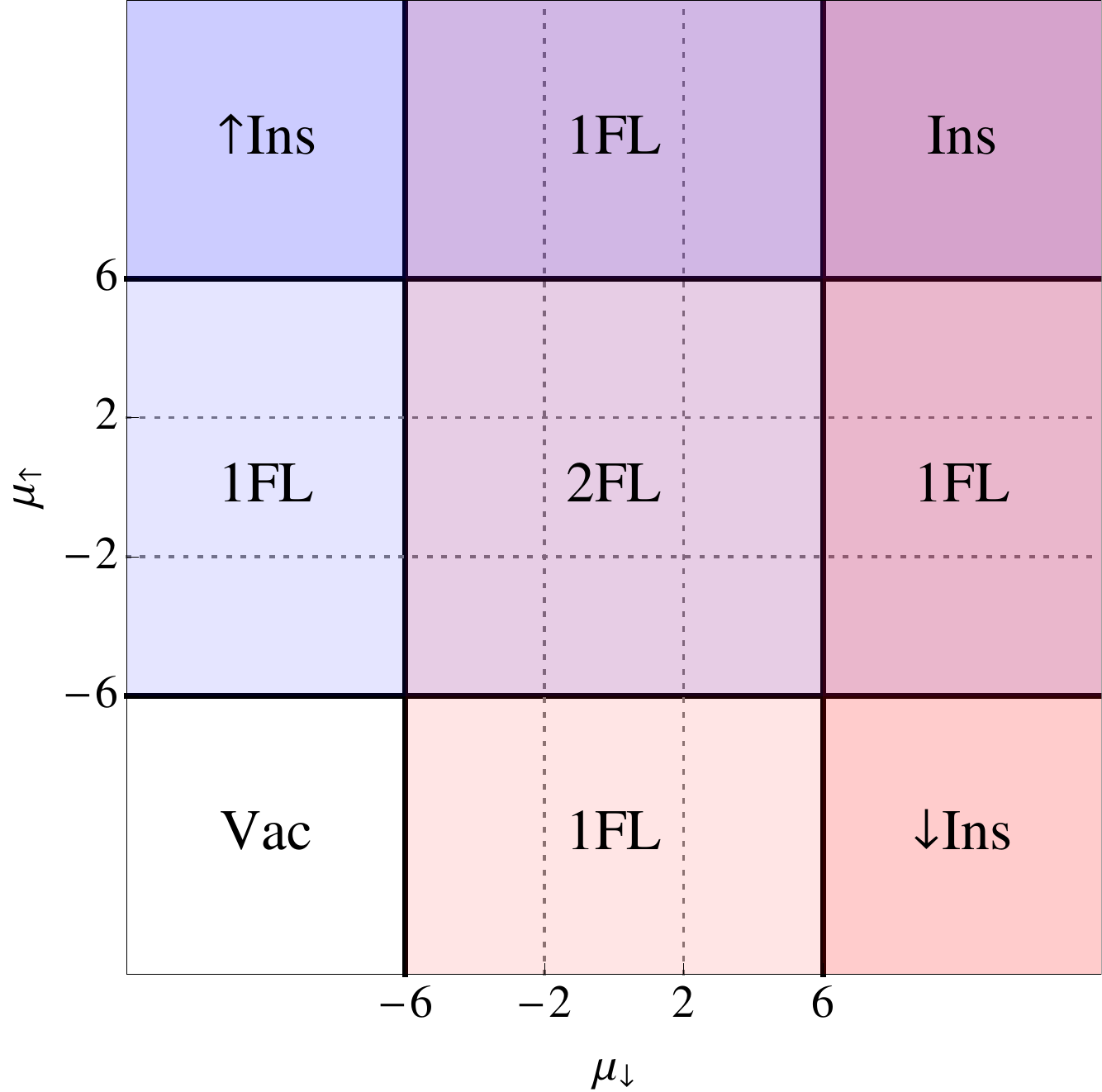} \label{TBPDmumu}
	}
	\subfigure[
		]{
		\includegraphics[width=0.33\textwidth]{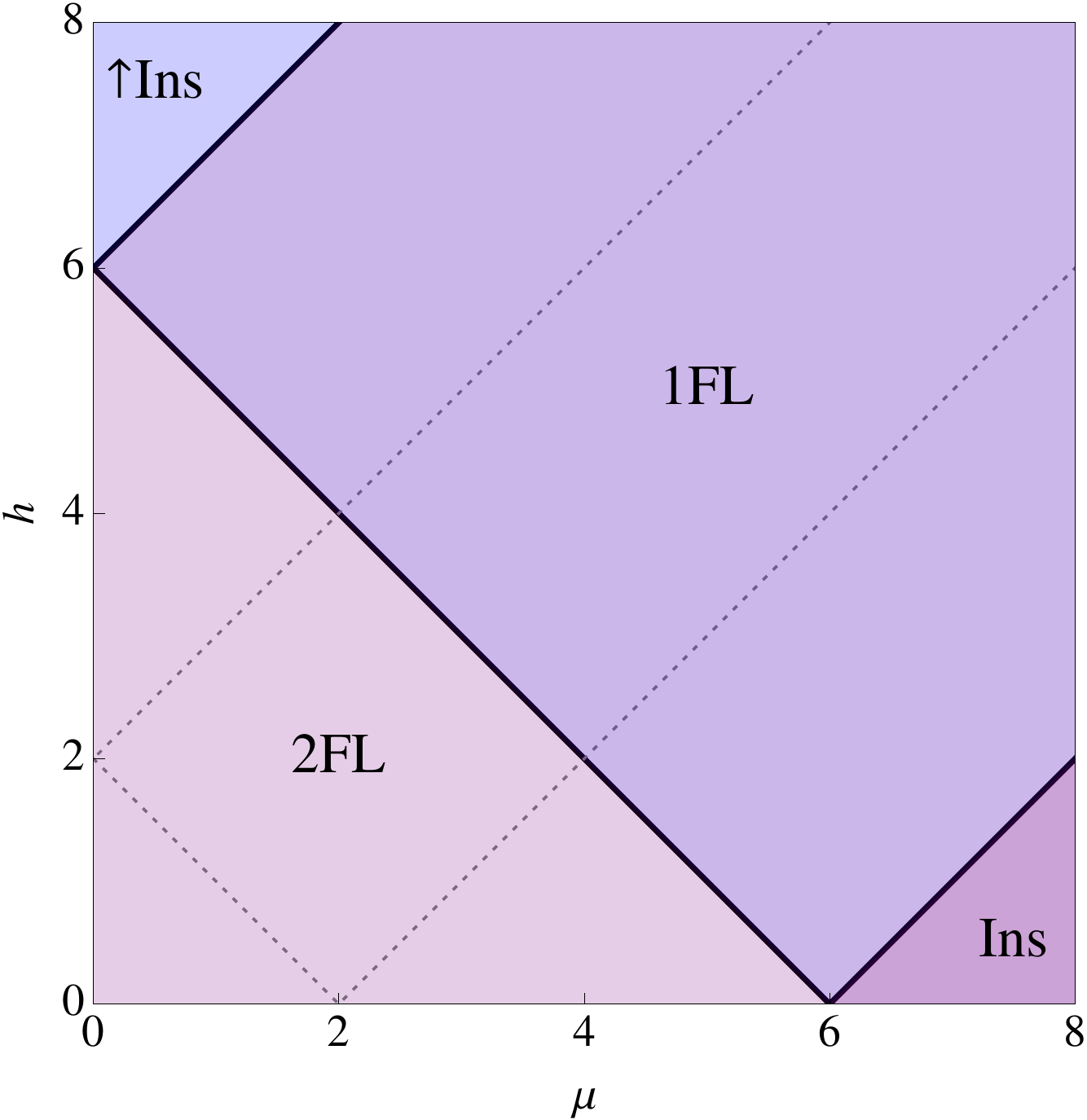} \label{TBPDmuh}
	}
	\subfigure[
		]{
		\includegraphics[width=0.33\textwidth]{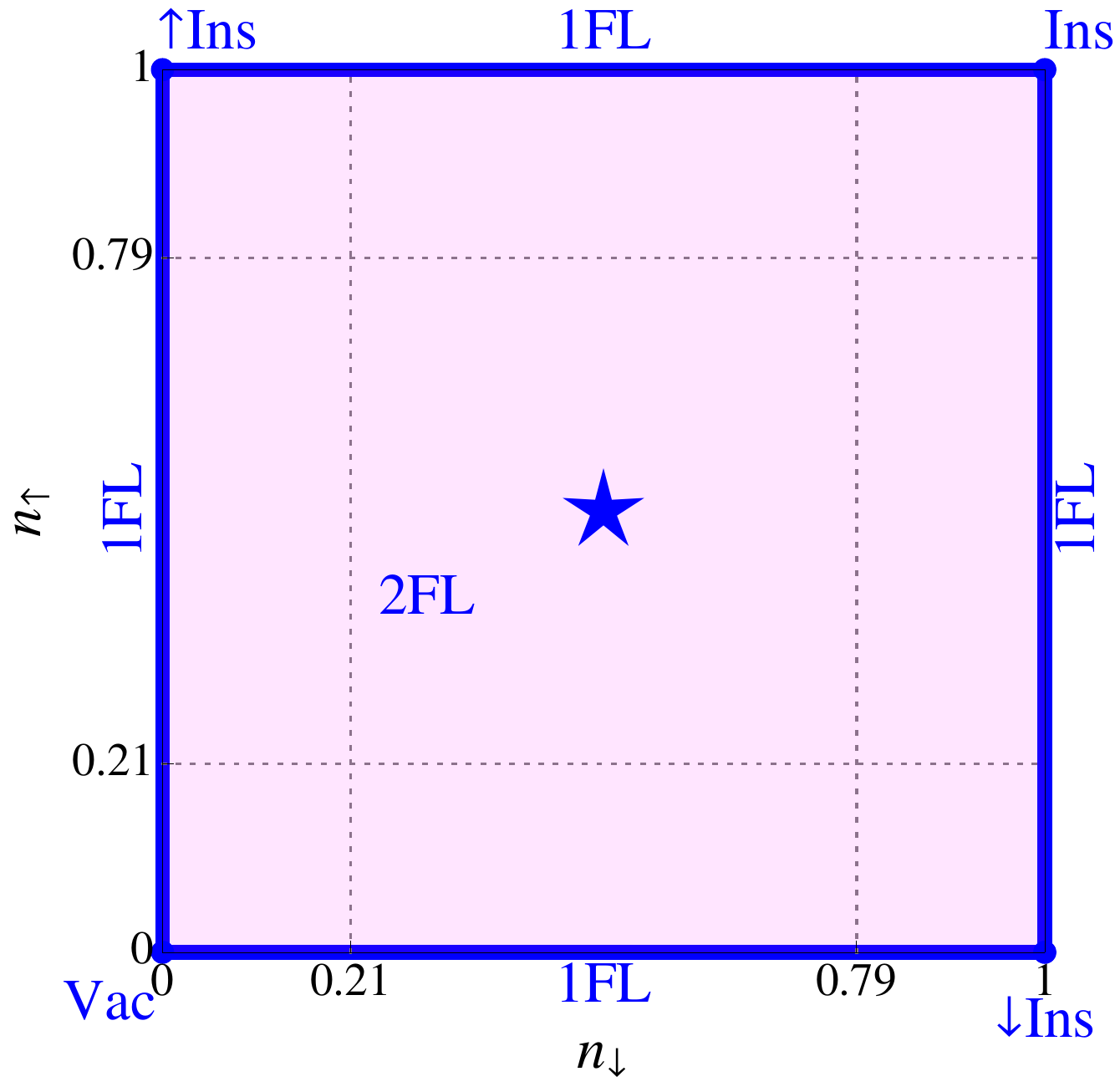} \label{TBPDnn}
	}
	\subfigure[
		]{
		\includegraphics[width=0.33\textwidth]{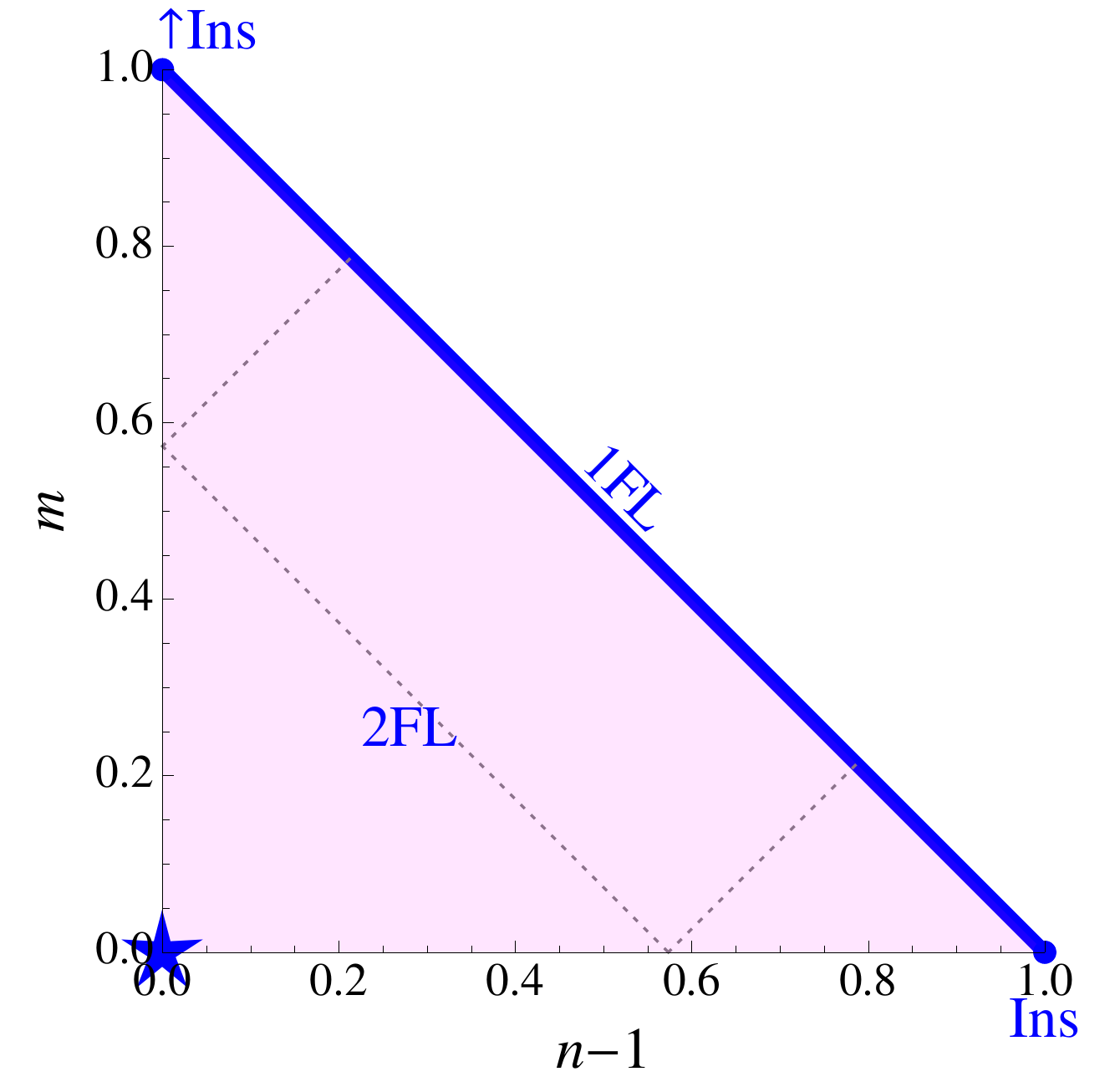} \label{TBPDnm}
	}
	\caption{
		\label{TightBindingPhaseDiagrams}
		Phase diagram of the cubic lattice tight-binding model with tunneling $t=1$
		as functions of chemical potentials $(\mu_\up,\mu_\dn)$
		and densities $(n_\up,n_\dn)$.
		Phase diagrams in $(\mu,h)$ space and $(n,m)$ space are also shown
		for $\mu,h,n-1,m \geq 0$; the phase diagrams are symmetrical under sign changes
		of these parameters.
		Dashed lines correspond to van Hove singularities within the band.
		The star indicates the high-symmetry point where both species are at half-filling.
	}
	\end{figure}

\subsubsection{A. Cubic lattice tight-binding model}
To orient this discussion, first consider free fermions in a cubic lattice.  Take the tunneling amplitude as the unit of energy, i.e., $t=1$.  The phase diagram is shown in Figure~\ref{TightBindingPhaseDiagrams}.  There are van Hove singularities at $\mu=-6,-2,2,6$, corresponding to densities $n=0, 0.213, 0.787, 1$.  Thus for two species of fermions the zero-temperature phase diagram in terms of chemical potentials $(\mu_\up,\mu_\dn)$  or densities $(n_\up,n_\dn)$ has a grid structure, as shown in Figs.~\ref{TBPDmumu} and \ref{TBPDnn}.  It is often customary to work in terms of the average chemical potential $\mu=\half(\mu_\up+\mu_\dn)$ and the Zeeman field $h=\half(\mu_\up-\mu_\dn)$, such that $\mu_\sigma = \mu + h\sigma$, or with the average density $n=\half(n_\up+n_\dn)$ and magnetization $m=\half(n_\up-n_\dn)$; this simply corresponds to rotating the phase diagrams by $45^\circ$.
Extreme values of $\mu$ and $h$ lead to insulating phases with 0 (Vac), 1 ($\up$I and $\dn$I), or 2 (BI) fermions per site, whereas moderate values give fully polarized (1FL) or partially polarized (2FL) Fermi liquid states. 

	
	\begin{figure}[htb] \centering
		\includegraphics[width=0.45\textwidth]{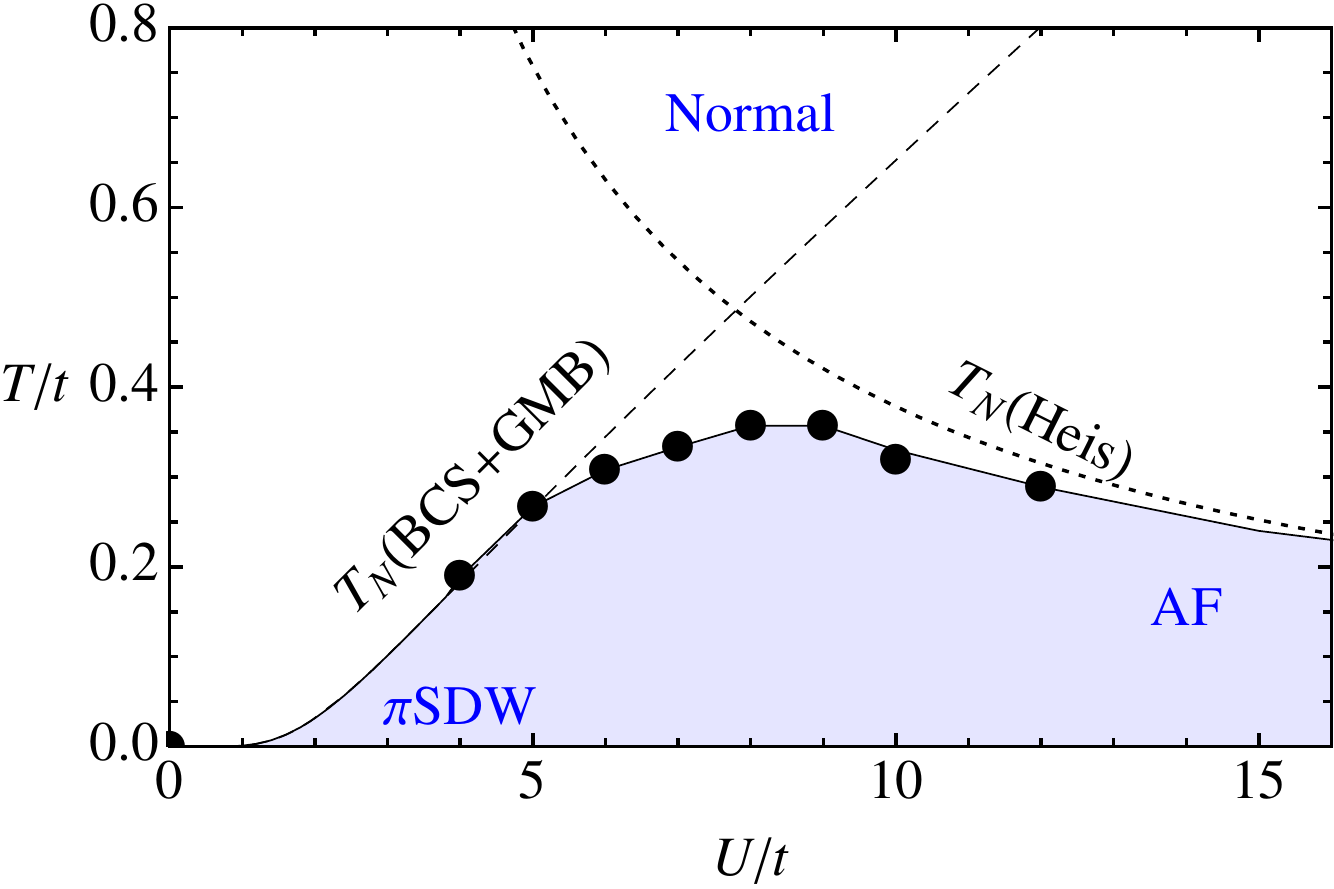}
	\caption{
		\label{HubbardPD-UT}
		Phase diagram of the cubic lattice Hubbard model with tunneling $t=1$
		as a function of repulsion $U$ and temperature $T$
		and densities $(n_\up,n_\dn)$.
		Solid points were obtained from QMC simulations; 
		see original publication \cite{staudt2000} for error bars and details.
		For attractive $U$ the phase diagram is the same,
		with BCS and BEC in place of $\pi$SDW and AF phases.
	}
	\end{figure}

\subsubsection{B. Half-filled Hubbard model: $(U,T)$ phase diagram}
Let us consider the high-symmetry parameter point in the tight-binding model ($\mu=h=m=0$, $n=1$).  
The Fermi surface at half-filling (Figure~\ref{FermiSurfaceMu0}) is perfectly nested with wavevector $(\pi,\pi,\pi)$.  Thus an infinitesimal repulsion $U$ is sufficient to produce spin-density-wave (SDW) ordering at zero temperature, albeit with an exponentially small critical temperature.  The dashed line is the mean-field critical temperature calculated using an integral over the cubic lattice density of states (an improvement to Eq.~\eqref{TcBCS})
times a GMB correction factor assumed to be 0.282 \cite{vandongen1994} (see Sec.~\ref{FermiPerturbationTheory}).
As $U$ increases from $0$ to $\infty$ there is a smooth crossover from a SDW regime with a weakly modulated spin density towards a well-developed Heisenberg antiferromagnet and a critical temperature $T_c \sim 0.346t^2/U$ (see Methods section).
Quantum Monte Carlo simulations (Staudt 2000) with finite-size scaling give estimates of $T_c(U)$ that are consistent with these two limits.   
The maximum critical temperature is approximately
$T_c(U=8,t=1,\mu=0,h=0) \approx 0.333(7)$ \cite{fuchs2011,staudt2000}.
See Figure~\ref{HubbardPD-UT}.

For attractive $U$ the phase diagram has exactly the same shape, with superfluid order instead of magnetic order.  An infinitesimal attraction renders the Fermi surface unstable to BCS pairing, producing a BCS superfluid, albeit with an exponentially small gap and critical temperature.

	\begin{figure}[htb] \centering
	\subfigure[
		]{
		\includegraphics[width=0.45\textwidth]{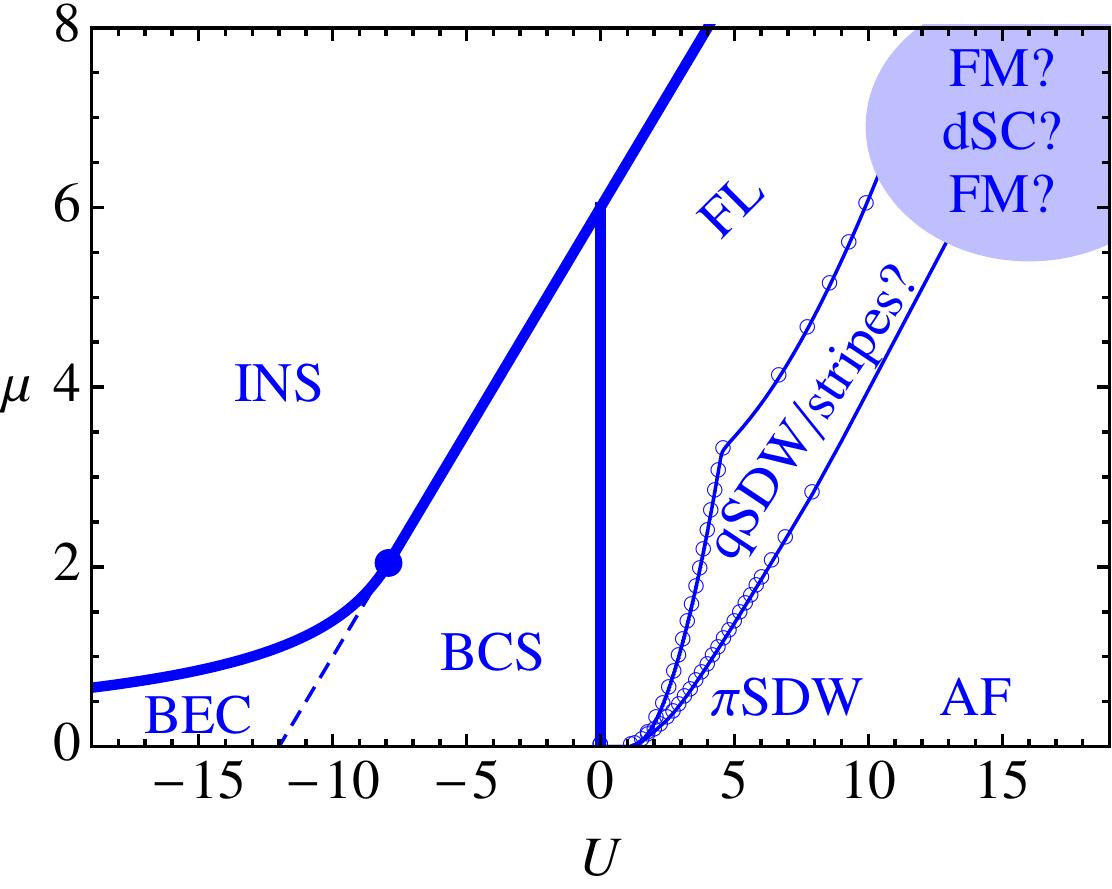} \label{HubbardPDUmu}
	}
	\subfigure[
		]{
		\includegraphics[width=0.45\textwidth]{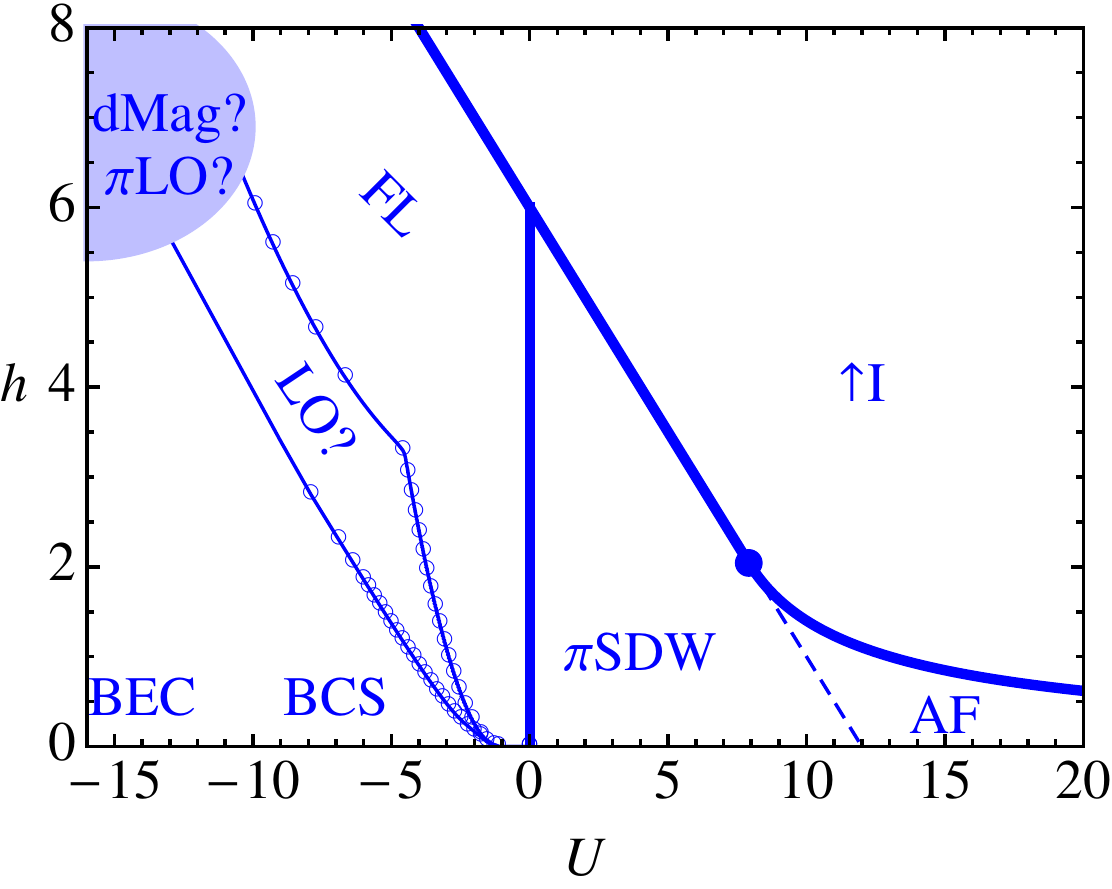} \label{HubbardPDUh}
	}
	\caption{
		\label{HubbardPDUmuAndUh}
		(a) Phase diagram of the cubic lattice Hubbard model with tunneling $t=1$
		as a function of repulsion $U$ and chemical potential $\mu$ for $h=T=0$.
		(b) Phase diagram as a function of $U$ and $h$ for $\mu=T=0$.
		Thick curves indicate exact phase boundaries.
		Empty circles indicate estimates within inhomogeneous mean field theory (BdG/HF).
		Abbreviations: 
		INS=band insulator, 
		$\up$I=fully polarized insulator,
		BEC=Bose-Einstein condensate,
		BCS=Bardeen-Cooper-Schrieffer s-wave superfluid,
		FL=Fermi liquid,
		$\pi$SDW=spin density wave at $(\pi,\pi,\pi)$,
		$q$SDW=spin density wave at other wavevector,
		AF=well-developed antiferromagnet at $(\pi,\pi,\pi)$,
		FM=Stoner ferromagnet (large $\mu$) or Nagaoka ferromagnet (smaller $\mu$),
		dSC=d-wave superfluid,
		LO=Larkin-Ovchinnikov modulated superfluid,
		dMag=d-wave bond magnetism.
	}
	\end{figure}

\subsubsection{C.  $(U,\mu,h=0,T=0)$ phase diagram}
First let us consider the attractive Hubbard model away from half-filling, i.e., the left half of Figure~\ref{HubbardPDUmu}.  
Obviously, for large positive $\mu$ there is a band insulator (INS) with two fermions per site.  The phase diagram is symmetrical under $\mu\rightarrow -\mu$, such that at large negative $\mu$ one has a vacuum with zero fermions per site.  The interesting physics occurs at intermediate chemical potentials with partial occupancy.

As long as the chemical potential (including the Hartree shift) lies within the band, there is an underlying Fermi surface with a pairing instability, and the system is a BCS superfluid at zero temperature, characterized by overlapping weakly bound Cooper pairs.  At strong attraction $\abs{U} > U_c = 7.91355$, even the band insulator undergoes a pairing instability according to the Stoner criterion $\Pi(\qqq=\0) = 1/\abs{U}$.  It can easily be shown that the INS-BEC phase boundary is given by
\newcommand{\mueff}{\widetilde{\mu}}
	\begin{align}
	U &= -\frac{2}{\Re G(\mueff)},
		\qquad
	\mu = \mueff - \frac{1}{\Re G(\mueff)}
	\end{align}	
in terms of the parameter $\mueff$ (where $\mueff>6$).
In this regime the system is best described as small, strongly bound fermion pairs -- molecules -- that behave somewhat like hard-core bosons, forming a superfluid Bose-Einstein condensate (BEC).  There is only a crossover from the BEC regime to the BCS regime; nevertheless the point $(-U_c,\mu_c)$ where $\mu_c=2.04322$ is a quantum critical point of some interest \cite{nikolic2007}.

The repulsive Hubbard model away from half-filling (the right half of the phase diagram) is a much more complicated beast.  At weak coupling and low hole densities (i.e., close to the band insulator), the ground state is a Fermi liquid.  For $\mu \approx 0$ the system has magnetic order at wavevector $(\pi,\pi,\pi)$.  However, for intermediate doping, there are many competing orders leading to almost-degenerate ground states and a tendency towards glassy behavior; analytic methods suffer from uncontrolled approximation, whereas numerical methods are severely hampered by the fermion sign problem.  
Despite several decades of work \cite{affleck1988,andersonScience1987,kishine2001,varma2000,casey2011}
the phase diagram is not understood.  Nevertheless, we attempt to describe some candidate phases to give an idea of the types of physics that can arise in the Hubbard model.

\paragraph{\underline {Spontaneously modulated magnets:}}
Starting in the Fermi liquid (FL) phase and going towards half-filling, perturbative or mean-field calculations -- whether or not one trusts them -- indicate that the system eventually experiences a magnetic instability \emph{not} at the antiferromagnetic wavevector, but typically at $\qqq=(\pi-\delta, \pi, \pi)$ and symmetry-related points, where $\delta$ depends on $U$.  
Near the FL state, or at weak coupling, this suggests the existence of incommensurate spin density wave (qSDW) phases characterized by modulations of the spin density at $\qqq$ (and consequent modulations of the number density at $(2\delta,0,0)$).
Going towards to the AF state, or towards strong coupling, there is a crossover to stripe phases, which are often described as rivers of charge flowing along domain walls in an antiferromagnetic background (for the 3D Hubbard model ``layered phases'' would be a more appropriate term).
See schematics in Figure~\ref{LMTCartoons}.
These states are likely to be vulnerable to quantum fluctuations.
Nevertheless, even if long-range order does not occur, fluctuating or glassy short-range order may still play an important role \cite{kohsaka2007}.

\paragraph{\underline{Spontaneously modulated superfluids:}}
Using the Lieb-Mattis transformation, we see that the $(U,h,\mu=0,T=0)$ phase diagram (Figure~\ref{HubbardPDUh}) is simply a mirror image of Figure~\ref{HubbardPDUmu}.  
Thus, SDW/stripe phases in the repulsive Hubbard model have their counterparts in the attractive Hubbard model, which are spontaneously modulated superfluids -- Larkin-Ovchinnikov (LO) phases \cite{singh1991}.
Within MFT, as one starts from a BCS superfluid and increases the Zeeman field $h$, it eventually becomes favorable for excess up-spin fermions to penetrate the superfluid in the form of domain walls.
Conversely, starting in the high-field normal Fermi liquid state and reducing the field, one encounters a pairing instability at a nonzero wavevector, which suggests a transition to a LO state.  
See schematics in Figure~\ref{LMTCartoons} and BdG configurations in Figure~\ref{BdGConfigs}.

The problem of superconductivity in a Zeeman field has in fact been studied since the 1960's.
For 3D continuum fermions,
homogeneous mean-field theory suggests a first-order transition from a superconductor to a high-field normal metal at the Chandrasekhar-Clogston critical field $h_{CC} = \Delta_0/\sqrt{2} \approx 0.71 \Delta_0$, where $\Delta_0$ is the zero-temperature zero-field gap.
\cite{chandrasekhar1962,clogston1962,sarma1963,gorkovrusinov1964}.
However, inhomogeneous mean-field theory reveals that in an intermediate range of fields the system can lower its free energy by forming periodic patterns known as Fulde-Ferrell-Larkin-Ovchinnikov (FFLO) states.
Fulde and Ferrell \cite{fulde1964} used the ansatz $\Delta(\rrr) \propto e^{i\QQQ\cdot\rrr}$, which breaks time-reversal symmetry, whereas Larkin and Ovchinnikov studied additional modulation patterns of the form $\Delta(\rrr) \propto \sum_\QQQ \cos \QQQ\cdot\rrr$ \cite{larkin1964}, which break translational symmetry.
It is generally found that LO states are favored over FF states, so we will henceforth simply refer to LO states.  
An LO state is quite similar to a smectic liquid crystal.
A proper treatment of the LO state requires some amount of numerical work. \cite{machida1984,burkhardt1994,yoshida:063601,koponen:120403,loh2010}
See Ref.~\onlinecite{casalbuoni2004} for a review.
In the 3D continuum, FFLO only occupies a tiny sliver of the mean-field phase diagram \cite{sheehy2006}; quantum and thermal fluctuations destroy even this sliver \cite{radzihovsky2009}.
In 2D, the mean-field FFLO region is larger, but fluctuations are even more severe.
The phase diagram is exquisitely sensitive to strong-coupling corrections, material properties, and calculational methods, and many aspects of FFLO physics continue to be debated;\cite{shimahara1998,matsuo1998,mora2005,bulgac2008}) for example, some authors find first-order transitions to a crystalline LO state with minority spins localized in a superconducting background.

The search for FFLO superconductivity in condensed matter systems is complicated by orbital depairing, spin-orbit coupling, impurities, etc.  
Pairing amplitude oscillations (induced FFLO) occur near superconductor-ferromagnet boundaries.
Where spontaneous oscillations (bulk FFLO) are concerned, thermodynamic signatures (heat capacity) have been reported in layered organic and heavy-fermion superconductors\cite{radovan2003,correa2007,koutroulakis2008,cho2009,koutroulakis2010}, 
and spectral signatures (excess zero-bias tunneling conductance) are seen in thin Al films \cite{adams2004,loh2011}, but evidence is otherwise slim.

In contrast, cold Fermi gases in optical lattices are a promising arena in which to search for FFLO physics \cite{parish:250403,zhao:063605,koponen:120403,loh2010}.  
The cubic lattice phase diagram has a much larger LO region than the 3D continuum phase diagram within BdG (due to better nesting and Hartree corrections).  Furthermore, an optical lattice may pin the spacing and direction of the ``domain walls'' within an LO state, gapping the translational and/or rotational Goldstone modes and allowing long-range order.
Ultimately, though, the most favorable systems appear to be quasi-one-dimensional systems -- coupled tubes or anisotropic lattices at weak-to-intermediate coupling.
There is an experimental effort to search for FFLO in such systems \cite{liao2010}.
Another intriguing possibility is to look for FFLO in $p$-orbital bands in 2D and 3D optical lattices \cite{cai2011}, which benefit from a combination of good nesting (because single-particle hopping is confined to the direction in which the orbitals overlap) and good phase coherence (because Josephson tunneling occurs in all directions).

\paragraph{\underline{Stoner ferromagnetism:}}

The fate of a FL upon turning on repulsive interactions and the possibility of ferromagnetism is a problem of fundamental interest.
This problem has a long history beginning with Stoner's mean field calculation in 1939. 
At strong repulsion ($U/t \gtrsim 16$), starting in the FL and going toward half-filling, the Stoner criterion suggests that the dominant instability is ferromagnetic ($\qqq=\0$).
However, Stoner's instability is obtained 
at a dimensionless coupling strength of order unity where the validity of mean field calculations becomes questionable. 
This is also alluded to in the Methods section, that perturbative or MFT results cannot be trusted at such a large value of $U$.  MFT assumes that the wavefunction is a ``rigid'' Slater determinant such that the only way to reduce double occupancy is through magnetism, whereas actual wavefunctions can ``deform'' such that fermions avoid each other without invoking magnetic ordering.

State-of-the-art variational and Green function or projective Monte Carlo techniques with trial wave functions that include backflow corrections have shown that there is indeed a ferromagnetic instability 
for a Fermi gas with hard sphere interactions~\cite{chang2011}. 


Note that at very large $U$ and low density the Hubbard model maps to the continuum Fermi gas with repulsive short-range scattering.
Based on the variational calculations discussed above that would indicate the existence of \emph{ferromagnetism at low densities} (i.e., in the top right corner of Figure~\ref{HubbardPDUmu}).

Recent experiments that created repulsive interactions between atoms in the upper branch
have observed signatures of ferromagnetism \cite{jo2009} 
but these are probably due to nonequilibrium quench dynamics\cite{pekker2011}.
However, it is possible that if three-body processes
leading to molecule formation can be suppressed, there may
be a window of time-scales where equilibrium physics in the upper
branch will be observable. 

%

\paragraph{\underline{Nagaoka ferromagnetism:}}
Some rigorous theorems have been proven about ferromagnetism \emph{near half-filling}\cite{Nagaoka1966,thouless1965}: for example, that adding (or removing) a single fermion to the half-filled $U=\infty$ Hubbard model on a finite cubic lattice changes the ground state from antiferromagnetic to ferromagnetic.  
There is some theoretical evidence that a finite \emph{concentration} of holes can produce ferromagnetic order over a finite region of the 2D Hubbard phase diagram 
at very strong repulsion ($U/t \gtrsim 80$) sufficiently close to half-filling ($0.8 \lesssim n \lesssim 1.2$) \cite{carleo2011,liu2012Hubbard,Shastry1990}.
This phenomenon is represented schematically as the lower FM state in Figure~\ref{HubbardPDUmu}.
 

\paragraph{\underline{Exotic superfluids:}}
In the doped 2D repulsive Hubbard model (and possibly also in the 3D Hubbard model) there may be $d$-wave superfluidity at fairly large $U/t$.  Strictly speaking, the pairing symmetry is not $d$-wave, but rather some representation of the lattice symmetry group
\cite{kotliar2001,maier2005,senechal2005}.
The Lieb-Mattis transformation then implies that the attractive Hubbard model in a Zeeman field has an exotic form of magnetic order.

	\begin{figure}[htb] \centering
	\subfigure[
		]{
		\includegraphics[height=0.33\textwidth]{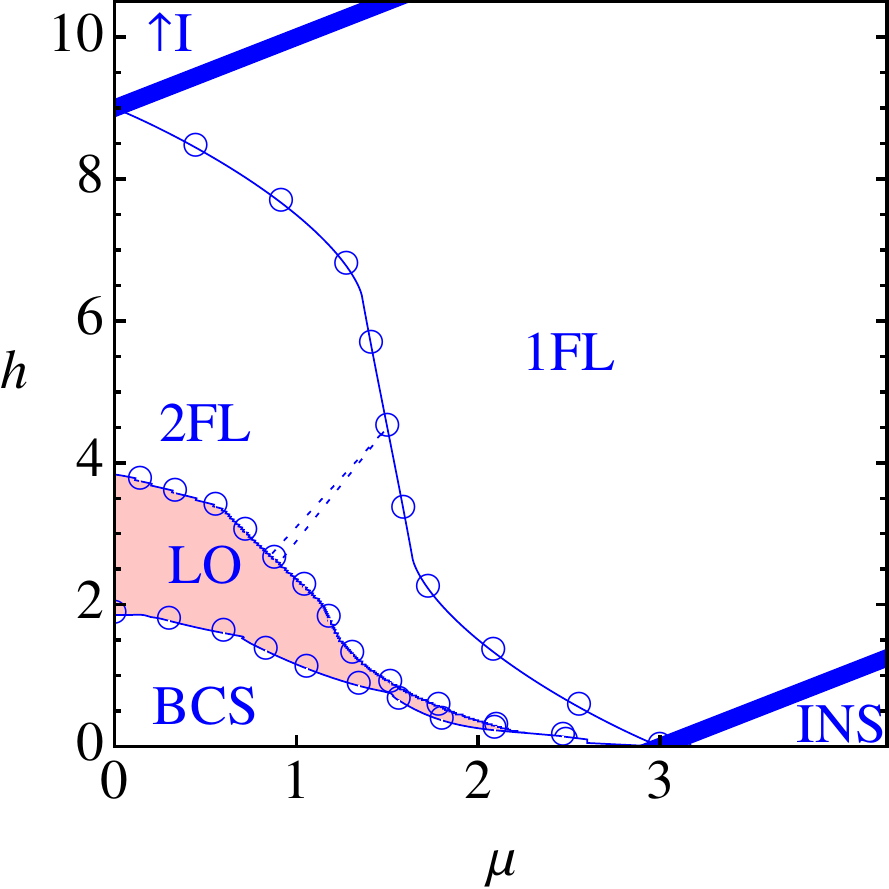} \label{HubbardPD-mu-h-U6-T0}
	}
	\subfigure[
		]{
		\includegraphics[height=0.33\textwidth]{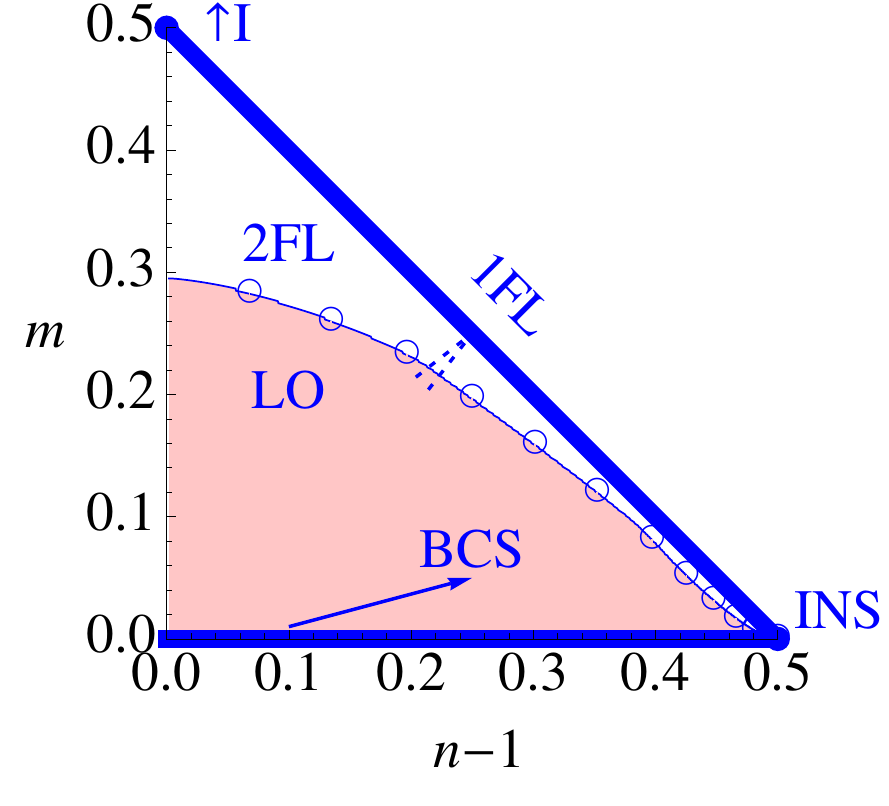} \label{HubbardPD-n-m-U6-T0}
	}
	\caption{
		\label{HubbardPDmuh}
		(a) Approximate phase diagram of the cubic lattice Hubbard model with tunneling $t=1$
		at $U=-6$ and $T=0$ in $(\mu,h)$ space.
		(b) Equivalent phase diagram in $(n,m)$ space.
		Thick lines indicate exact phase boundaries.  
		Empty circles indicate phase boundaries within the BdG approximation (see text).
	}
	\end{figure}

\subsubsection{D. $(h,\mu=0,U=6,T=0)$ phase diagram}
Another slice through the phase diagram is shown in Figure~\ref{HubbardPD-n-m-U6-T0}.
At extreme values of $\mu$ or $h$ the ground state is an insulator or one-component Fermi liquid (1FL); the 1FL phase in the figure has hole-like quasiparticles carrying down-spin.
For moderate $\mu$ and $h$ the ground state is a two-component Fermi liquid (2FL).

The figure shows the 1FL-2FL boundary within mean-field theory, including only Hartree corrections -- i.e., with a Slater determinant ansatz for the wavefunction.  In actual fact, when a single minority spin ($\dn$) is added, the majority spin ($\up$) Fermi liquid develops a screening cloud (polaron) around the minority spin, lowering the total energy in the process and favoring the 2FL state.  Therefore the true 1FL-2FL boundary is at larger $\mu$ and $h$ than shown.

The 2FL-LO boundary in the figure is the locus of the finite-wavevector pairing instability within the ladder approximation.  In actual fact, higher-order quantum corrections may shrink(or even destroy) the ordered LO phase; nevertheless, LO physics may still play a role.

The LO-BCS boundary in the figure is determined by the point at which it becomes energetically  favorable to insert spin-polarized domain walls into a BCS state.  Again, it is very difficult to estimate the effect of quantum fluctuations.

When $n_\dn=1$, i.e., $n-1=m$, the down-spin Fermi surface is perfectly nested, and it becomes favorable to form a charge-density-wave (CDW) phase with wavevector close to the nesting vector $(\pi,\pi,\pi)$.  This is a rigorous result in the weak-coupling limit, but the CDW phase has a tiny critical temperature and occupies a tiny sliver of the phase diagram (shown schematically as dotted lines), so it is irrelevant to real experiments.

\subsubsection{E. Thermodynamics and correlations}
For pedagogical purposes and to give orientation, we collate some results for the half-filled repulsive Hubbard model here ($U>0, \mu=h=0$).  

\begin{figure}[!htb]
	\subfigure{
		\includegraphics[width=0.45\columnwidth]{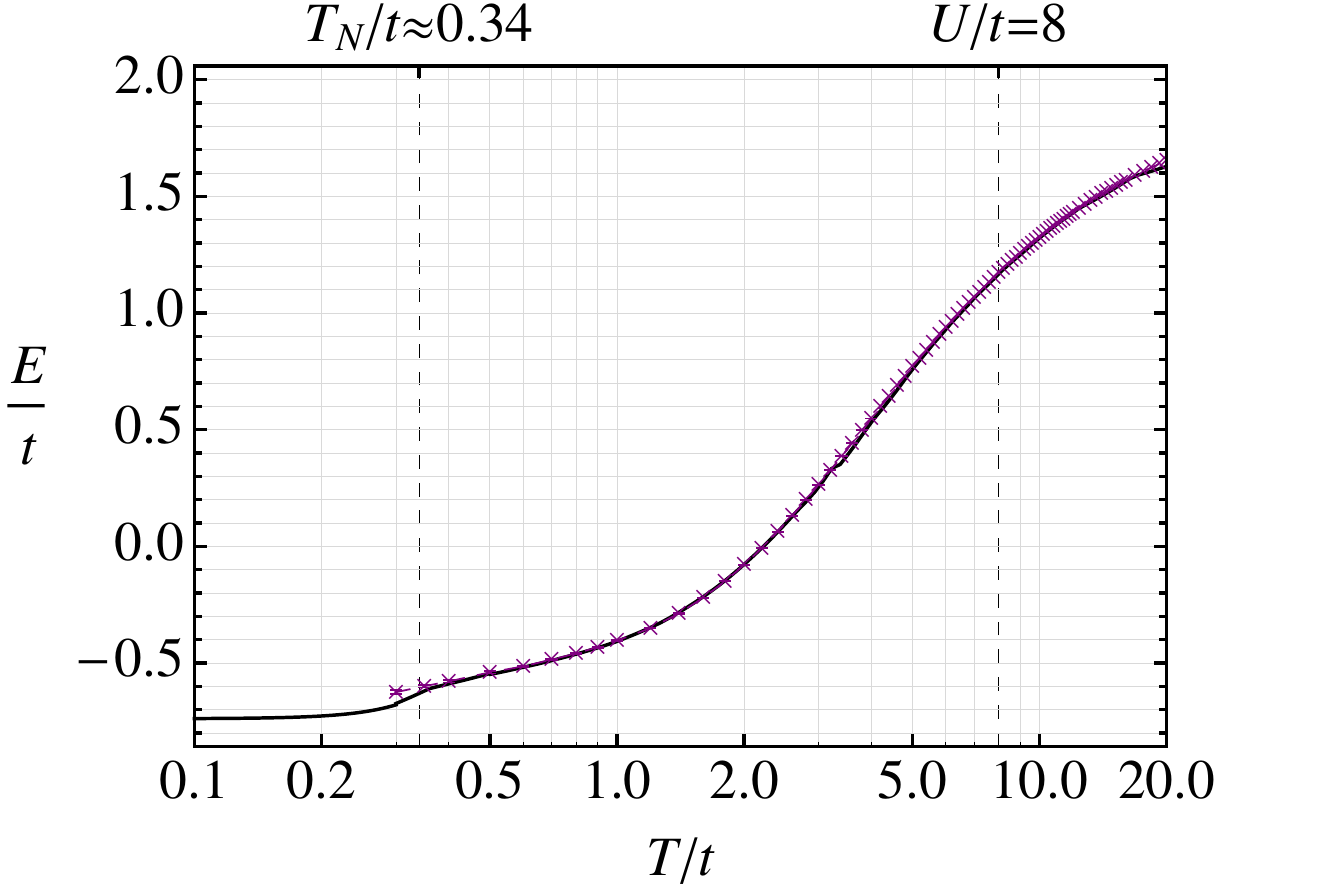}
		\label{HalfFillingET}
	}
	\vspace{-1cm}
	\subfigure{
		\includegraphics[width=0.45\columnwidth]{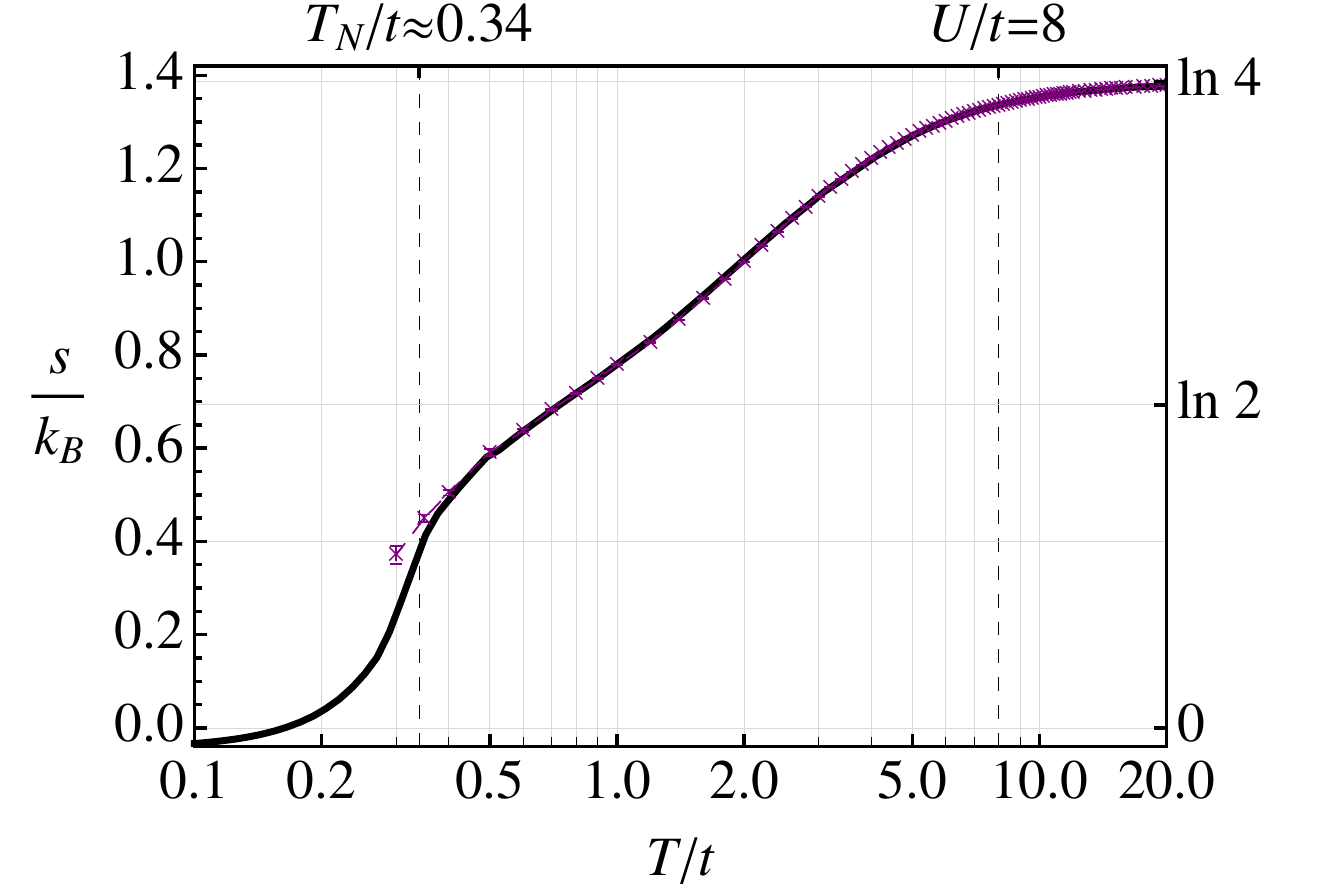}
		\label{HalfFillingST}
	}
	\vspace{5mm}
	\caption{
		\label{ETsTFinal}
	(a) Energy per site of half-filled cubic lattice Hubbard model	at $U/t=8$
	according to extrapolated DCA (symbols) and extrapolated and fitted DQMC (curve).
	(b) Entropy per site,
	showing a shoulder at the Mott scale $T_\text{Mott} \simeq U$
	and a distinct feature at the N\'eel temperature $T_N$ due to critical fluctuations.
	}
	\end{figure}

\paragraph{\underline{Temperature dependence:}}
Figure~\ref{ETsTFinal} shows the energy and entropy according to DCA and DQMC simulations (after appropriate scaling and extrapolation).  See original publications \cite{fuchs2011,paiva2011}
for error estimates and data tables.
At high temperatures $T \gtrsim U$ each Hubbard site is equally likely to be in any of the four possible states, so $s \approx \ln 4$.  As $T$ is decreased the system enters a Mott regime where  singly occupied sites are favored and $s \approx \ln 2$.  Antiferromagnetic order appears only at a much lower temperature $T_N$.
\footnote{The DCA calculations assume no magnetic ordering.}
\footnote{The slightly negative value of $s(T)$ at low $T$, due to error accumulated in the integral formula for the entropy, serves as a check on the size of systematic errors.}

\begin{figure}[!htb]
	\subfigure{
		\includegraphics[width=0.31\columnwidth]{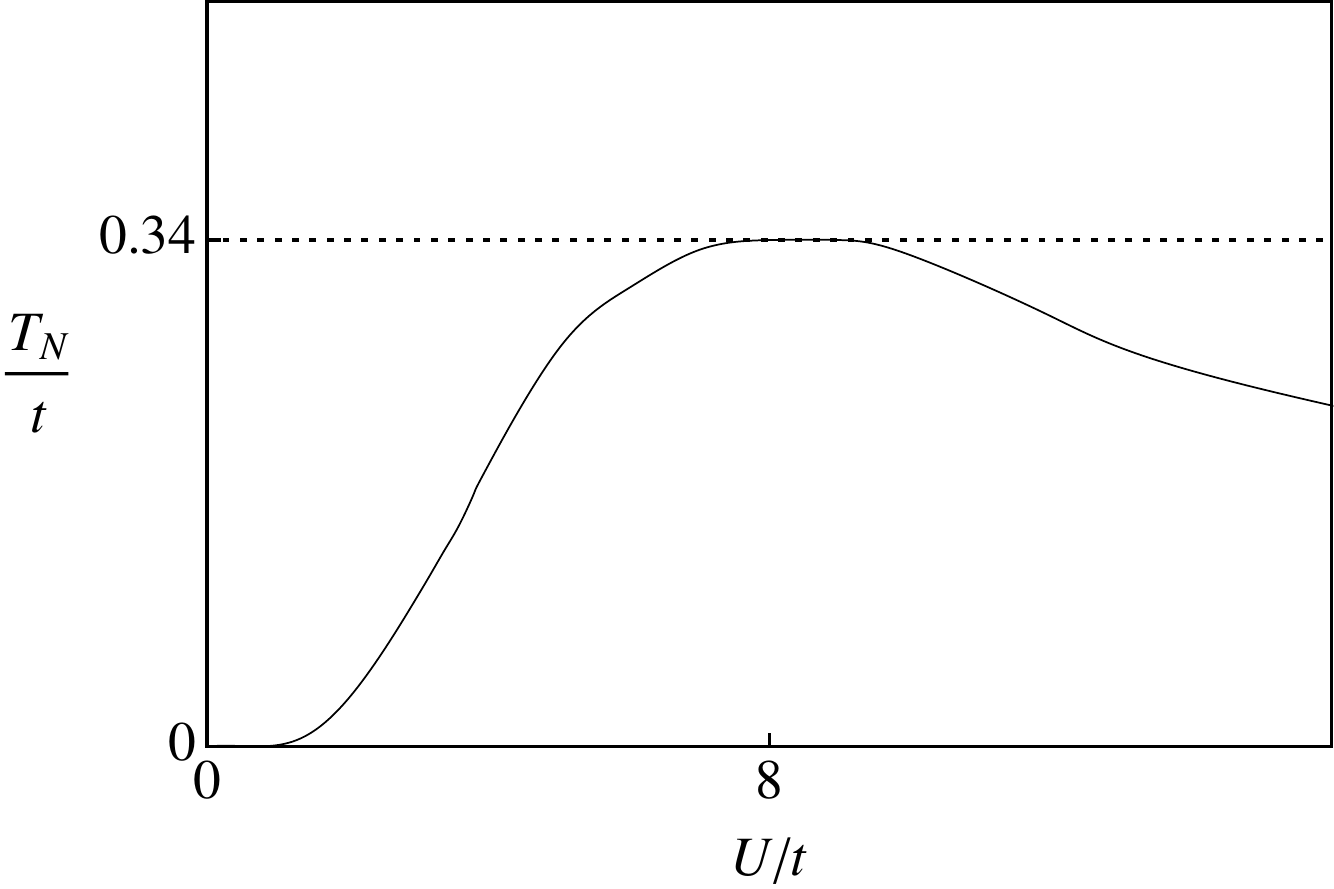}
		\label{HalfFillingTU}
	}
	\subfigure{
		\includegraphics[width=0.31\columnwidth]{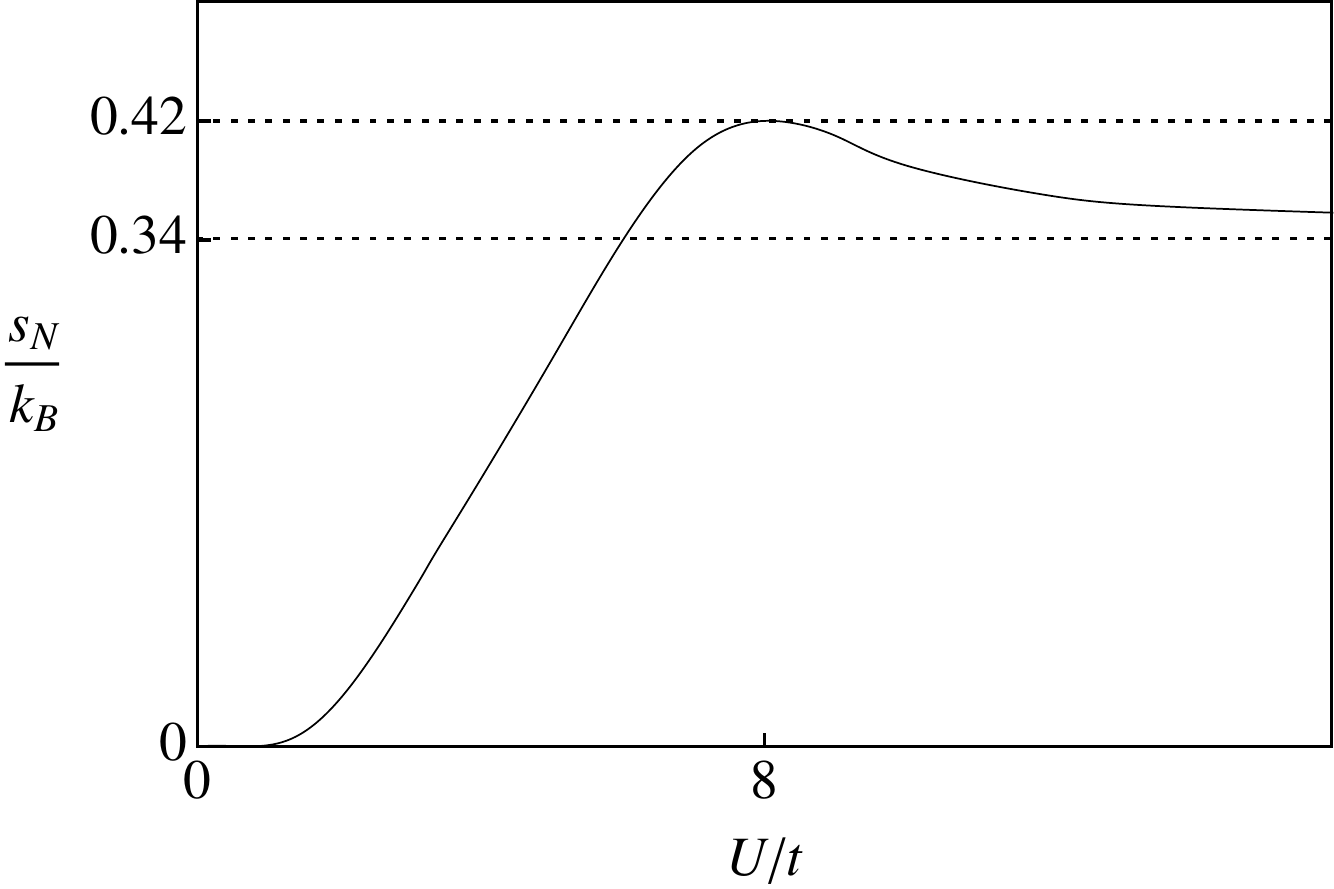}
		\label{HalfFillingSU}
	}
	\subfigure{
		\includegraphics[width=0.31\columnwidth]{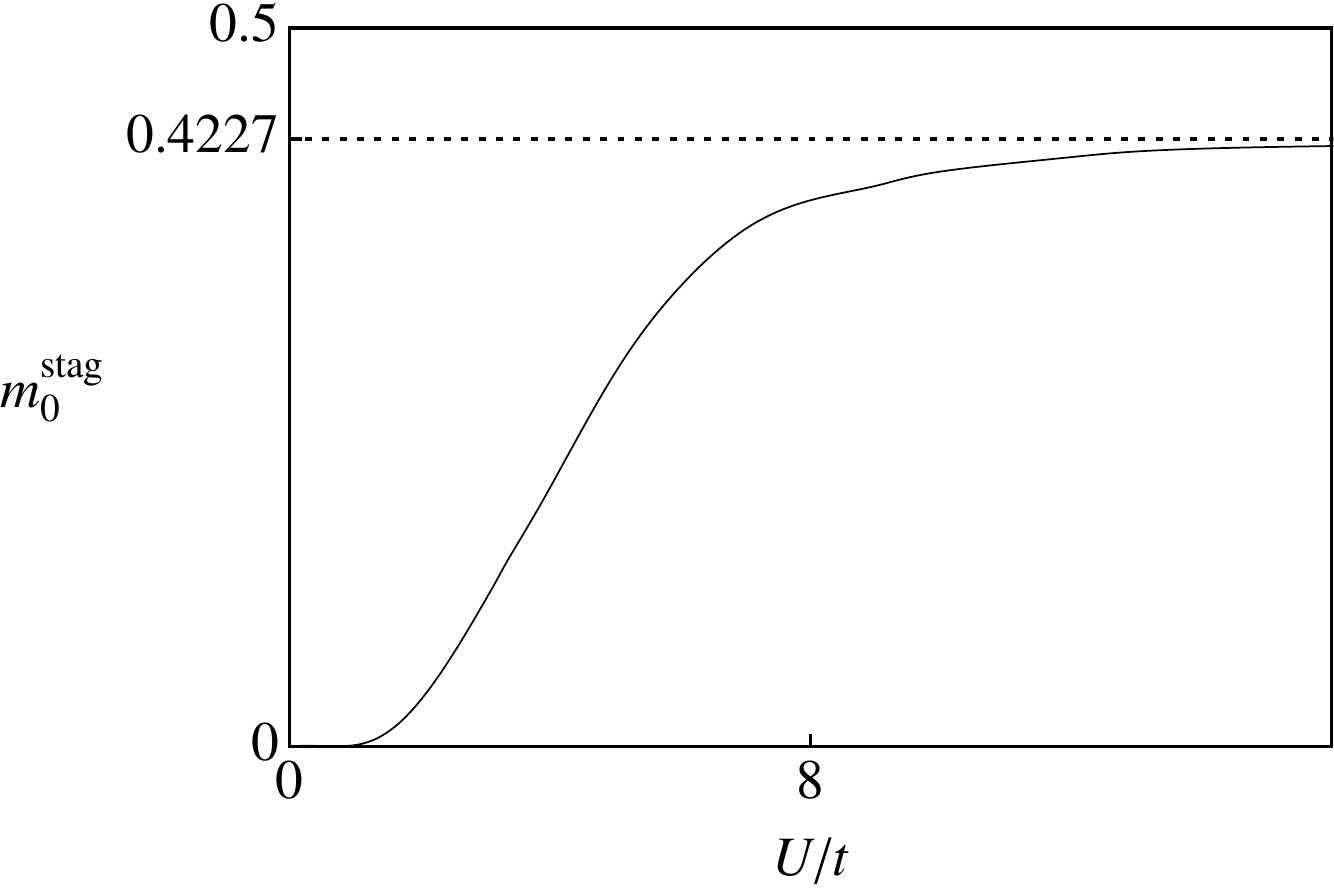}
		\label{HalfFillingMU}
	}
	\caption{
		\label{HalfFillingUDependence}
		Schematic of various properties 
		of the half-filled cubic lattice Hubbard model
		as functions of repulsion $U$.
	}
	\end{figure}

\paragraph{\underline{Interaction dependence:}}
Figure~\ref{HalfFillingTU} is a schematic of Figure~\ref{HubbardPD-UT}, reproduced here for convenience.
Figure~\ref{HalfFillingSU} shows a schematic of the critical entropy per site $s_N(U)$,
defined to be the entropy at the N\'eel temperature for a particular value of $U$. 
At $U/t=8$, $s_N/k_B=0.42(2)$ 
according to DCA and DQMC simulations \cite{fuchs2011,paiva2011,staudt2000},
whereas in the limit $U\rightarrow \infty$, $s_N/k_B \rightarrow 0.341(5)$ 
according to SSE QMC simulations \cite{wessel2010}.
Figure~\ref{HalfFillingMU} is a schematic of the zero-temperature staggered magnetization 
$M^\text{stag}_0(U)$, which tends to 0.4227 \cite{oitmaa1994} in the Heisenberg limit (slightly reduced from the classical value 0.5 by quantum fluctuations).

The Lieb-Mattis transformation gives corresponding properties of the attractive Hubbard model at half-filling.  The staggered magnetization maps to the anomalous density $F$.  For superfluids, it is conventional to quote the pairing amplitude $\Delta=\abs{U} F$ instead of $F$ itself.  BCS theory predicts that $T_N$, $s_N$, and $\Delta$ are all exponentially small as $U\rightarrow 0$.

\subsubsection {F. Other lattices and dimensions}


We have focused on the cubic lattice Hubbard model.  Myriads of other interesting phenomena arise from different dimensionalities and lattice structures and are based on the effects of stronger fluctuations in lower dimensions, 
unusual band structure such as on a honey comb lattice with Dirac cones\cite{geim2007} that has 
been successfully emulated with cold atoms \cite{tarruell2012}, and frustration as in traingular and kagome lattices\cite{yan2011,nakano2011} which has also been realized in optical lattices~\cite{jo2012}.
The attractive-$U$ Hubbard model generally has a superfluid ground state. 
For the repulsive-$U$ Hubbard model there are more possibilities.

On frustrated non-bipartite lattices quantum Monte Carlo methods suffer from the sign problem due to geometric frustration.  
It has been shown using variational methods that the Hubbard model on a triangular lattice maps onto a Heisenberg model with additional ring exchange 
at large $U$. The effect of the ring exchange is to melt the three-sublattice N{\'e}el order with coplanar moments at $120^\circ$ angles to each other into a variety of possible spin liquids that can be described as projected Fermi liquids~\cite{motrunich2005,motrunich2006}, or nematic d-wave 
gapless spin liquids with nodal spinons~\cite{grover2010}.

\subsection{Some experimental considerations}
\begin{figure}[htb] \centering
	\subfigure[
		]{
		\includegraphics[width=0.35\textwidth]{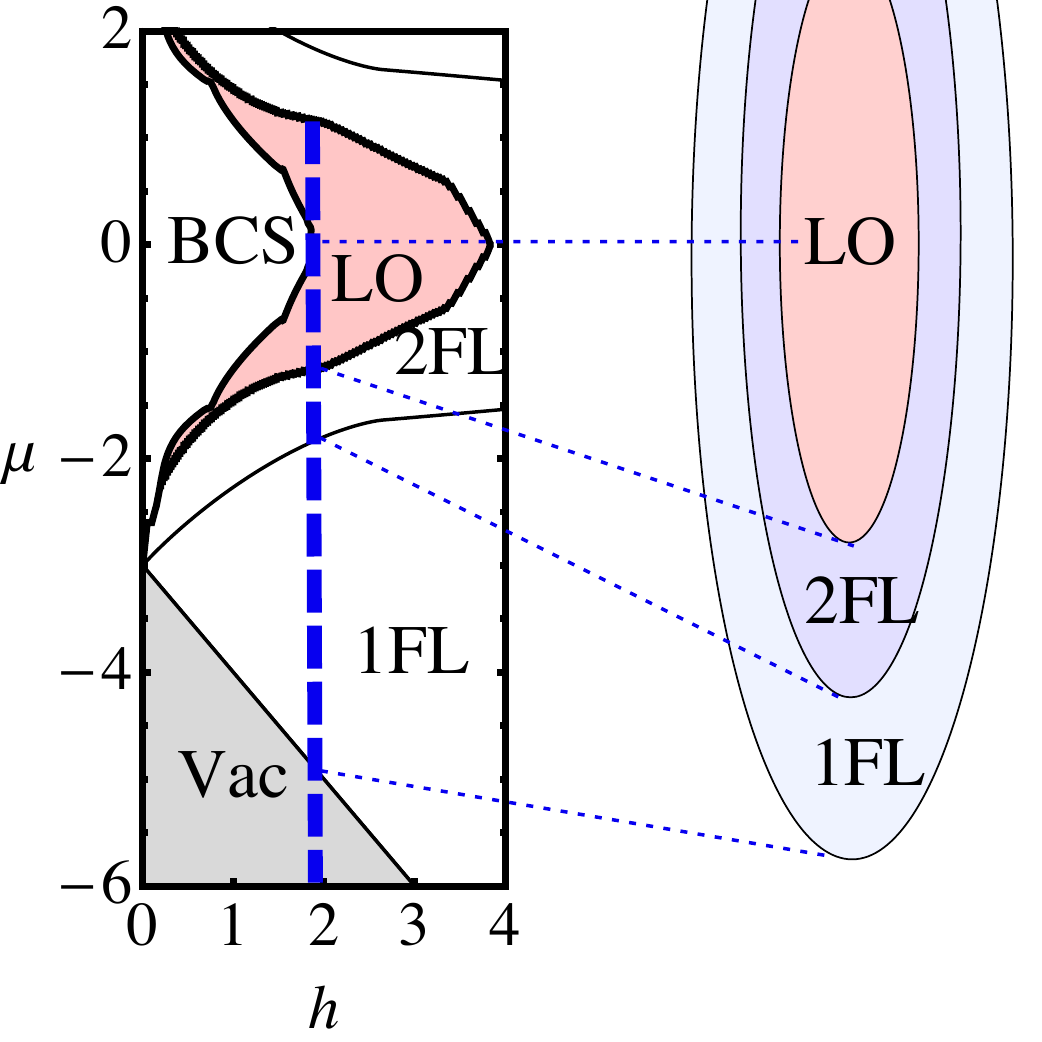} \label{LDAExample}
	}
	\subfigure[
		]{
		\includegraphics[width=0.55\textwidth,clip=true,trim=160mm 0mm 0mm 0mm]
			{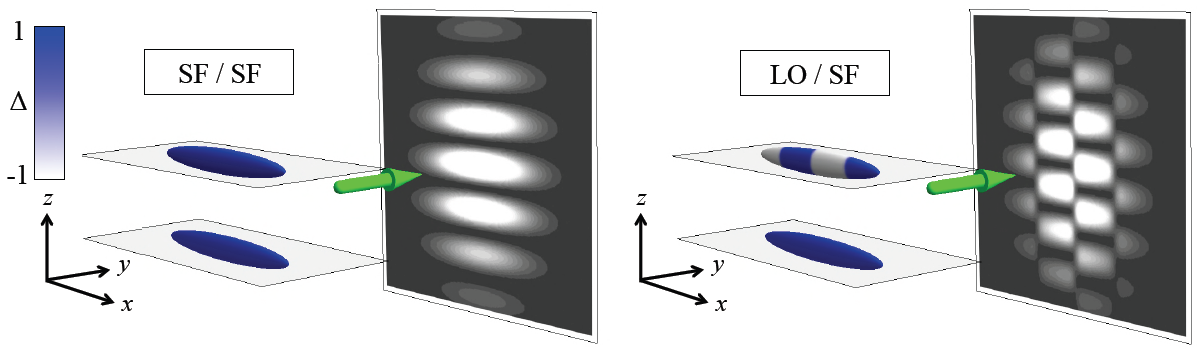} \label{MasonsFig2}
	}
	\caption{
		(a)
		Local density approximation applied to the attractive Hubbard model
			with a population imbalance.
		Left: BdG phase diagram of the cubic lattice Hubbard model 
		at $U=-6t, T=0$ in the $(\mu, h)$ plane.  
		The dashed blue line indicates a slice through the phase diagram at constant $h$
			and changing $\mu$.
		Right: Shell structure in a harmonic trap within the LDA.
		For the chosen value of $h$,
		$80\percent$ of the atoms are in the LO phase.
		(b)
		A proposed interferometric method for the detection of LO phases (modulated superfluids)		  	
		\cite{swanson2012}.
	}
	\end{figure}

\paragraph{\underline{Trap-induced inhomogeneity:}}
All cold atom experiments are performed in traps.  
The trapping potential can typically be approximated by a harmonic potential of the form
$V(x,y,z) = V_0 - \frac{m}{2} (\omega_x{}^2 x^2 + \omega_y{}^2 y^2 + \omega_z{}^2 z^2)$.
The local density approximation (LDA) assumes that local thermodynamic properties are determined by the local chemical potential $\mu(\rrr) = \mu_0 - V(\rrr)$; this generally gives good intuition about distribution of phases in the trap.
As one goes away from the center of the trap, the chemical potential decreases, whereas the Zeeman field and temperature remain constant (assuming thermodynamic equilibrium), so the locus of the $(\mu,h,T,U)$ parameter point is a line segment on the $(\mu,h,T,U)$ phase diagram starting at $\mu_0$ and going towards $\mu=-\infty$.  The intersections of this line segment with phase boundaries in the $(\mu,h,T,U)$ phase diagram correspond to phase boundaries in space, i.e., a shell structure, which will of course be slightly smeared out.
See Figure~\ref{LDAExample}.

\paragraph{\underline{Cooling:}}
We have presented a theoretical discussion of various novel quantum phases of fermions in optical lattices at low temperatures but cooling to such low temperatures is a major experimental challenge \cite{mckay2011review}.
Although we discussed repulsive and attractive Hubbard models on an equal footing, the actual experimental situation is quite asymmetric.  
In the continuum, Fermi superfluids are relatively easy to produce and equilibrate because the fermion mass transport is fast and because heat can be efficiently transferred out of the system (e.g., into a boson bath in a sympathetic cooling setup).
Ramping up an optical lattice imposed on a Fermi gas naturally leads toward a superfluid ground state of the attractive Hubbard model.
However, the insulating antiferromagnetic phases of the repulsive Hubbard model are much more difficult to realize because of the slower timescales associated with the mass transport and the superexchange required for equilibration.
Fermionic ${}^{40}$K atoms in optical lattices have been cooled to entropies of about $1 k_B$/atom \cite{ulrichSchneiderThesis}, which is still signifcantly higher than the $\sim0.42 k_B$ needed.
An effort is underway with ${}^{6}$Li atoms to exploit automatic entropy redistribution in a dimpled trap \cite{bernier2009,deleo2011,paiva2011,fuchs2011} and thereby lower the entropy in the center of the trap \cite{mathy2012}, which is hoped to be sufficient for antiferromagnetism.

\paragraph{\underline{Detection:}}
To make connections between theory and experiment, the simplest observable is the real-space density profile $n(\rrr)$, inferred from shadow imaging or fluorescence imaging.  A shell structure will show up as kinks in the density profile.  Measurements of density profiles of a quasi-1D imbalanced Fermi gas are indeed consistent with the phase diagram obtained using the Bethe ansatz \cite{liao2010}, which lends confidence to other predictions of the theory such as the existence of a phase with FFLO correlations.  Using large-aperture optical microscopy it is even possible to measure densities with single-site resolution \cite{bakr2010}.
A complementary observable is the momentum distribution $n(\kkk)$, which can be inferred by turning off the trap potential and the interactions (ramping the magnetic field to a value such that the scattering length is zero), allowing the atom cloud to expand freely for a sufficiently long time-of-flight, and then imaging it.
Spectral functions $A_\sigma(\kkk,E)$ and local densities of states $n_\sigma(\rrr,E)$ can also be inferred using radiofrequency spectroscopy
~\cite{jin2008,ketterle2008,zwierlein2009} -- applying a rf pulse to transfer either species of fermions ($\ket{\up}$ or $\ket{\dn}$) into a third hyperfine state if they have the correct energy for resonance.  
Besides single-fermion properties,
one can also study two-fermion properties and collective modes by measuring
density, spin, or pairing correlations.  
Ways of measuring correlation functions include Bragg scattering (elastic or inelastic scattering of laser light from a periodic array of atoms) \cite{corcovilos2010}, noise correlation methods (performing the same experiment millions of times and accumulating spatial autocorrelations of images), and atom interferometry \cite{swanson2012}.  
An example of the latter is illustrated in Figure~\ref{MasonsFig2}. 
Two Fermi superfluid layers are ramped to the BEC side of the Feshbach resonance to ``shrink'' Cooper pairs into molecules while leaving spatial variations in the pairing amplitude roughly unchanged.  The resulting BECs are then allowed to expand and interfere.  If one of the initial superfluids contains pairing amplitude modulations due to long-range or short-range LO order, the interference fringes form a tire-tread pattern.

The ingenuity of cold atom researchers seems unlimited; there are a large number of other tools that we have no space to review.

\section{Concluding remarks}
In this article we have focused on the equilibrium thermodynamics of bosons and fermions in optical lattices in the single-band Hubbard regime, with an emphasis on interesting magnetic, superfluid, and spin liquid ground states.
The quantum Monte Carlo methods that we have discussed for bosons and fermions are quite different in construction and implementation. For the Bose Hubbard model we described the state-of-the-art worm algorithm,
a path integral formulation of boson world lines in space and imaginary time that can simulate ``exactly", with only statistical error, on the order of a million  
bosons in an optical lattice with an overall confining trap potential.
We showed how the energy, entropy, compressibility, superfluid density and momentum distribution 
can be calculated within the phases and in the quantum critical region. 

For the Fermi Hubbard model we discussed the quantum Monte Carlo simulations using the Hubbard-Stratonovich auxiliary field method. At half-filling, the particle-hole symmetry renders this method numerically ``exact'', while away from half filling there is a ``sign problem'' that can however be controlled down to temperatures a fraction of the bandwidth. We showed that for a general Hamiltonian, the finite temperature phase diagram can be determined directly from experiments using only the density profile in the trap as the input. The density profile essentially generates a chemical potential scan of the phase diagram from a single measurement. Kinks or singularities in the local compressibility, arising from critical fluctuations, demarcate the boundaries between superfluid and normal phases in the trap. 

For both the Bose and Fermi Hubbard model we also discussed mean field theory, variational approaches and perturbation theory that complement the quantum Monte Carlo methods. We hope the computational methods, phase diagrams and quantum phase transitions, and the variety of detection methods can guide future investigations of other 
interesting quantum many body Hamiltonians. 
We also hope that this chapter will provide common ground for readers branching out into rapidly evolving topics such as shallow lattices,
disorder potentials,
lattices for emulating topological insulators,
dipolar molecules confined to parallel planes,
nonequilibrium phenomena \cite{ulrichSchneiderThesis},
and so on.


We acknowledge support from NSF Grant DMR-0907275
and from DARPA Grant 60025344 under the Optical Lattice Emulator program.
We also acknowledge computational resources from the Ohio Supercomputer Center.


\end{document}